\documentclass[lettersize,journal]{IEEEtran}
\usepackage{amsmath,amssymb,amsfonts}
\usepackage{algorithm}
\usepackage{array}
\usepackage[caption=false,font=normalsize,labelfont=sf,textfont=sf]{subfig}
\usepackage{textcomp}
\usepackage{stfloats}
\usepackage{url}
\usepackage{verbatim}
\usepackage{graphicx}
\usepackage{cite}
\usepackage{multirow}
\usepackage{mdwmath}

\DeclareMathOperator*{\argmax}{argmax}
\DeclareMathOperator*{\argmin}{argmin}

\DeclareMathOperator{\diag}{diag}

\newcommand{\Cfinal}{\operatorname{\mathbf{C}}_{\mathrm{final}}}
\newcommand{\Cqam}{\operatorname{\mathbf{C}}_{\mathrm{QAM}}}
\newcommand{\Phiqam}{\operatorname{\mathbf{\Phi}}_{\mathrm{QAM}}}
\newcommand{\euler}{e}
\usepackage{algpseudocode}
\DeclareMathOperator{\histo}{histogram}
\DeclareMathOperator{\var}{variance}
\newcommand{\vr}{var}
\DeclareMathOperator{\bound}{boundary}
\DeclareMathOperator{\peri}{perimeter}
\DeclareMathOperator{\ar}{area}
\DeclareMathOperator{\circu}{circularity}
\newcommand{\imagi}{j}

\begin{document}

\title{Pilotless Uplink for Massive MIMO Systems}

\author{\IEEEauthorblockN{P Aswathylakshmi and Radha Krishna Ganti\\}
    \IEEEauthorblockA{Department of Electrical Engineering\\
                      Indian Institute of Technology Madras\\
                      Chennai, India 600036\\
                      \{aswathylakshmi, rganti\}@ee.iitm.ac.in}
\thanks{A preliminary version of this work was presented at the IEEE Global Communications Conference (Globecom) 2023 \cite{aswathylakshmi2023pilotless}.}}



\maketitle

\begin{abstract}
Massive MIMO OFDM waveforms help support a large number of users in the same time-frequency resource and also provide significant array gain for uplink reception in cellular systems.  However,  channel estimation in such large antenna systems can be tricky as pilot assignment for multiple users becomes more challenging with increasing number of users. Additionally, the pilot overhead especially for wideband rapidly changing channels can diminish the system throughput quite significantly. In this paper, we propose an iterative matrix decomposition algorithm for the blind demodulation of massive MIMO OFDM signals without using any pilots. This new decomposition technique provides estimates of both the user symbols and the user channel in the frequency domain simultaneously (to a scaling factor) without any pilots. We discuss methods for finding the appropriate initial points for the algorithm that ensure its convergence in different types of wireless channels. We also propose new methods for resolving the scaling factor in the estimated signal that do not increase pilot overhead. We show how the method can be adapted to both single-user and multi-user systems. Simulation results demonstrate that the lack of pilots does not affect the error performance of the proposed algorithm when compared to the conventional pilot-based channel estimation and equalization methods across a wide range of channels for both single and multi-user cases. We also demonstrate techniques to reduce the complexity of the estimation algorithm over multiple OFDM symbols in a 5G MIMO system by leveraging the temporal correlations in the channel.
\end{abstract}

\begin{IEEEkeywords}
Pilotless communication, massive MIMO, alternating minimization, iterative matrix decomposition, blind estimation.
\end{IEEEkeywords}

\section{Introduction}\label{intro}

Massive multiple-input-multiple-output (MIMO) orthogonal frequency division multiplexing (OFDM) systems use a large number of antennas at the base station to multiplex multiple users on the same time-frequency resources \cite{bjornson2017}. This maximises spectrum usage while providing very high data rates in 5G new radio (NR) and 6G \cite{nextG}. The successful implementation of such multi-user massive MIMO relies on the accurate estimation of the channel of each user \cite{bjornson2017}. In 5G NR, channel estimation is accomplished using pilots, known to both the transmitter and the receiver, which are assigned to predesignated subcarriers \cite{38101}. The symbols in these subcarriers do not carry information and therefore do not contribute to the system throughput. Since the pilot sequences assigned to simultaneously scheduled users must be orthogonal to each other to avoid interference, longer pilot sequences become necessary as the number of users supported by the massive MIMO system increases. Moreover, in order to avoid pilot-data interference, subcarriers assigned to pilots for one user cannot be assigned as data subcarrier to another user simultaneously. These factors make pilot assignment in large multi-user systems challenging, as well as impact the system throughput significantly.

\par{Various strategies for reducing pilot overhead have been explored in \cite{simeone2004adaptive}, \cite{vsimko2013design}, \cite{tomasi2015pilot}, and \cite{ksairi2016pilot}. Adaptive pilot patterns depending on users' Quality-of-Service (QoS) requirements or delay spread of the channel are proposed in \cite{simeone2004adaptive} and \cite{vsimko2013design}, respectively. However, they do not consider the interference from simultaneously scheduled users in a multi-user scenario. For a multi-user setting, a pilot length minimization algorithm is proposed in \cite{tomasi2015pilot} that makes use of the spatial correlation between the users' channels, but this requires knowledge of user channel covariance matrices. A pilot pattern adaptation scheme for users grouped and scheduled together on the basis of similar channel conditions is introduced in \cite{ksairi2016pilot}. However, this imposes constraints on the scheduler and needs second order statistics of all the users at the base station. }

\par{Blind demodulation techniques avoid these overhead and scheduling issues by eliminating the very need of pilots for channel estimation. A blind amplitude and phase estimation technique is proposed for radio surveillance in \cite{leblind} that detects several blocks of OFDM symbols using the cyclic prefix (CP). It estimates the modulation order by assuming it is constant across several OFDM symbols and calculating the fourth order cumulants. It then uses this information to perform modulation stripping and calculate the phase offset. It estimates the channel by assuming that the transmitted constellation exhibits unit variance on each subcarrier across several OFDM symbols. The technique  works well for line-of-sight (LOS) and strongly correlated channels but becomes unstable for non-LOS (NLOS) channels. An analytical multimodulus algorithm is proposed in \cite{daumont2010analytical} for blind demodulation in time-varying MIMO channels. In this technique, a batch analysis needs to be performed at the end of a blind source separation algorithm in order to resolve phase ambiguities and source permutations. A blind detection algorithm modeled for non-selective/flat-fading that exploits the sparsity of MIMO channels is presented in \cite{zhang2017blind}. It factorizes the received signal matrix into transmitted signal and channel by formulating it as a dictionary learning problem and uses a modified approximate message passing algorithm to solve it. However, the method cannot be generalized to MIMO channels with frequency selectivity/multi-path. It also requires knowledge of prior distribution of the transmitted symbols and some pilots to resolve phase and permutation ambiguities.}

\par{A matrix decomposition technique  is proposed in \cite{aswathy2023ojcoms} for fronthaul compression in massive MIMO systems, where a matrix of received signals is modelled as the product of a user signal matrix and a low rank channel matrix, which are iteratively estimated using an alternating minimization algorithm. However, unlike \cite{zhang2017blind}, the method in \cite{aswathy2023ojcoms} does not require knowledge of any prior distribution of the channel. In this paper, we propose an algorithm, inspired by \cite{aswathy2023ojcoms}, that performs blind estimation of OFDM symbols using an alternating minimization framework.}

\par{The idea of using an alternating minimization approach for blind estimation was first explored in \cite{you1995blind}. It carries out data estimation and channel identification alternately for binary pulse amplitude modulated (PAM) signals in a single carrier band-limited channel. It uses the steepest descent algorithm to minimize the objective function in each step. Although the complexity of the algorithm is low, it suffers a 3-5dB SNR loss in error performance when compared to the Viterbi algorithm for maximum likelihood sequence estimation (MLSE). The method becomes too complex for other higher order modulation types such as quadrature amplitude modulation (QAM) and cannot be scaled easily for highly frequency-selective channels or for MIMO-OFDM systems. A semi-blind detection algorithm for massive MIMO systems is introduced in \cite{liang2019semi} which uses regularized alternating least squares for a low rank matrix completion model of the problem. However, it assumes knowledge of the large-scale fading characteristics of the channel and is modeled for and applicable only to a single-tap/flat-fading channel.}

\par{The method we propose in this paper offers the following advantages over the existing blind estimation methods:
\begin{enumerate}
    \item Provides estimates of both the user signal and the channel simultaneously. Therefore, the channel estimated using the algorithm for one OFDM symbol can be used for subsequent symbols, depending on the coherence time of the channel. 
    \item Requires only a single reference signal/pilot to resolve scaling and phase ambiguities of the user signal and channel estimates, irrespective of the length of the OFDM symbol, thus improving the goodput of the system.
    \item Does not require any knowledge of the signal or channel characteristics at the receiver.
    \item Can be applied to higher QAM modulation orders.
    \item Designed for use in frequency-selective/multi-path fading channels.
    \item Can simultaneously decode several users with different multi-path fading profiles.
\end{enumerate}
}

In the following sections, we explain how matrix decomposition algorithm of the received signal matrix can be adapted to meet the objectives of this blind estimation problem, which are different from those of the fronthaul compression problem in \cite{aswathy2023ojcoms}. We pay special attention to the numerical stability and convergence of the algorithm, which were not a concern for the compression case but essential for blind estimation. We propose modifications to the signal matrix decomposition algorithm that ensure that these new criteria are met. We also discuss the choice and calculation of an initial point for the algorithm ensuring convergence which is non-trivial, unlike in the compression case.

The organization of the paper is as follows: Section \ref{sysmod} outlines the signal model for the massive MIMO OFDM system, Section \ref{MD} describes the proposed pilot-free  receiver that uses matrix decomposition to blindly estimate the transmitted signal and channel for a single user, including algorithms for finding the appropriate initial point and for the resolution of scaling/phase ambiguities, Section \ref{BER1user} presents the error performance results of the link-level simulations for a single user, Section \ref{MU-MIMO} extends the method to multi-user scenarios with a discussion of its error performance in Section \ref{BER_MU}, and Section \ref{Gain} quantifies the spectral gains from the proposed method compared to the conventional pilot-based system. In Section \ref{Tempcorr} we analyze the performance of the method over time, across multiple OFDM symbols, and discuss how temporal correlations in the channel can be leveraged to reduce the algorithm complexity, which is derived in Section \ref{complexity}. Finally, Section \ref{concl} concludes the paper.

\section{System Model}\label{sysmod}

In this paper, we consider the signals received at a massive MIMO base station (BS) with $N_r$ antennas, i.e., the uplink. In an OFDM system, each user maps its bit-stream to an $M$-QAM constellation, embeds them onto its allocated subcarriers, performs inverse fast Fourier transform (IFFT) and adds the cyclic prefix (CP) before transmission \cite{38101}. We assume the channel between each user and the BS has a maximum of $L$ significant resolvable multi-path taps. At the receiver, after sampling and CP removal, the signal received at antenna $r$ at a sampling instant $n$ is given by
\begin{equation*}
y_{r}[n] =  x[n] \circledast h_{r}[n] + w_{r}[n],  
\end{equation*}
where $x[n]$ is the  time domain OFDM symbol transmitted by the user, $h_{r}[n]$ is the multi-path channel response between the  user  and  the  BS antenna $r$, $\circledast$ represents circular convolution resulting from the CP in OFDM, and $w_r$ is the additive white circularly symmetric complex Gaussian noise (AWGN) at antenna $r$ with variance $\sigma^2$. 

\par{We consider one OFDM symbol of length $N$  (after removal of the CP) received across the $N_r$ antennas and arrange the received IQ symbols into the $N\times N_r$ matrix,
\begin{equation*}
\mathbf{Y}=
  \begin{bmatrix}
   y_{1}[1] & y_{2}[1] & . & . & . & y_{N_r}[1] \\
    y_{1}[2] & y_{2}[2] & . & . & . & y_{N_r}[2]\\
    . & . & .  &   &   &. \\
    . & . &   &  . &   &. \\
    . & . &   &   & .  &. \\
    y_{1}[N] & y_{2}[N] & . & . & . & y_{N_r}[N]\\
  \end{bmatrix}_{N\times N_r} .
\end{equation*}
Here, each column of $\mathbf{Y}$ represents the received signal at an antenna over $N$ sampling instants. The goal is to blindly estimate the symbols/bit-stream of each user from this received signal matrix $\mathbf{Y}$.}  

\section{pilot-free Receiver using Matrix Decomposition}\label{MD}

In this paper, we propose a method to decompose the received signal matrix, $\mathbf{Y}$ into a user signal component and a channel component without the use of pilot symbols for estimating the channel. This decomposition is based on the low rank signal model for $\mathbf{Y}$ discussed in \cite{aswathy2023ojcoms}. 

\subsection{Low Rank Signal Model for Single User/Stream}\label{lowrank}

We begin our analysis with the single user/data stream case, where only one single-antenna user is allocated all the $N$ subcarriers in the OFDM symbol for uplink transmission. This model will later be extended to the multi-user MIMO (MU-MIMO) case, where multiple users transmit their data streams simultaneously using the same subcarriers. 

\par{Taking FFT of the matrix $\mathbf{Y}$ across each column, we can express $\mathbf{Y}$ in the frequency domain as  
\begin{equation}
    \mathbf{Y_f} =\mathbf{X_f}\mathbf{H_f} + \mathbf{W_f},
\label{eq.1}
\end{equation}
where $\mathbf{X_f}$ is the $N\times N$ diagonal matrix with the $N$-length frequency-domain user data (which are the $M$-QAM symbols) as its diagonal, $\mathbf{H_f}$ denotes the channel in the frequency domain, and $\mathbf{W_f}$ is the noise in the frequency domain. The frequency-domain channel, $\mathbf{H_f}$ is the Fourier transform of the time-domain channel for the $N_r$ antennas. If we assume that there are at most $L$ significant multi-paths for all the MIMO channels, then the time-domain channel matrix, denoted as $\mathbf{H_t}$ has only $L$ non-zero rows corresponding to these paths. Therefore, $\mathbf{H_f}$ can be compactly expressed as the product of $L$ columns of the $N\times N$ Discrete Fourier Transform (DFT) matrix, $\mathbf{F_L}$ and the $L\times Nr$ time-domain channel matrix, $\mathbf{H_t}$, i.e., $\mathbf{H_f} = \mathbf{F_{L}H_t}$. Thus, we arrive at the low rank signal model for $\mathbf{Y_f}$ as
\begin{equation}
\mathbf{Y_f} = \mathbf{X_{f}F_{L}H_{t}} + \mathbf{W_f}, 
\label{eq.2}\end{equation}
where $\mathbf{F_L}$ denotes the columns of the $N\times N$ DFT matrix corresponding to the delay of the $L$ multi-paths in units of the sampling time.}

\par{Here, we note that the number of significant multi-paths is typically in the order of tens of taps, even in rich scattering environments, since long paths and/or multiple reflections result in very high attenuation. This is much lower than $N$, the length of the OFDM symbol/FFT (which is of the order of thousands in 5G and 6G \cite{38101}).  It would also be lower than the number of antennas, $N_r$ in a massive MIMO setting with hundreds of antennas at the base station. Therefore, the above signal model suggests that while the rank of the received signal matrix, $\mathbf{Y_f}$ may be $N_r$, the signal component of interest, $\mathbf{X_{f}F_{L}H_{t}}$ in it resides in a much smaller subspace of dimension $L$. This allows us to estimate the correct signal subspace iteratively from $\mathbf{Y_f}$ using the method described in the next section, which also leads to elimination of the noise in the other $N_r-L$ dimensions.}

\subsection{Blind Estimation of User symbols and Channel}\label{blind}

In a conventional system, pilots inserted in $\mathbf{X_f}$ are used to estimate the frequency-domain channel, $\mathbf{H_f} $ from $ \mathbf{Y_f}$. Then these channel estimates can be used to decode the data $\mathbf{X_f}$ from $\mathbf{Y_f}$ in subsequent transmission.

\par{In this paper, we obtain the estimates of $\mathbf{X_f}$ and $\mathbf{H_t}$ without any pilots. These estimates $\mathbf{\hat{X}}$ and $\mathbf{\hat{H}}$ can be obtained by solving
\begin{equation}
    (\mathbf{\hat{X}}, \mathbf{\hat{H}}) = \argmin_{\substack{\mathbf{X}:N\times N \text{ diagonal}; \\ \mathbf{H}:L\times N_r}} \Vert \mathbf{Y_f} - \mathbf{X}\mathbf{{F_{L}H}} \Vert^{2}_{F}.
    \label{eq.3}
\end{equation}
This is a non-convex problem in $(\mathbf{X},\mathbf{{H}})$, but convex in individual variables. Alternating minimization technique can simplify the problem in \eqref{eq.3} by solving for one variable at a time using the previous iterate for the second variable. We propose the following iterative algorithm that begins with an initial guess for $\mathbf{\hat{X}}$ and use it to find
\begin{equation}
\mathbf{\hat{H}} = \argmin_{\mathbf{H}} \Vert \mathbf{Y_f} - \mathbf{\hat{X}}\mathbf{{F_{L}{H}}} \Vert^{2}_{F}.
 \label{eq.4}
\end{equation}
We use the linear least squares solution to \eqref{eq.4} to estimate $\mathbf{H}$ as
\begin{equation}
    \mathbf{\hat{H}} = (\mathbf{\hat{X}F_{L}})^{\dagger}\mathbf{Y_f},
    \label{eq.5}
\end{equation}
where $\dagger$ denotes the Moore-Penrose inverse.
The $\mathbf{\hat{H}}$ found in \eqref{eq.5} is then used to solve
\begin{equation}
    \mathbf{\hat{X}} = \argmin_{\mathbf{X}} \Vert \mathbf{Y_f} - \mathbf{X F_{L}\hat{H}} \Vert^{2}_{F}.
    \label{eq.6}
\end{equation}
Let $\mathbf{B = F_{L}\hat{H}}$, and $\mathbf{a_{n}^{T}}$ denote the $n-$th row of the matrix $\mathbf{A}$. Using the fact that $\mathbf{\hat{X}}$ needs to be diagonal, \eqref{eq.6} is simplified as
\begin{align}
    \mathbf{\hat{X}} &= \argmin_{\mathbf{X}} \Sigma_{n=1}^{N} \Vert \mathbf{y_{n}^{T}} - x(n)\mathbf{b_{n}^{T}} \Vert^{2}, \\
    &= \argmin_{\mathbf{X}} \Sigma_{n=1}^{N} \Sigma_{r=1}^{N_r} |y(n,r) - x(n)b(n,r)|^{2}, 
    \label{eq.7&8}
\end{align}
where $x(n)$ denotes the $n-$th diagonal element of $\mathbf{X}$, $y(n,r)$ is the element at row $n$ and column $r$ of $\mathbf{Y_f}$, $b(n,r)$ is the element at row $n$ and column $r$ of $\mathbf{B}$. This gives us the solution
\begin{equation}
    \hat{x}(n) = \frac{\Big(\sum_{r=1}^{N_r}y(n,r)b^{*}(n,r)\Big)}{\Big(\sum_{r=1}^{N_r}|b(n,r)|^2\Big)},
    \label{eq.9}
\end{equation}
where $\hat{x}(n)$ denotes the $n-$th diagonal element of $\mathbf{\hat{X}}$ and $b^{*}(n,r)$ denotes the complex conjugate of $b(n,r)$. We observe that equation \eqref{eq.5} is channel estimation and equation \eqref{eq.9} is maximal-ratio-combining (MRC). This technique is then iterated till the solution converges or for a fixed number of iterations.}

\par{The above alternating minimization algorithm with solutions given by \eqref{eq.5} and \eqref{eq.9} is similar to the fronthaul compression algorithm proposed in \cite{aswathy2023ojcoms}. However, the algorithm in \cite{aswathy2023ojcoms} can begin with any initial point since convergence of the estimate $\mathbf{\hat{X}}$ to the actual transmitted $\mathbf{X_f}$ is not necessary for the compression case, whereas for a blind estimation algorithm, this convergence is paramount. Therefore, the above algorithm has to satisfy certain initial point conditions and numerical stability conditions for convergence. The Lemma 1 in \cite{aswathy2023ojcoms} also proves that the solutions, \{$\mathbf{\hat{H},\hat{X}}$\} of the alternating minimization obtained through \eqref{eq.5} and \eqref{eq.9} are related to the true channel, $\mathbf{H_t}$ and the true user signal, $\mathbf{X_f}$, respectively, by a complex scaling factor, i.e., 
\begin{equation}
    \mathbf{\hat{X}} = \Big(\frac{1}{\lambda}\Big)\mathbf{X_f}, \hspace{3mm} \text{and} \hspace{3mm} \mathbf{\hat{H}} = \lambda \mathbf{H_t},
\end{equation}
where $\lambda \in \mathbb{C}$, the set of complex numbers. Therefore, we also need to resolve this scaling/rotation ambiguity introduced by $\lambda$ in the estimated signal and channel as a result of the alternating minimization. In the following sections, we propose additional modules and modifications to this alternating minimization framework that satisfy these conditions in order to solve the blind estimation problem.}

\subsection{Finding the Initial Point}\label{initpoint}

The alternating minimization algorithm can begin with an initial guess for either $\mathbf{\hat{X}}$ or $\mathbf{\hat{H}}$; it then solves for the other using \eqref{eq.5} or \eqref{eq.9}, respectively. If the initial point of the algorithm is orthogonal or nearly orthogonal to the quantity being estimated, the algorithm will have difficulty converging to $\mathbf{X_f}$ or $\mathbf{H_t}$, as no amount of scaling or rotation can yield a vector in the orthogonal subspace \cite{AMconvergence2015}. Since the quantity being estimated here is $\mathbf{X_f}$, we need to find an initial point for $\mathbf{\hat{X}}$ which is non-orthogonal to $\mathbf{X_f}$. The only information available in this blind estimation scenario is the received signal matrix $\mathbf{Y_f}$; therefore, we examine the relationship between $\mathbf{X_f}$ and $\mathbf{Y_f}$ to find the appropriate initial point for the algorithm.

We expand the signal part of \eqref{eq.1} in terms of the components in $\mathbf{X_f}$ as follows, using the fact that $\mathbf{X_f}$ is a diagonal matrix:
\begin{align}
    \mathbf{Y_f} &= \mathbf{X_{f}H_{f} + W_{f}}, \\
    &= \begin{bmatrix}
        x(1)\mathbf{h_{1}^{T}(f)} \\ x(2)\mathbf{h_{2}^{T}(f)} \\
        \vdots \\ x(N)\mathbf{h_{N}^{T}(f)}
    \end{bmatrix} + \mathbf{W_f}, \label{Yfcol}
\end{align}
where $x(i)$ is the $i$-th diagonal element of $\mathbf{X_f}$ and $\mathbf{h_{i}^{T}(f)}$ is the $i$-th row of the frequency-domain channel matrix $\mathbf{H_f}$, for $i = 1, 2, \hdots, N$. We observe from equation \eqref{Yfcol} that each column of the signal part of $\mathbf{Y_f}$ is multiplied by the vector of user symbols in the diagonal of $\mathbf{X_f}$, therefore, the latter can be spanned by the column space of $\mathbf{Y_f}$. Since the left singular vectors of a matrix provide a basis for its column space, we inspect the singular value decomposition (SVD) of $\mathbf{Y_f}$ to devise a good initial point for the algorithm. 

\par{Let the SVD of $\mathbf{Y_f}$ be
\begin{equation}
    \mathbf{Y_f} = \mathbf{U\Sigma V^{H}},
    \label{eq.10}
\end{equation}
where $\mathbf{U}$ is the $N\times N$ orthogonal matrix of left singular vectors of $\mathbf{Y_f}$, $\mathbf{\Sigma}$ is the $N\times N_r$ truncated diagonal matrix of the corresponding singular values arranged in decreasing order, $\mathbf{V}$ is the $N_{r}\times N_{r}$ orthogonal matrix of the corresponding right singular vectors, and $\mathbf{V^{H}}$ denotes the conjugate transpose of $\mathbf{V}$. In this work, we focus on finding an initial point for $\mathbf{\hat{X}}$ to begin the alternating minimization algorithm, therefore we examine the relationship between $\mathbf{X_f}$ and the left singular vectors in $\mathbf{U}$. Specifically, we are interested in the top singular vector in $\mathbf{U}$, which is the direction of maximum variance in the data and therefore captures the most information.}

\par{Suppose the average powers of the $L$ taps of the time-domain channel are $\rho_{1}, \rho_{2}, ..., \rho_{L}$. Then the time-domain channel, $\mathbf{H_t}$ can be expressed as
\begin{equation}
    \mathbf{H_t} =  {\scriptstyle\begin{bmatrix}
                    \sqrt{\rho_{1}} & & \\ 
                    & \ddots & \\
                    & & \sqrt{\rho_{L}} \\
                    \end{bmatrix}
    \begin{bmatrix}
    h_{1,1} & h_{1,2} & \hdots & h_{1,N_r} \\
    h_{2,1} & h_{2,2} & \hdots & h_{2,N_r}\\
    \vdots & \vdots & \ddots & \vdots \\
    h_{L,1} & h_{L,2} & \hdots & h_{L,N_r}\\
  \end{bmatrix}_{L\times N_r}},
  \label{eq.11}
\end{equation}
where $h_{i,j}$ is the channel coefficient for multi-path $i$ and antenna $j$, representing the small-scale fading. Then the matrix $\mathbf{Y_f}$ can be expressed as 
\begin{align}
    \mathbf{Y_f} &= \mathbf{X_{f}F_{L}H_{t} + W_{f}} \\
    &= \mathbf{X_{f}}
    {\scriptstyle\begin{bmatrix}
    \mathbf{f_1} & \mathbf{f_2} & \hdots \mathbf{f_L}
    \end{bmatrix}
    \begin{bmatrix}
    \sqrt{\rho_{1}} & & \\ & \ddots & \\ & & \sqrt{\rho_{L}} \\
    \end{bmatrix}
    \begin{bmatrix}
    \mathbf{h_1} \\ \mathbf{h_2} \\ \vdots \\ \mathbf{h_L} \\
    \end{bmatrix}} + \mathbf{W_f}\\
    &= \mathbf{X_{f}} {\scriptstyle\begin{bmatrix}
    \sqrt{\rho_{1}}\mathbf{f_{1}h_{1}} + \sqrt{\rho_{2}}\mathbf{f_{2}h_{2}} + \hdots + \sqrt{\rho_{L}}\mathbf{f_{L}h_{L}}
    \end{bmatrix}} + \mathbf{W_f}, 
    \label{eq.15new}
\end{align}
where $\mathbf{f_i}$ (of dimension $N\times 1$) denotes column $i$ of $\mathbf{F_{L}}$, and $\mathbf{h_i} = [h_{i,1} \hspace{1mm} h_{i,2} \hspace{1mm} \hdots \hspace{1mm} h_{i,N_r}]$ for $i = 1, 2, ..., L$. If we assume that the channel taps are uncorrelated to each other, then equation \eqref{eq.15new} expresses the frequency-domain channel, $\mathbf{H_{f} = F_{L}H_{t}}$ as a weighted summation of uncorrelated rank-1 $N\times N_r$ matrices, $\mathbf{f_{i}h_{i}}$. 
When we take the SVD of $\mathbf{Y_f}$, the top singular value and the corresponding left and right singular vectors together ($\mathbf{u_{1}\sigma_{1}v^{H}_{1}}$) capture the rank-1 $N\times N_r$ matrix that is closest to $\mathbf{Y_f}$. Equation \eqref{eq.15new} also allows us to view $\mathbf{Y_f}$ as a weighted average of the data $\mathbf{X_f}$ distorted by the channel in different directions by the matrices $\mathbf{f_{i}h_{i}}$, weighted by the corresponding tap powers $\rho_{i}$, for $i = 1, 2, ..., L$.  If the time-domain channel has one clearly dominant tap (path) $k$ that captures the majority of the channel energy, i.e., 
\begin{equation*}
    \rho_{k} \gg \rho_{i}, \hspace{2mm} i = 1, 2, ..., L, i \neq k,
\end{equation*} 
then $\mathbf{u_{1}\sigma_{1}v^{H}_{1}}$ will be heavily skewed towards the rank-1 matrix $\sqrt{\rho_k}\mathbf{X_{f}f_{k}h_{k}}$. This allows us to make the following approximation for the top left singular vector, $\mathbf{u_1}$:
\begin{equation}
    \mathbf{u_1} \approx \alpha_{k}\sqrt{\rho_k}\mathbf{X_{f}f_{k}},
    \label{eq.13}
\end{equation}
where $\alpha_{k} \in \mathbb{C}$. If there are multiple dominant taps with similar powers, then $\mathbf{u_1}$ will be approximately equal to the average of these directions, weighted by the corresponding tap powers.}

\par{By simulation, we observe that the alternating minimization algorithm converges if the initial point satisfies the following condition:
\begin{itemize}
    \item $\langle\mathbf{X_{f}, \hat{X}}\rangle \neq 0 $: $\mathbf{\hat{X}}$ should not be orthogonal to $\mathbf{X_f}$, since the estimators in the algorithm do not yield a vector in the subspace orthogonal to the initial point. 
\end{itemize}
In the scenario where \eqref{eq.13} holds true, assuming tap $k$ has the highest power, we can construct an initial point for the algorithm as follows
\begin{equation}
    \mathbf{\hat{X}} = \diag(\mathbf{u_{1}\odot f^{*}_{k}}), 
    \label{eq.14}
\end{equation}
where $\odot$ denotes element-wise multiplication and $\mathbf{f^{*}_{k}}$ is the complex conjugate of $\mathbf{f_k}$. Since $\mathbf{f_{k}\odot f^{*}_{k}} = [1 \hspace{1mm} 1 \hspace{1mm} ... \hspace{1mm} 1]^{T}_{1\times N}$, substituting equation \eqref{eq.13} in equation \eqref{eq.14} gives us
\begin{equation}
    \mathbf{\hat{X}} \approx \diag(\alpha_{k}\sqrt{\rho_{k}}\mathbf{X_{f}f_{k}\odot f^{*}_{k}}) \approx \alpha_{k}\sqrt{\rho_{k}}\mathbf{X_f}.
    \label{eq.IP}
\end{equation}}

\par{We observe that such an initial point not only satisfies the non-orthogonality condition for convergence, but it is also close to a complex scaled version of the quantity being estimated ($\mathbf{X_f}$), i.e., $\Vert \mathbf{\hat{X}} - c \mathbf{X_f} \Vert < \delta, c \in \mathbb{C}, \delta > 0 $, which ensures faster convergence. In cases where the wireless channel has one or very few dominant tap(s), i.e., the majority of the power in the channel's power delay profile (PDP) is concentrated in very few taps, using $\mathbf{u_1}$, the top singular vector of $\mathbf{Y_f}$ in equation \eqref{eq.14} will lead to an initial point of the form given in equation \eqref{eq.IP}. However, if the channel power is distributed uniformly over a very large number of taps, then $\mathbf{u_1}$ will be the average of many different directions corresponding to these taps with equal power, and will not lead to the form in equation \eqref{eq.IP}. Most physical channels such as LOS or pedestrian channels, as well as the channel models given by 
the 3GPP fall into the former category \cite{38101}. Therefore, in this work, we focus on these cases so that the initialization in equation \eqref{eq.IP} can be used, and analyze the performance of the blind estimator using the alternating minimization algorithm. We note that channels of the latter type, with a large number of equal power taps, are rare since all the scatterers corresponding to the equal power taps need to have the same physical/reflecting characteristics for this to be possible. In such cases, the initial point constructed using equation \eqref{eq.14} may not approach the approximation given in equation \eqref{eq.IP}, resulting in the algorithm taking too long to converge or not converging.}

\par{Construction of the initial point as given in equation \eqref{eq.14} requires the base station to know which channel tap has the highest relative power as well as the delay of that tap (in units of the sampling time) in order to choose the correct column of $\mathbf{F_L}$ to be used. This can be obtained either from the Physical Random Access Channel (PRACH) where the user's PDP can be estimated, or estimated from the top left singular vector $\mathbf{u_1}$ using one of the following methods:}

\par
\begin{figure}
    \centering
    \includegraphics[width = 0.95\columnwidth]{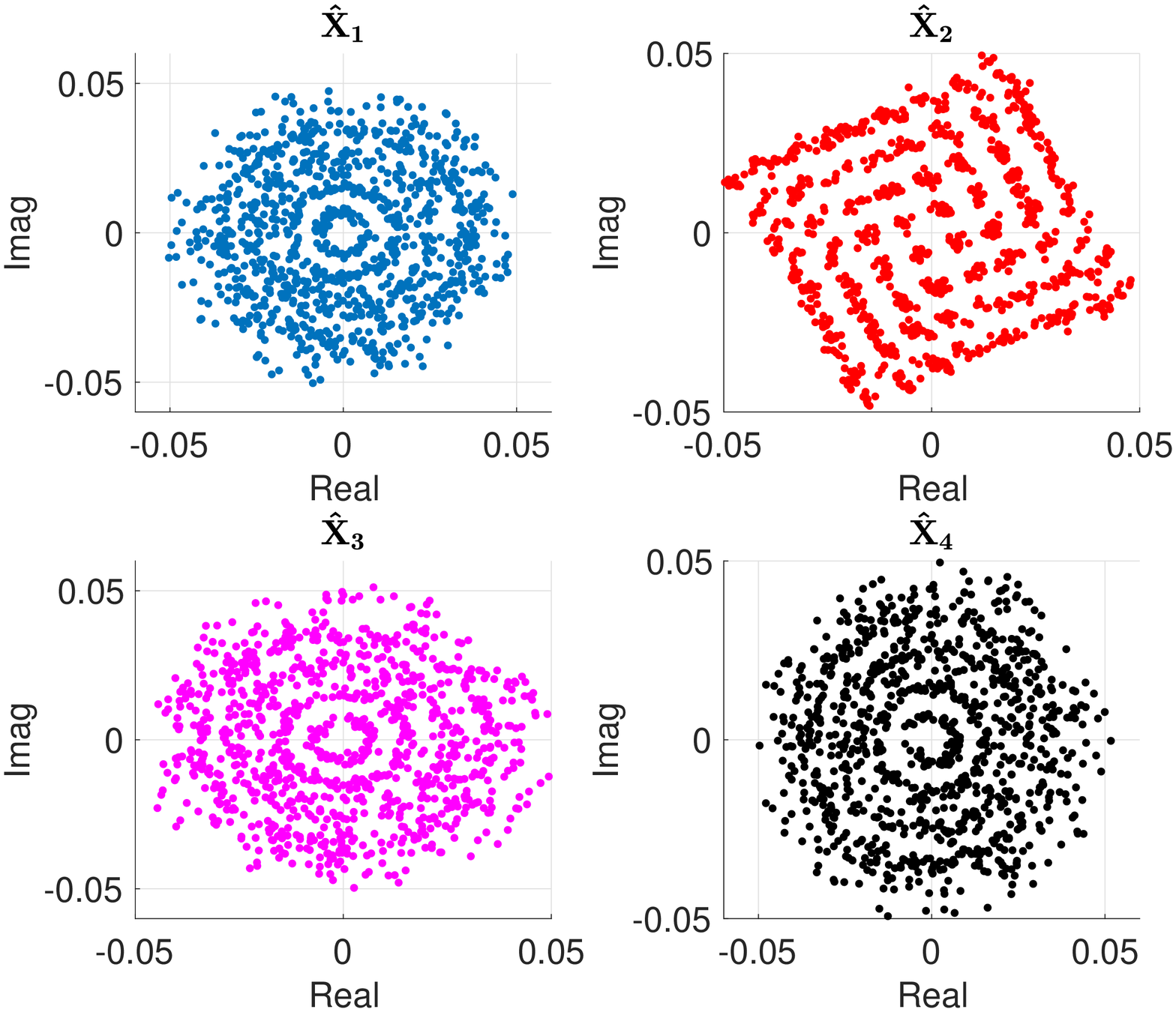}
    \caption{Scatter-plot of potential initial points constructed from $\mathbf{u_1}$ and different columns of the DFT matrix, $\mathbf{F_L}$ as per \eqref{eq.15} for a 4-tap channel with the second tap having the highest relative power. When $\mathbf{u_1}$ is multiplied with the conjugate of the column of $\mathbf{F_L}$ corresponding to the dominant tap (second tap in the above case), the initial point strongly resembles a QAM constellation.}
    \label{Blind_initpts}
\end{figure}

{We construct a set of \textit{potential} initial points, $\mathbf{\hat{X}_{1}, \hat{X}_{2}, \hdots, \hat{X}_{L}}$ as
\begin{equation}
    \mathbf{\hat{X}_i} = \diag(\mathbf{u_{1}\odot f^{*}_{i}}), \hspace{2mm} i = 1, 2, ..., L.
    \label{eq.15}
\end{equation}
Assuming tap $k$ has the highest power, we substitute the approximation of equation \eqref{eq.13} in equation \eqref{eq.15} to obtain
\begin{align}
    \mathbf{\hat{X}_i} &\approx \diag(\alpha_{k}\sqrt{\rho_k}\mathbf{X_{f}f_{k}\odot f^{*}_{i}}) \\ 
    &\approx \diag(\alpha_{k}\sqrt{\rho_k}\mathbf{X_{f}f^{(k-i)}}),
    \label{eq.Xi_hat_approx}
\end{align}
where $\mathbf{f^{(k-i)}}$ denotes the result of the element-wise multiplication $\mathbf{f_{k}\odot f^{*}_{i}}$,
\begin{equation}
    \mathbf{f_{k}\odot f^{*}_{i} = f^{(k-i)}} 
        = \begin{bmatrix}
            1 \\ \omega^{(k-i)} \\ \omega^{2(k-i)} \\ \vdots \\ \omega^{(N-1)(k-i)}
        \end{bmatrix},
\end{equation}
and $\omega = \euler^{-\imagi 2\pi /N}$. When $k \neq i$, the points in $\mathbf{f^{(k-i)}}$ which lie on the unit circle get scaled and rotated by the corresponding components of $\mathbf{X_f}$ in the expression \eqref{eq.Xi_hat_approx}. Therefore the points in $\mathbf{\hat{X}_i}$ will be distributed in concentric circles of various radii around the origin of the Argand plane. However, when $k = i$, $\mathbf{f^{(k-i)}}$ collapses to a vector of ones, resulting in the desired initial point given in equation \eqref{eq.IP}. Thus, the scatter-plot of $\mathbf{\hat{X}_k}$ will strongly resemble the QAM constellation of $\mathbf{X_f}$ (although scaled and rotated). This is illustrated in Fig. \ref{Blind_initpts}.} 

We need to quantify the resemblance of the potential initial points to a QAM constellation in order to pick the correct one. This can be done in multiple ways. One of the simplest methods is to compare the angular distribution of the potential initial points. Since the points in $\mathbf{\hat{X}_i}, i \neq k$ will be distributed in concentric circles around the origin, the angles of these points will be distributed almost uniformly in all directions. However, when resembling a QAM constellation, the points in $\mathbf{\hat{X}_k}$ will be concentrated in certain directions more than the others, resulting in a higher variance for the angular distribution. Therefore, the tap corresponding to the $\mathbf{\hat{X}_i}$ with the highest variance in angular distribution is deemed the dominant tap and used for the initial point of the algorithm. This procedure for estimation of the initial point is summarized in Algorithm \ref{BlindIPvar}. 

\renewcommand\footnoterule{} 
\begin{algorithm}
\caption{Variance-based Initial Point Estimation}\label{BlindIPvar}\footnotetext{$\hat{x}_{i}(n)$ denotes the $n$-th diagonal element of $\mathbf{\hat{X}_i}, \angle \hat{x}_{i}(n)$ denotes the complex angle of $\hat{x}_{i}(n)$.}
\begin{algorithmic}[1]
\State Input $\mathbf{u_{1}, F_{L} = [f_{1} \hspace{1mm} f_{2} \hspace{1mm} \hdots \hspace{1mm} f_{L}]}$.
\For {$i \gets 1$ to $L$}
\State Construct ${\mathbf{\hat{X}_{i}}}\gets \diag(\mathbf{u_{1}\odot f_{i}})$.
\State Calculate angular distribution, \newline 
    $\hspace*{3em} A_{i} \gets \histo\{\angle \hat{x}_{i}(n), n = 1, 2, \hdots, N$\}.
\State Calculate $\vr (i) = \var \{A_{i}\}$
\EndFor
\State $k \gets \argmax \vr$\\
\Return Initial point, $\mathbf{\hat{X} = \hat{X}_{k}}$
\end{algorithmic}
\end{algorithm}

In scenarios where a more robust estimation method is needed, for example in the multi-user case (discussed later) where interference can increase estimation errors, binary image classification-based methods can be applied for finding the dominant tap. For this, we note from the discussion above that the outline of the scatter-plots of $\mathbf{\hat{X}_{i}}$ where $i \neq k$ would resemble circles whereas the outline of the scatter-plot of the correct initial point, $\mathbf{\hat{X}_{k}}$ would resemble a square (for input QAM constellations). Therefore, we can measure the \textit{circularity} of the scatter-plots using the fact that the shape which maximizes area for a given perimeter is the circle. The tap which leads to the least circular scatter-plot in equation \eqref{eq.15} is estimated as the dominant tap. This procedure is summarized in Algorithm \ref{BlindIPcirc}.

\begin{algorithm}
\caption{Circularity-based Initial Point Estimation}\label{BlindIPcirc}\footnotetext{$\hat{x}_{i}(n)$ denotes the $n$-th diagonal element of $\mathbf{\hat{X}_i}, \angle \hat{x}_{i}(n)$ denotes the complex angle of $\hat{x}_{i}(n)$.}
\begin{algorithmic}[1]
\State Input $\mathbf{u_{1}, F_{L} = [f_{1} \hspace{1mm} f_{2} \hspace{1mm} \hdots \hspace{1mm} f_{L}]}$.
\For {$i \gets 1$ to $L$}
 \State Construct $\mathbf{\hat{X}_{i}} \gets \diag(\mathbf{u_{1}\odot f_{i}})$.
 \State Find the boundary points of $\mathbf{\hat{X}_{i}} \rightarrow \bound (\mathbf{\hat{X}_{i}})$.
 \State Calculate $\peri (i)$ from $\bound (\mathbf{\hat{X}_{i}})$.
 \State Calculate $\ar (i)$ enclosed by $\bound (\mathbf{\hat{X}_{i}})$.
 \State Calculate $\circu (i) = \frac{4\pi \ar (i)}{(\peri (i))^2}$.
\EndFor
\State $k \gets \argmin \circu$\\
\Return Initial point, $\mathbf{\hat{X} = \hat{X}_{k}}$
\end{algorithmic}
\end{algorithm}

\begin{figure}
    \centering
    \includegraphics[width = 0.9\columnwidth]{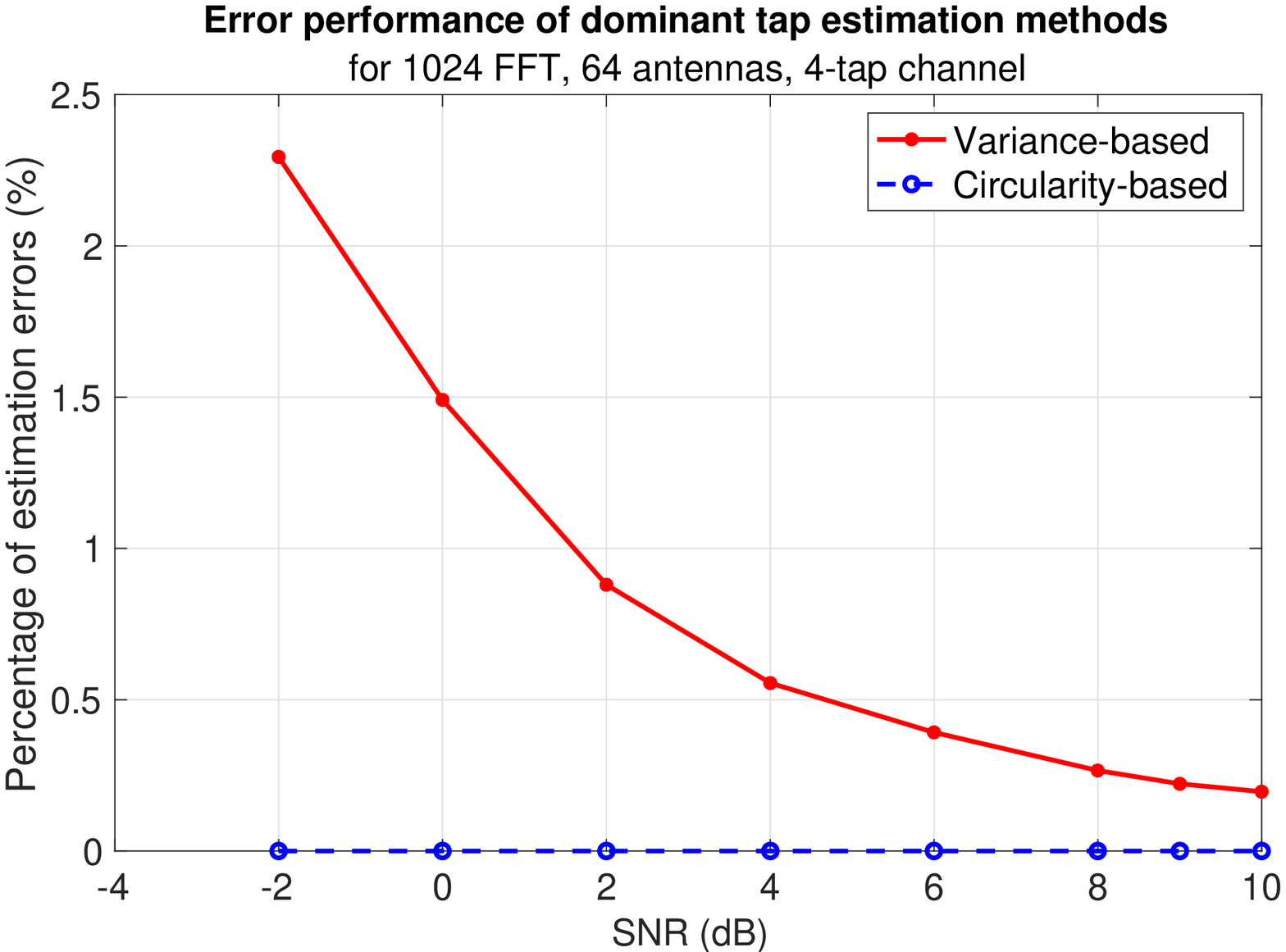}
    \caption{Error performance of the variance-based (Algorithm \ref{BlindIPvar}) and circularity-based (Algorithm \ref{BlindIPcirc}) initial point estimation methods for a single user.}
    \label{Blind_domtap_err}
\end{figure}

A comparison of the performance of the above two dominant tap estimation methods is shown in Fig. \ref{Blind_domtap_err}, where we plot the percentage of estimation errors with respect to the signal-to-noise ratio (SNR). Although the variance-based estimation method performs worse than the circularity-based method, the former is much simpler to implement than the latter. Therefore, we use the former for the single user case where we have to deal with only noise in the estimation, and the latter for the multi-user case, where a more robust method is needed due to the presence of both noise and interference. 

\subsection{Regularization for Numerical Stability}\label{reg}

\begin{figure}
    \centering
    \includegraphics[width = 0.9\columnwidth]{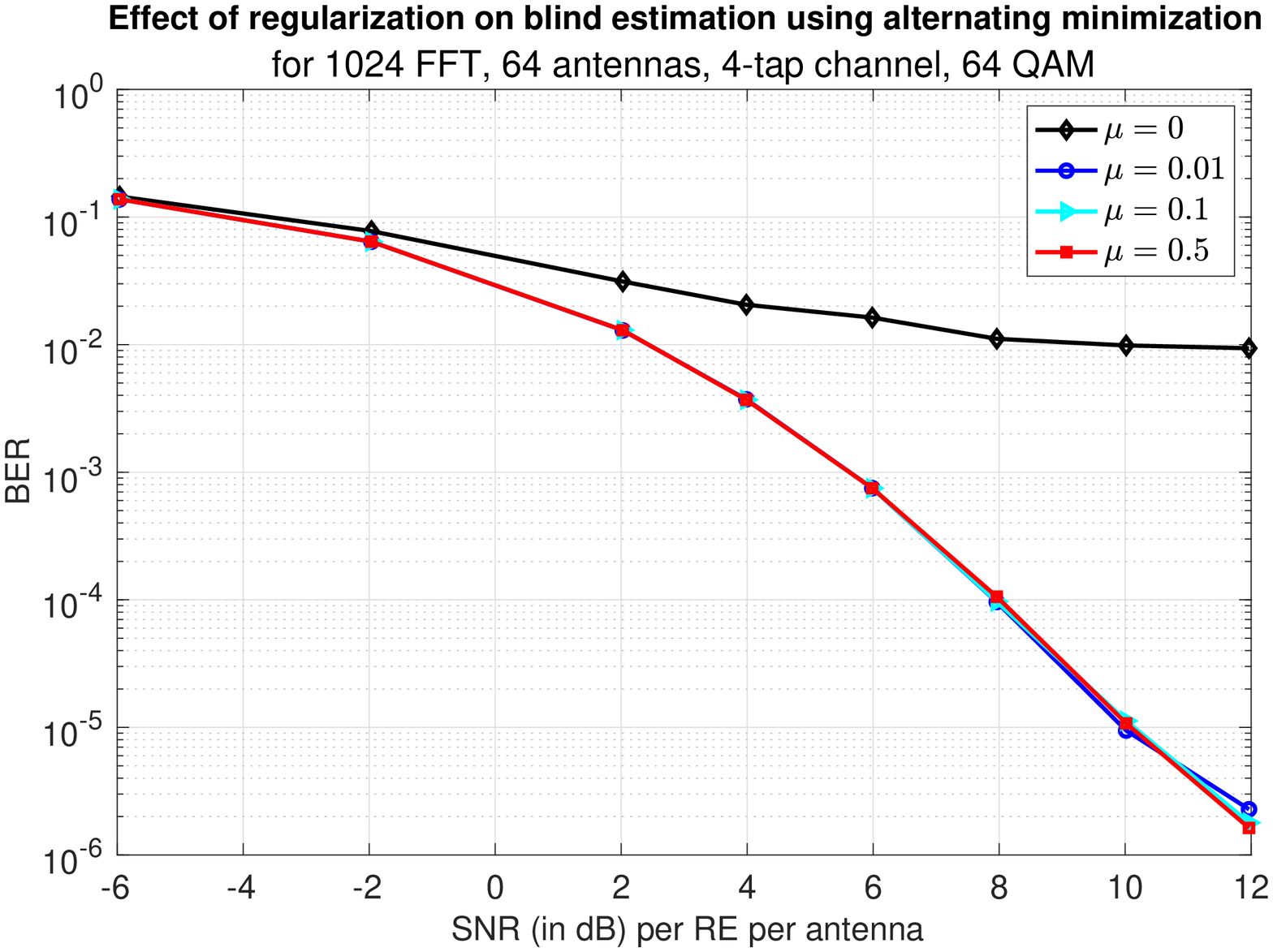}
    \caption{Bit error rates of blind estimation using alternating minimization algorithm (10 iterations) for different values of the regularization parameter, $\mu$. }
    \label{Blind_reg}
\end{figure}

The implementation of the matrix pseudo-inverse in the linear least squares solution given in \eqref{eq.5} is prone to numerical instabilities. Therefore, we break up the matrix pseudo-inverse in \eqref{eq.5} and include a regularization term as follows:
\begin{equation*}
    (\mathbf{\hat{X}F_{L}})^{\dagger} = (\mathbf{F_{L}^{H}\hat{X}^{H}\hat{X}F_{L}} + \mu\mathbf{I_{L}})^{-1}\mathbf{F_{L}^{H}\hat{X}^{H}},
\end{equation*}
where $\mu$ is the regularization parameter, $\mathbf{I_{L}}$ is the identity matrix of dimension $L$ and the superscript $\mathbf{A}^H$ denotes conjugate transpose of the matrix $\mathbf{A}$. Fig. \ref{Blind_reg} shows the bit error rates of the alternating minimization algorithm for different values of the regularization parameter. We observe that any choice of $\mu$ in the range $0<\mu<1$ ensures numerical stability of the $\mathbf{\hat{H}}$ estimates \cite{vogel2002computational} without affecting the error performance of the algorithm till reasonable SNR.

Using this iterative method, and a correct choice of initial point (as described above), the solution $\mathbf{\hat{X}}$  to \eqref{eq.3} can be numerically obtained. This obtained solution is a scaled version of the actually transmitted  constellation points. Fig. \ref{fig:md_1} shows the scatter-plot for the estimated $\mathbf{\hat{X}}$. We observe a rotation and scaling of the constellation; the exact complex scaling factor can be estimated using a single pilot as described in the next sub-section.

\subsection{Estimating the scaling factor}\label{lambda}
\label{sec:3c}
Lemma 1 in \cite{aswathy2023ojcoms} proves that any estimates $\mathbf{\hat{X}}$ and $\mathbf{\hat{H}}$ obtained as a solution to \eqref{eq.3}  are related to the actual $\mathbf{X_f}$ and $\mathbf{H_t}$ by a complex scalar transform, i.e.,
\begin{equation}
    \mathbf{\hat{X}} = \lambda\mathbf{X_f} \hspace{2mm}\text{and}\hspace{2mm} \mathbf{\hat{H}} = (\tfrac{1}{\lambda})\mathbf{H_t},
    \label{eq:6}
\end{equation}
where $\lambda \in \mathbb{C}$, the set of complex numbers. Thus, $\lambda$ causes amplitude scaling and phase rotation of the constellation, which needs to be estimated to fully decode the user symbols. This can be accomplished by the use of a single pilot symbol embedded in the frequency domain data that is known to the receiver. We can de-rotate $\mathbf{\hat{X}}$ by the $\lambda$ estimated using the pilot and perform QAM demodulation, completing the receiver chain.  

In this paper, we assume that there are $\eta$ pilots to estimate this scaling factor and will study the performance of the receiver as a function of $\eta$. We term these pilots  as  {\em rotational pilots}. We use $p_i \in \{1,\hdots, N\}$ to denote the location of the $i$-th rotational pilot and its complex value  known both at the transmitter and the receiver as $P_i$.  The scaling factor can be estimated as 
\[\hat{\lambda}=\frac{1}{\eta} \sum_{i=1}^\eta \frac{\hat{x}(p_i)}{P_i},\] where $\hat{x}(p_i)$ is the received pilot after completion of the alternating minimization algorithm.

\begin{figure*}[!t]
\centering
\subfloat[]{\includegraphics[width=0.9\columnwidth]{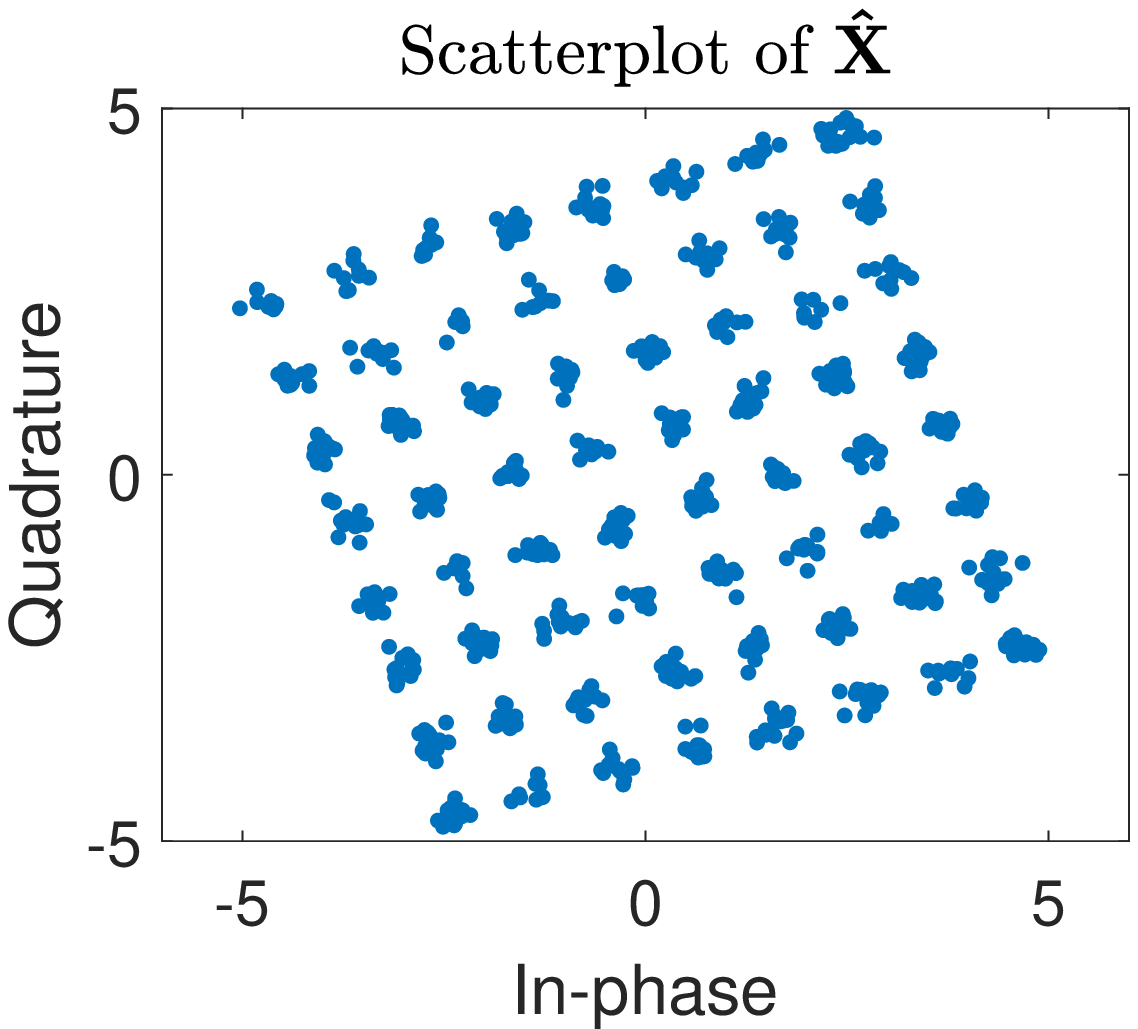}%
\label{fig:md_1}}
\hfil
\subfloat[]{\includegraphics[width=0.9\columnwidth]{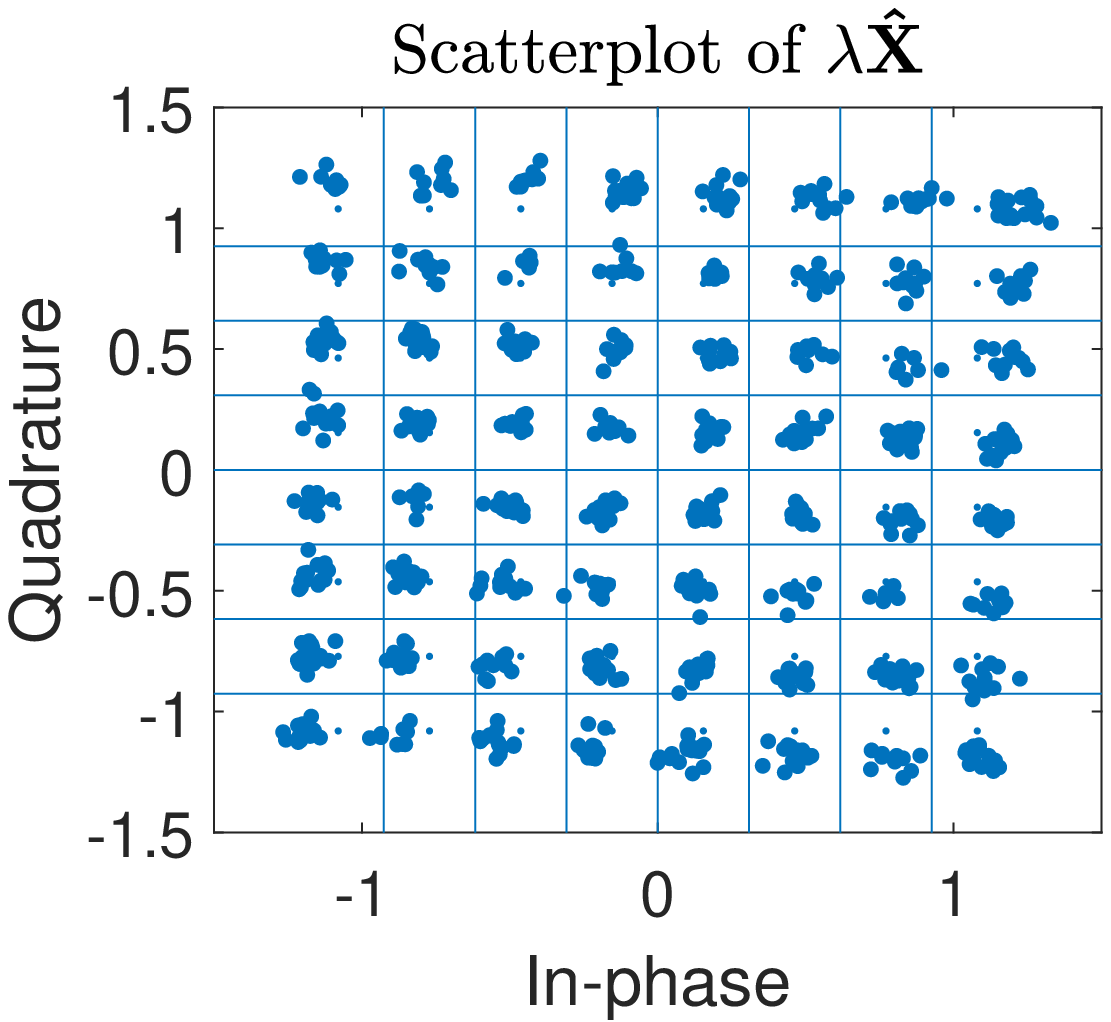}%
\label{fig:md_2}}
\caption{$\mathbf{\hat{X}}$ output of 10 iterations of the matrix decomposition (MD) algorithm for $10$ dB SNR, 1024 FFT, $64$ antennas, (a) before de-rotation, and (b) after  de-rotation using the scaling factor $\lambda$ estimated with one pilot shown against the Voronoi regions for 64-QAM.}
\label{AltMinOP}
\end{figure*}

\par{Fig. \ref{fig:md_2} shows the constellation from Fig. \ref{fig:md_1} corrected by the scaling factor $\lambda$ obtained using only one rotational pilot ($\eta =1$). When this is contrasted with the Voronoi regions for 64-QAM demodulation, we observe slight residual scaling and rotation in the constellation due to the error in the estimation of the scaling parameter $\lambda$. This will cause some points to be demodulated incorrectly if QAM demodulation is applied directly to this constellation.}

\par{It is well known that the estimator performance improves with increasing $\eta$, however at the cost of increased overhead. While a single pilot can provide an estimate  $\hat{\lambda}$, the MMSE of this estimate will decay only as the inverse of SNR causing an increase in symbol error rate for multi-antenna systems. Therefore, we turn to  additional non-pilot aided techniques to mitigate this residual scaling and rotation. }

\begin{figure*}[!t]
\centering
\subfloat[]{\includegraphics[width=0.9\columnwidth]{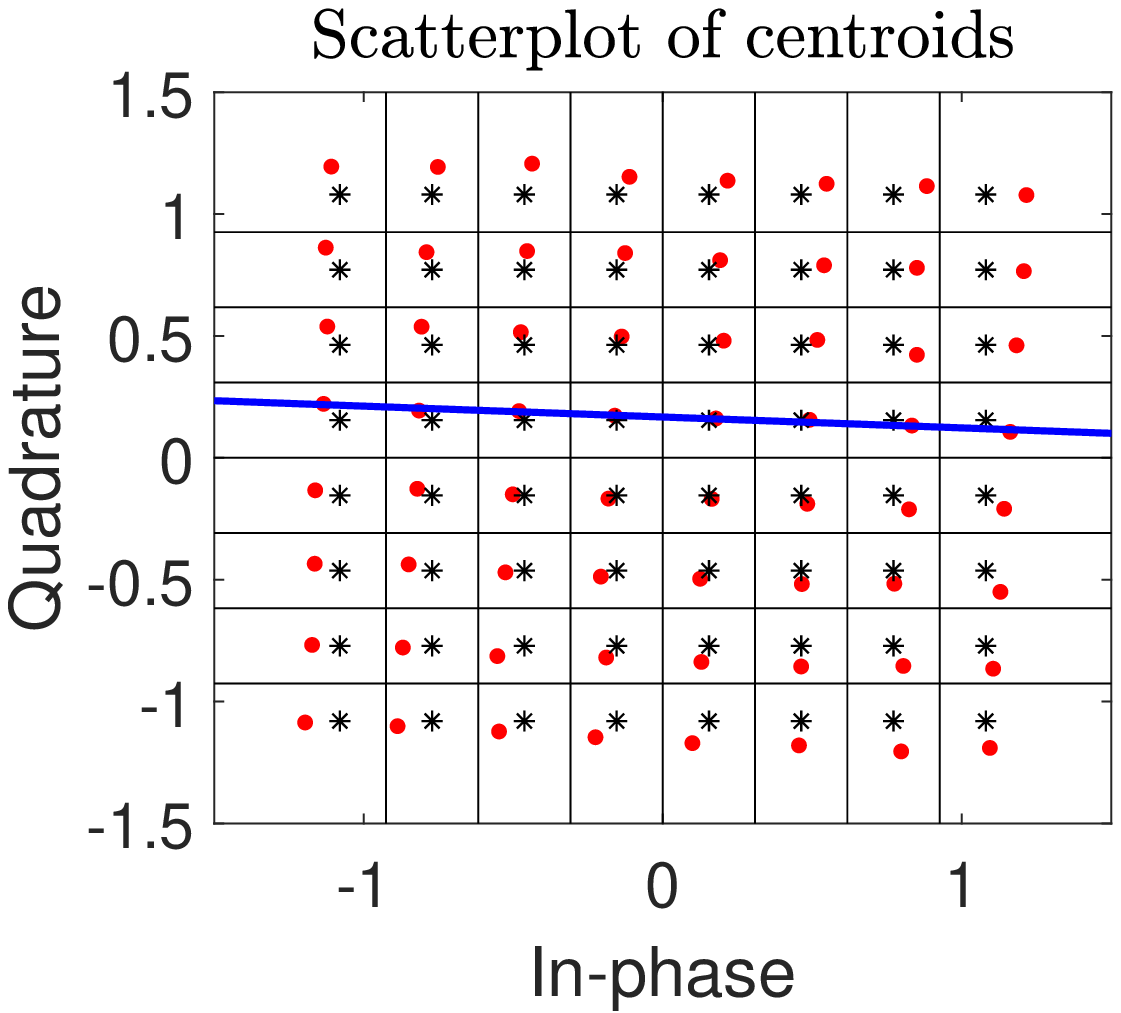}%
\label{fig:md_3}}
\hfil
\subfloat[]{\includegraphics[width=0.9\columnwidth]{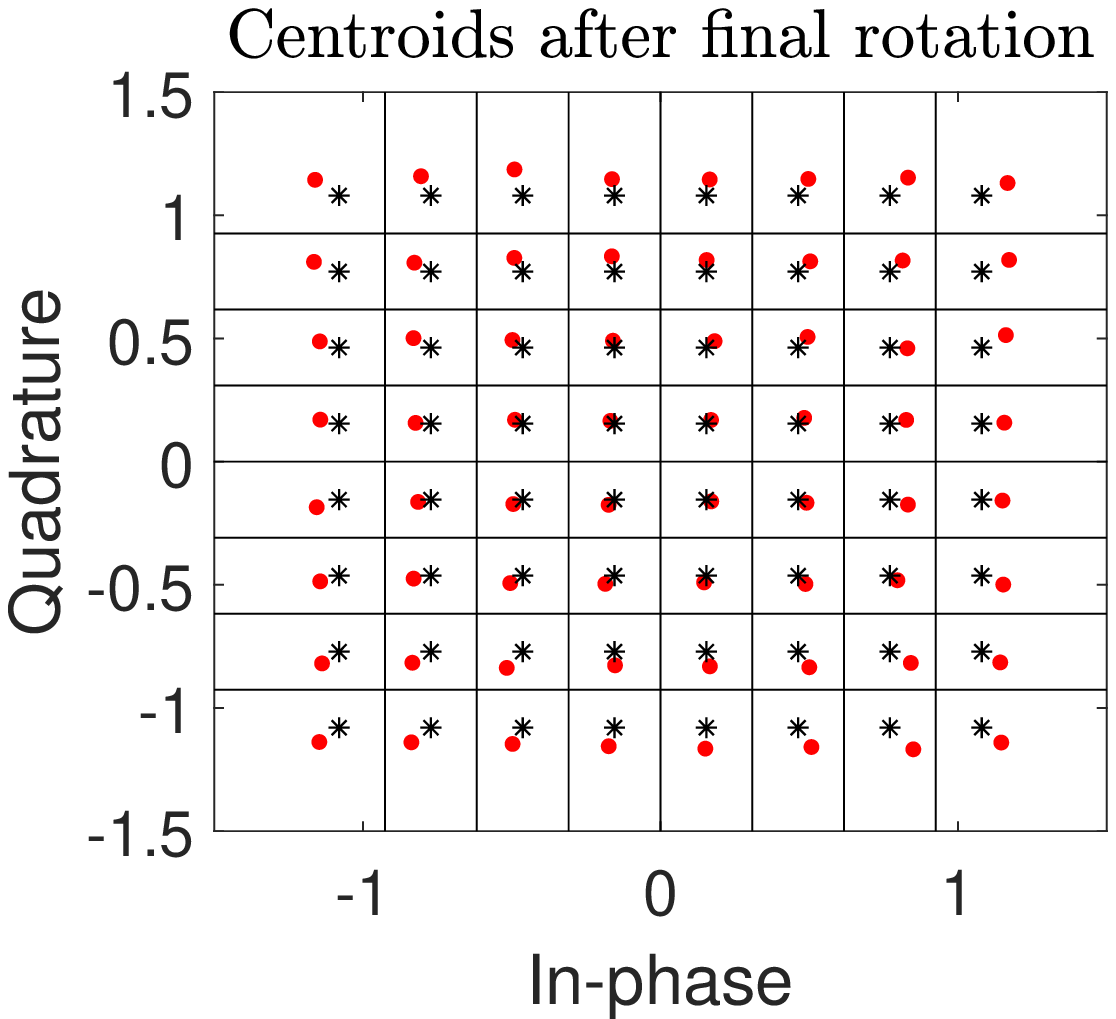}%
\label{fig:md_4}}
\caption{Centroids (in red) of the clusters obtained from the constellation in Fig. \ref{fig:md_2} using Lloyd's k-means algorithm, (a) with residual rotation. The slope of the line through a row of centroids (in blue) gives the residual rotation to be corrected. (b) Centroids after correction of the residual rotation using the slope of the blue line from (a). Points in black correspond to the standard 64-QAM constellation. }
\label{cenroids}
\end{figure*}

\par{The residual scaling (amplitude) can be corrected using the total energy of the constellation. The total energy of the transmitted constellation is typically normalized to unity and therefore, the received constellation must have unit energy before we perform demodulation. We can estimate the energy of the constellation obtained from the alternating minimization algorithm and use it as the normalization factor on $\mathbf{\hat{X}}$ to correct for the residual scaling. We propose two different methods to correct the residual rotation.}

\subsubsection{Method 1}

\par{The first method is a simple clustering technique \cite{kmeans2019} that can be run as a separate additional module at the end of the alternating minimization algorithm. In this module, we use Lloyd's k-means clustering algorithm with the number of clusters equal to the modulation order, $M$, and the standard $M$-QAM constellation rotated by $\lambda$ estimated from the single rotational pilot as the initial centroids. Fig. \ref{fig:md_3} shows the centroids of the clusters obtained after k-means clustering. We note that in cases where the single rotational pilot is heavily corrupted by noise, the residual rotation might be very high, causing even the centroids of some of the clusters to be demodulated incorrectly. To mitigate this, we need to compensate for the residual rotation. To this end, we can fit a line through a row of these centroids as illustrated in Fig. \ref{fig:md_3}, and the slope of such a line would give us the value of the residual rotation to be compensated. The scatter plot of the centroids thus corrected is given in Fig. \ref{fig:md_4}. This module is summarized in Algorithm \ref{ClusterAlgo}.}

\renewcommand\footnoterule{}      
\begin{algorithm}
\caption{Clustering Algorithm for Residual De-rotation}\label{ClusterAlgo}
\footnotetext{$\mathbf{c_r}$ denotes elements of the set $\mathbf{C_r}$.}
\label{alg:loop}
\begin{algorithmic}[1]
\State Input $\mathbf{\hat{X}}$ (after alternating minimization), pilot subcarrier $p$, pilot symbol $x(p)$, modulation order $M$
\State Calculate initial rotation, $\lambda = x(p)/\hat{x}(p)$
\State Define $\mathbf{Q} \gets M$-QAM  constellation
\State Rotate $(\frac{1}{\lambda})\mathbf{Q} \rightarrow \mathbf{Q_{r}}$
\State Input to Lloyd's k-means algorithm: $\mathbf{\hat{X}}$, $M$ clusters, initial centroids $\mathbf{Q_{r}}$
\State Output of Lloyd's algorithm: $\mathbf{\Phi} \gets$ Cluster indices for $\mathbf{\hat{X}}$, $\mathbf{C} \gets$ Final centroids
\State De-rotate $\lambda\mathbf{C} \rightarrow \mathbf{C_r}$
\State Find points $\mathbf{P} = \{\mathbf{c_r} \in $ closest Voronoi regions above real-axis for $\mathbf{Q}\}$
\State Calculate $\theta =$ slope of line fit through $\mathbf{P}$
\State Define rotation matrix, $\mathbf{\Theta} = \begin{bmatrix}
    \cos(\theta) & -\sin(\theta) \\ \sin(\theta) & \cos(\theta)
    \end{bmatrix}$
\State De-rotate $\mathbf{C_{r}\Theta} \rightarrow \Cfinal$
\State $M$-QAM demodulate $\Cfinal \rightarrow \Cqam$
\State Demodulate $\mathbf{\Phi}$ using $\Cqam \rightarrow \Phiqam$\\
\Return $\Phiqam$
\end{algorithmic}
\end{algorithm}

\subsubsection{Method 2}

\begin{figure}
    \centering
    \includegraphics[width=0.9\columnwidth]{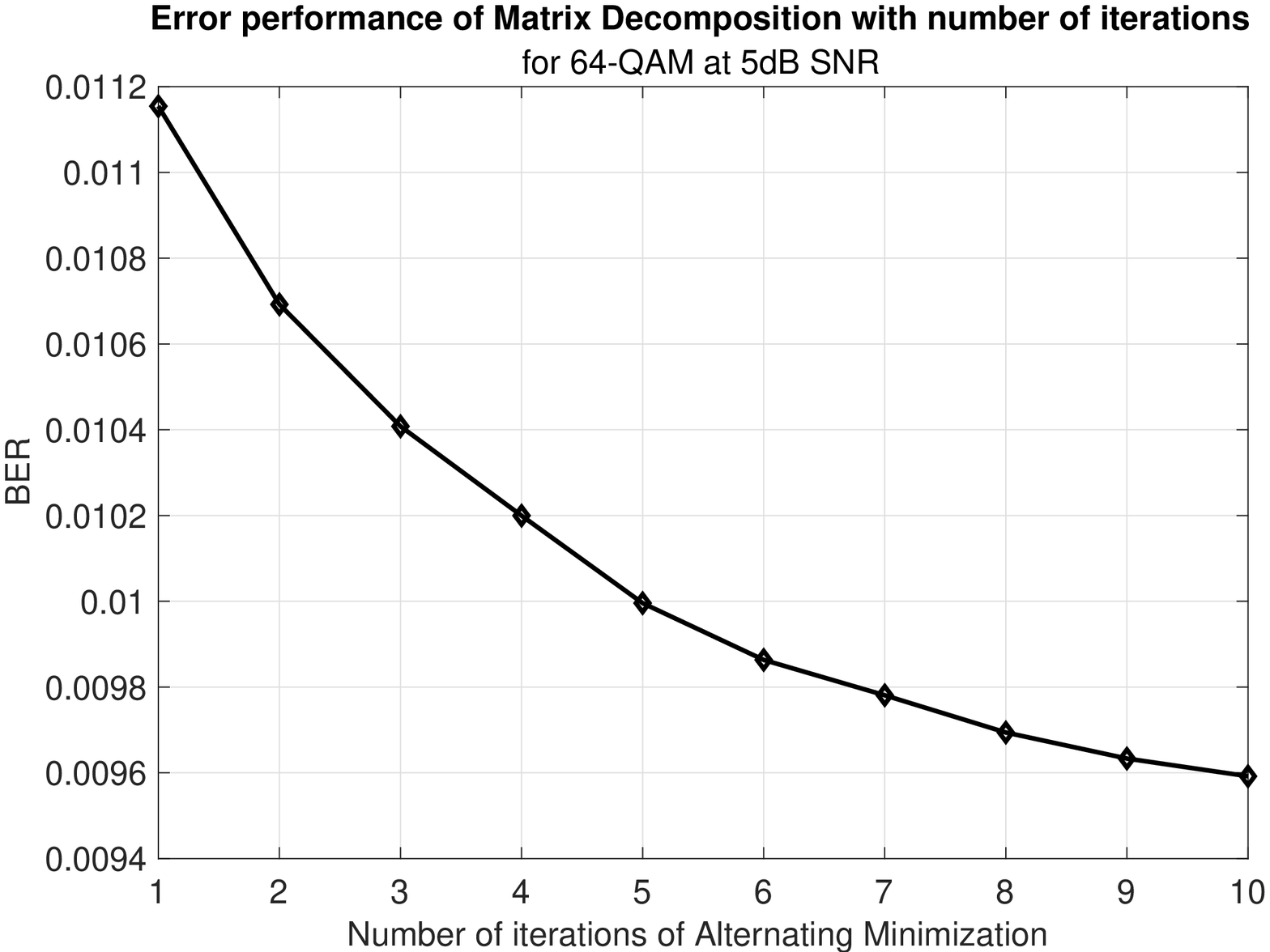}
    \caption{BER of alternating minimization algorithm at 5 dB SNR as a function of the number of iterations, with one rotational pilot used for estimating $\lambda$.}
    \label{Blind_AMitns}
\end{figure}

We now propose an alternative method for correcting the rotation that can be executed within the alternating minimization framework. This avoids the complexity of running a clustering algorithm, which is high for higher order modulations. Fig. \ref{Blind_AMitns} shows the error performance of the alternating minimization algorithm as a function of the number of iterations. We observe that the error rate drops steeply in the first few iterations after which the slope of the curve decreases. This happens because after the first few iterations, the QAM constellation has been almost fully recovered by the algorithm, however the remaining error is due to residual rotation of the constellation. Therefore, after the first few iterations of alternating minimization when the error drops to a certain threshold, say approximately one percent, which happens around the fourth iteration as seen in Fig. \ref{Blind_AMitns}, we calculate $\lambda$ using the one pilot and correct the estimated rotation on $\mathbf{\hat{X}}$. We then map all the points in $\mathbf{\hat{X}}$ after the initial de-rotation to the nearest points of the transmitted QAM constellation of the same modulation order, in all the subsequent iterations of alternating minimization to successively correct the residual rotation. Thus, this ``clean" QAM constellation closest to $\mathbf{\hat{X}}$ is fed into the next iteration of alternating minimization. This boosts the overall SNR of the constellation as most of the points are mapped correctly at this stage, eliminating the noise power in them, and only a few points are mapped incorrectly due to the residual rotation. The next iteration of alternating minimization is then able to correct these erroneous points due to the SNR boost from the correctly mapped points. We continue this process for a few more iterations to ensure that any residual rotation has been fully corrected. This modified alternating minimization algorithm is summarized in Algorithm \ref{BlindalgoSU}.

\renewcommand\footnoterule{}      
\begin{algorithm}
\caption{Blind Estimation and De-rotation for Single User}\label{BlindalgoSU}\footnotetext{$\mathbf{y_1}$ denotes the first column of matrix $\mathbf{Y_f}$, $\dagger$ denotes matrix pseudo-inverse, $*$ denotes complex conjugation, $a(n,r)$ denotes element at row $n$ and column $r$ of matrix $\mathbf{A}$ and $x_{k}(n)$ denotes $n-$th diagonal element of $\mathbf{\hat{X}_k}$.}
\label{alg:loop}
\begin{algorithmic}[1]
\State Input $\mathbf{Y_f}$, initial point $\mathbf{\hat{X}}$, number of iterations $T$, pilot subcarrier $p$, pilot symbol $x(p)$, QAM order $M$, regularization parameter $\mu$.
\State Initialize $k = 1, \mathbf{B} = [\hspace{1mm}]$
\Repeat
    \State {${\mathbf{\hat{H}}}\gets (\mathbf{F_{L}^{H}\hat{X}^{H}\hat{X}F_{L}} + \mu\mathbf{I})^{-1}\mathbf{F_{L}^{H}\hat{X}^{H}}\mathbf{Y_f}$}
    \State $\mathbf{B} \gets \mathbf{F_{L}\hat{H}}$
    \For {$n \gets 1$ to $N$}
        \State $\hat{x}(n) = \frac{\Big(\sum_{r=1}^{N_r} y(n,r)b^{*}(n,r)\Big)}{\Big(\sum_{r=1}^{N_r} |b(n,r)|^2\Big)}$
    \EndFor
    \If{$k > 3$}
        \If{$k = 4$}
            \State $\lambda \gets \hat{x}(p)/x(p)$
            \State $\mathbf{\hat{X} = \hat{X}}/\lambda$
        \EndIf
        \State $\mathbf{\hat{X}_q} \leftarrow$ $M$-QAM constellation points closest to $\mathbf{\hat{X}}$
        \State $\mathbf{\hat{X}} \leftarrow \mathbf{\hat{X}_q}$
    \EndIf
    \State $k \gets k+1$
\Until $k = T$\\
\Return $\mathbf{\hat{X}}, \mathbf{\hat{H}}$
\end{algorithmic}
\end{algorithm}

\begin{figure}
    \centering
    \subfloat[Initial $\mathbf{\hat{X}}$]{
        \includegraphics[width=0.65\columnwidth]{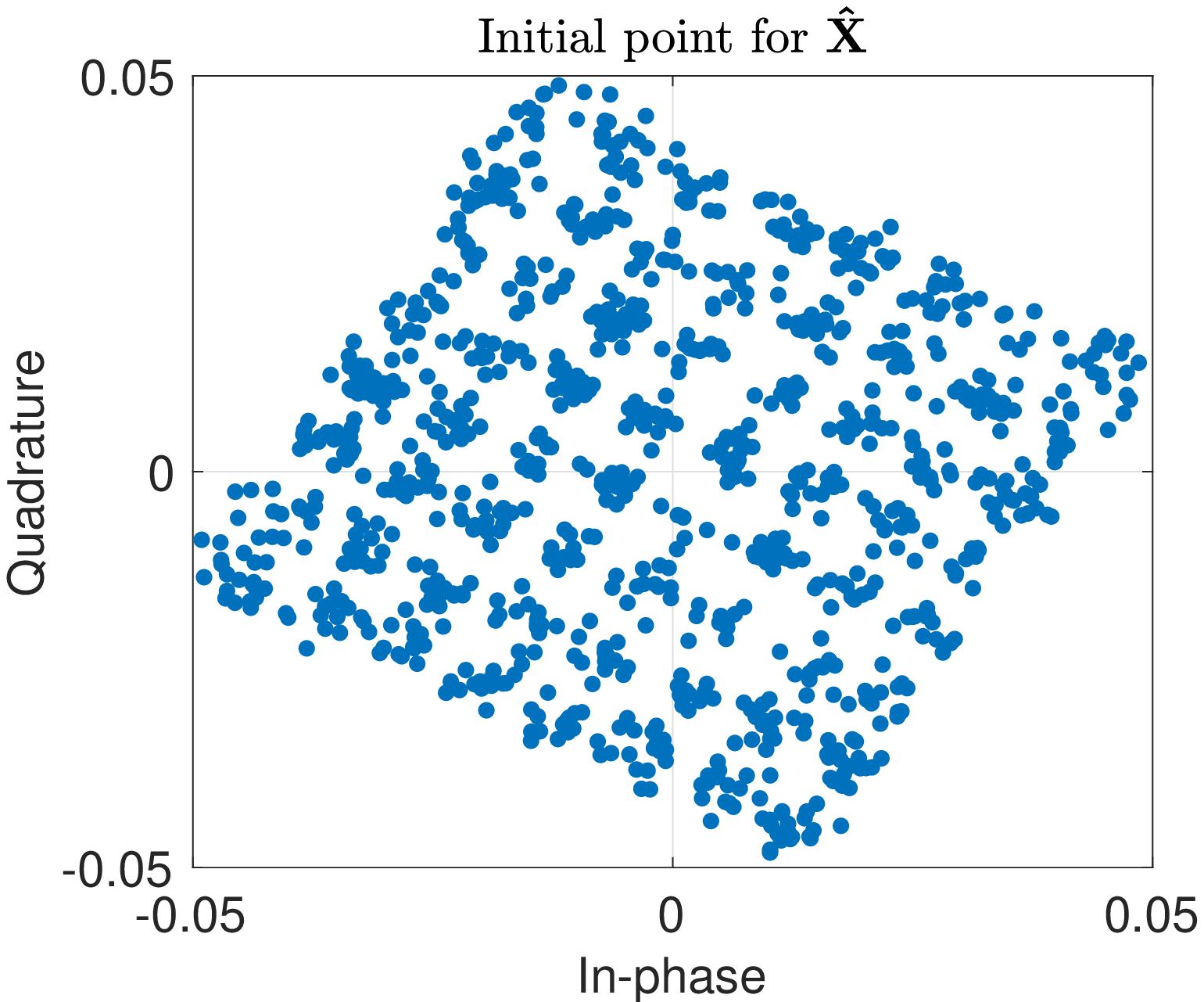}
        \label{X1}
    }
   \par
    \centering
    \subfloat[$\mathbf{\hat{X}}$ after 4 iterations]{
        \includegraphics[width=0.65\columnwidth]{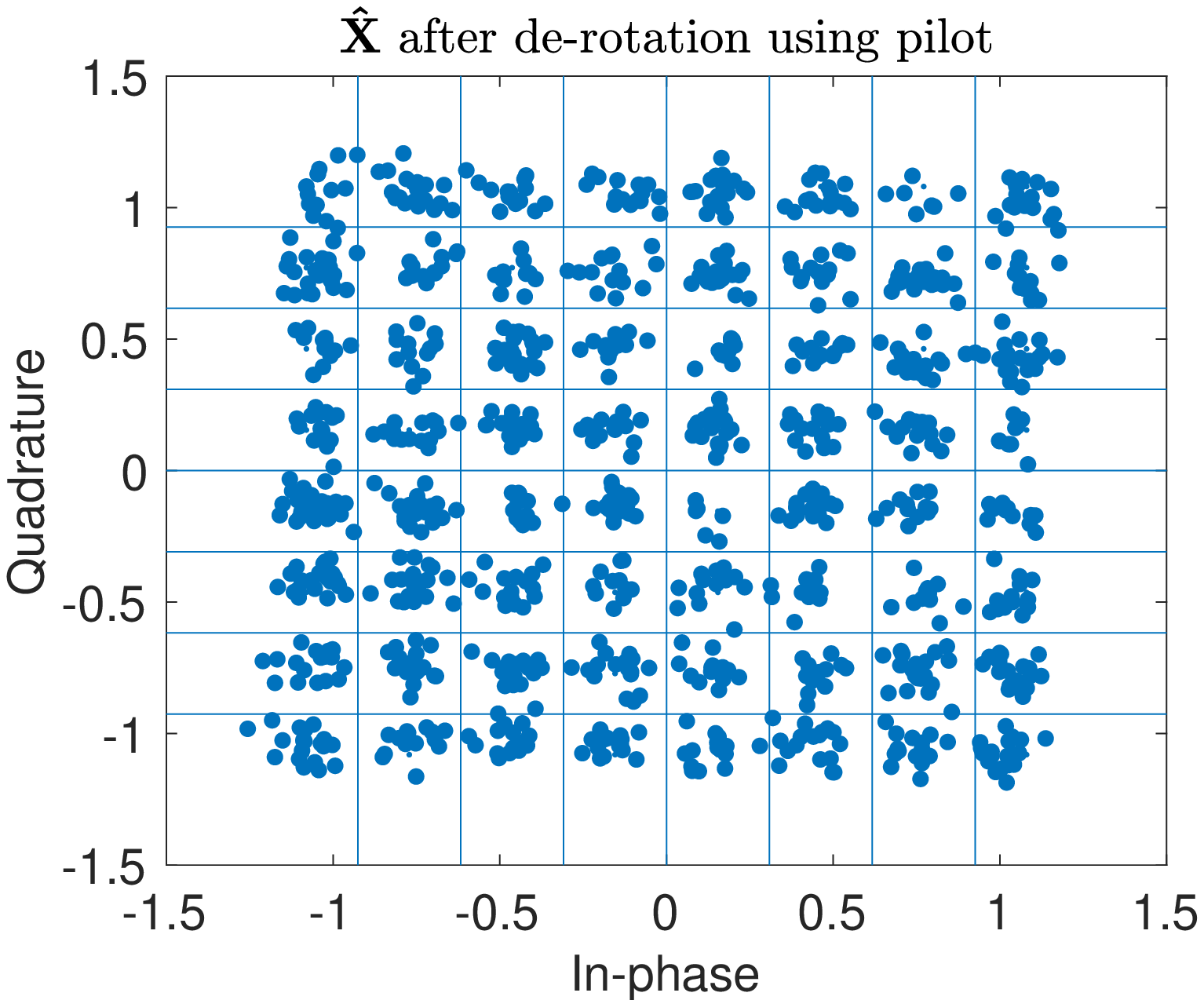}
        \label{X4}
    }
    \par
    \centering
    \subfloat[$\mathbf{\hat{X}}$ after 6 iterations]{
        \includegraphics[width=0.65\columnwidth]{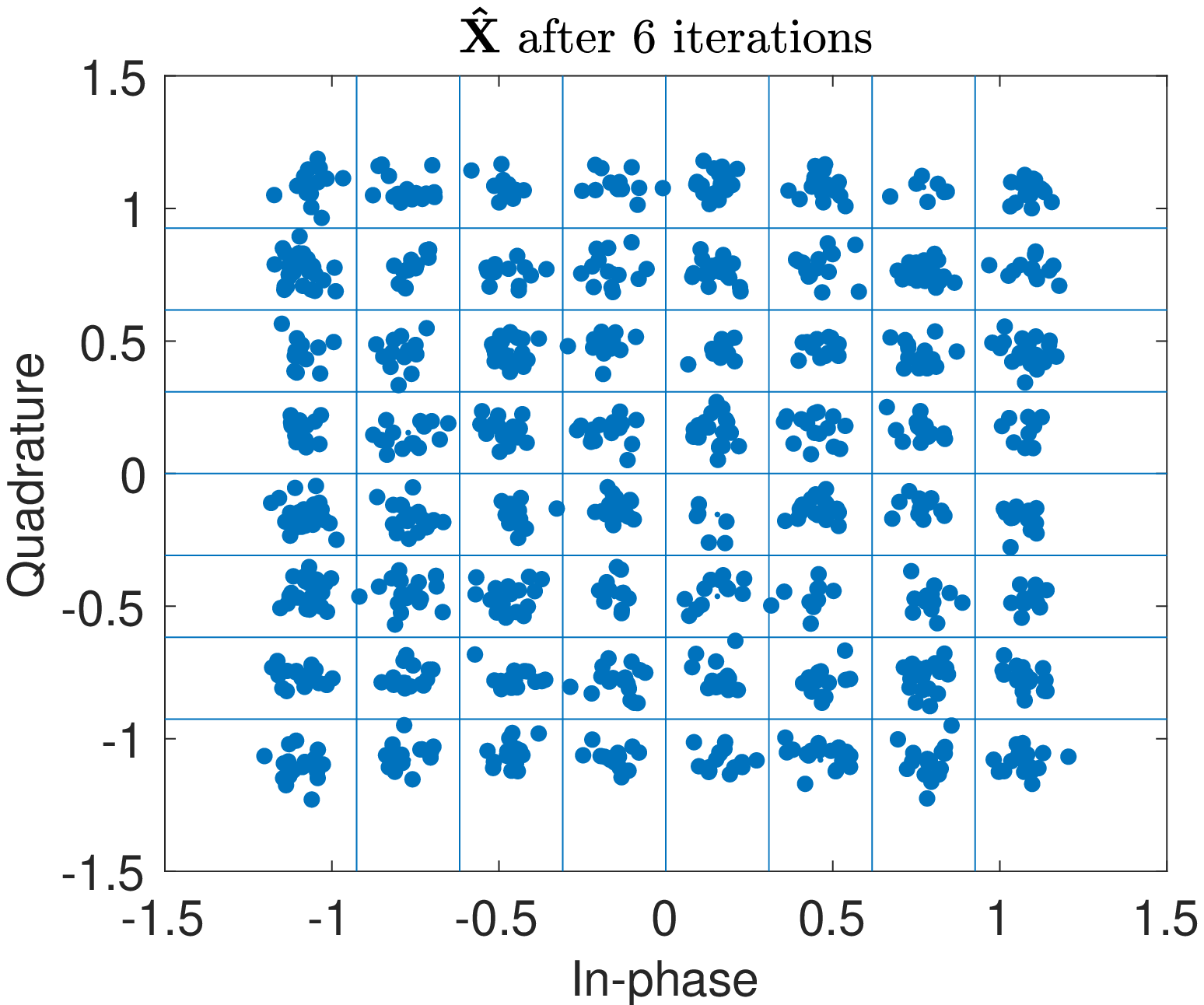}
        \label{X6}
    }
   \par
    \centering
    \subfloat[Final $\mathbf{\hat{X}}$ (after 10 iterations)]{
        \includegraphics[width=0.65\columnwidth]{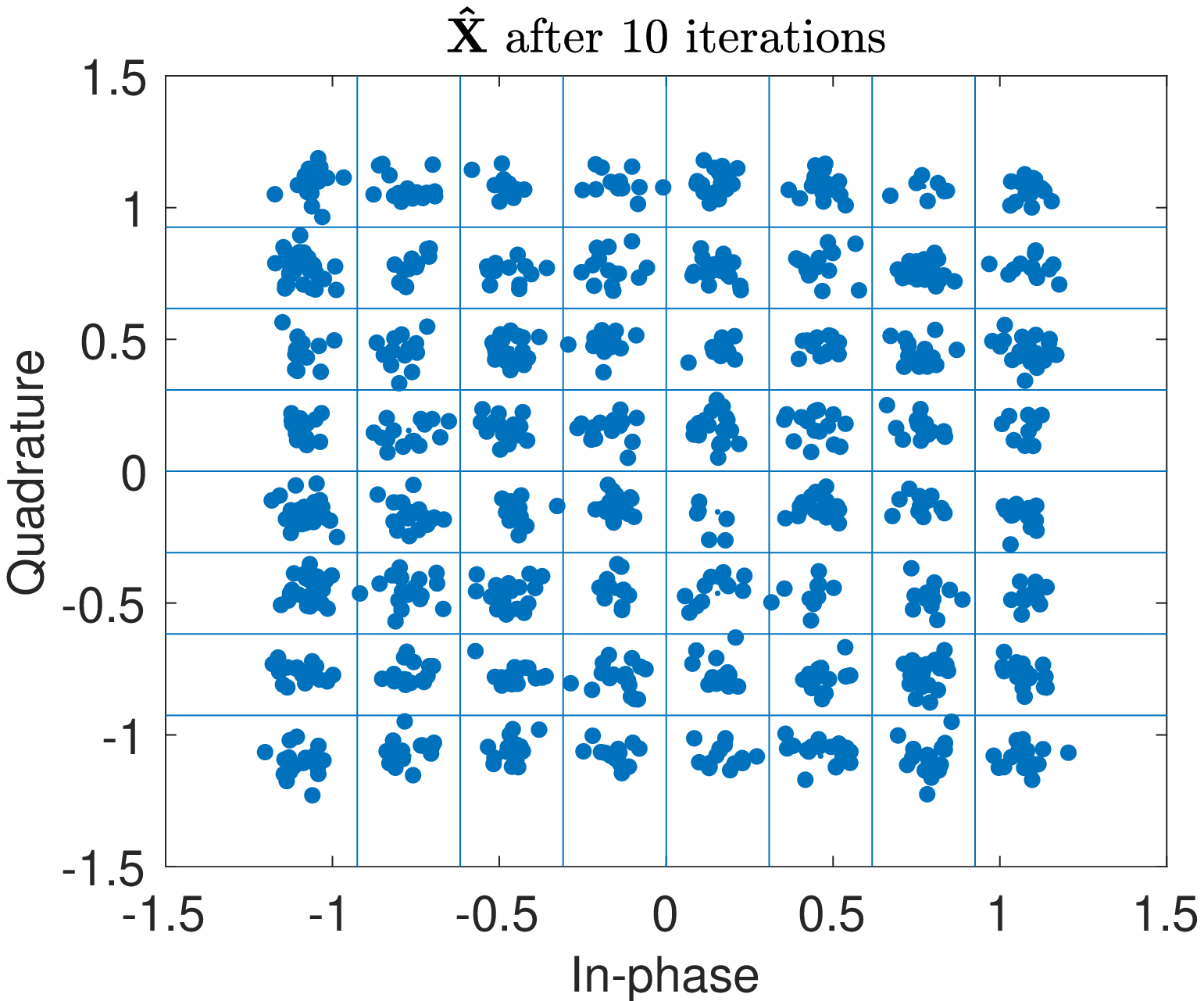}
        \label{X10}
    }
\caption{Scatter-plots of $\mathbf{\hat{X}}$ obtained at different stages of Algorithm \ref{BlindalgoSU} at 5dB SNR for 1024 FFT and 64 receive antennas.}
\label{BlindSUX}
\end{figure}

Fig. \ref{BlindSUX} presents the constellations of $\mathbf{\hat{X}}$ obtained during various stages of  Algorithm \ref{BlindalgoSU} at an SNR of 5dB. Fig. \ref{X1} is the initial point obtained from the left singular vector of $\mathbf{Y_f}$. After 4 iterations of alternating minimization, the scaling/rotation factor, $\lambda$ estimated using the pilot symbol as per Step 11 of the algorithm is used for de-rotating the constellation, which is given in Fig. \ref{X4}. Then we apply demodulation and modulation on this constellation as per steps 14 and 15 of the algorithm along with the alternating minimization. The constellation obtained after 6 iterations is given in Fig. \ref{X6}, where we can observe that the combination of demodulation and modulation continues to correct the residual rotation. Finally, after 10 iterations, we stop the algorithm, and the resulting constellation is given in Fig. \ref{X10}. 

\section{Error Performance for Single User}\label{BER1user}

\begin{table}
\caption{Simulation Parameters}
\centering
\begin{tabular}{ |l||c|  }
 \hline
 FFT size ($N$) & 1024\\\hline
 Number of RRH antennas ($N_r$) & 64\\\hline
 Multi-path Channel length ($L$) & 4\\\hline
 Modulation scheme ($M$) & 64-QAM \\\hline
\end{tabular}
\label{Blind_simpara_1u}
\end{table}

We evaluate the performance of our algorithm through link level simulations of an uncoded massive MIMO system with the parameters given in Table \ref{Blind_simpara_1u}, where a single user occupies all the subcarriers in the OFDM symbol. The number of subcarriers in the OFDM symbol are 1024 and there are 64 receive antennas at the base station. We use the pedestrian channel model with four taps and employ the variance-based dominant-tap estimation method described in Algorithm \ref{BlindIPvar} for calculation of the initial point. For comparison, we also evaluate the performance of the same system with conventional pilot-based channel estimation and maximal-ratio-combining (MRC). For MRC, we assume that each OFDM symbol has $10$\% equi-spaced pilots in the frequency domain, which in our configuration translates to 104 pilots. We compare our method against both linear channel interpolation and FFT-based channel interpolation methods in the MRC scheme \cite{moon2017fft}. 

\begin{figure}
    \centering
    \includegraphics[width=0.9\columnwidth]{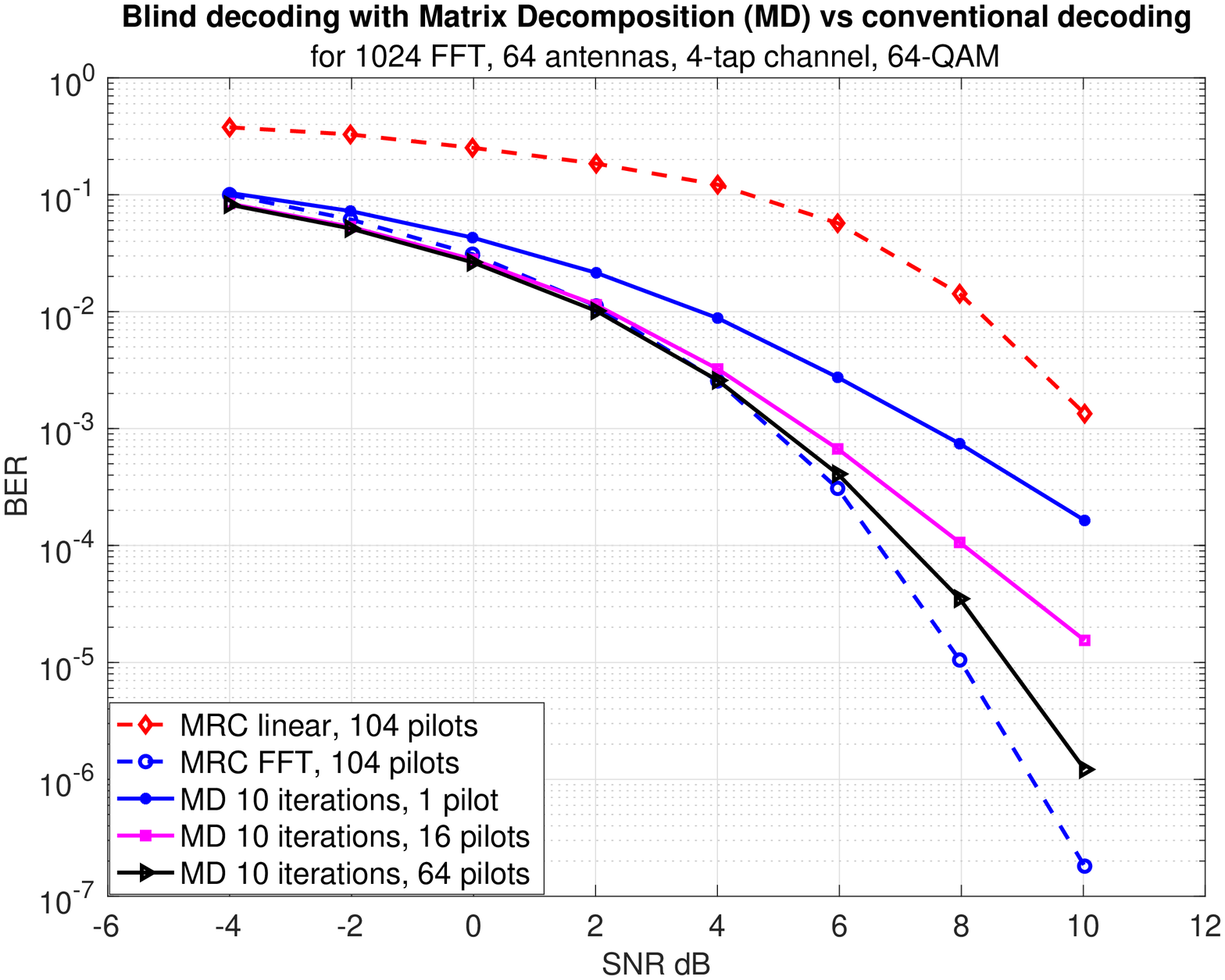}
    \caption{Comparison of uncoded BER of 10 iterations of the proposed matrix decomposition using alternating minimization method for different $\eta$ and MRC with linear and FFT channel interpolations using 104 pilots.}
    \label{Blind_BER_eta}
\end{figure}

Fig. \ref{Blind_BER_eta} shows the uncoded bit error rate (BER) for 10 iterations of alternating minimization, with the scaling/rotation factor $\hat{\lambda}$ calculated using  $\eta =1, 16, \text{and } 64$ rotational pilots against MRC with $104$ channel estimation pilots. We observe that the proposed blind matrix decomposition method (MD) matches the performance of the conventional MRC for low signal-to-noise ratios (SNR). However, at high SNR, we see that the BER of the MD method  does not decay as fast as that of MRC with respect to the SNR. This is mainly because of the estimation error in $\hat{\lambda}$ which results in a small residual rotation as discussed earlier. Also as expected, we observe the performance improving with increasing number $\eta$ of rotational pilots.

\begin{figure}
    \centering
    \includegraphics[width=0.9\columnwidth]{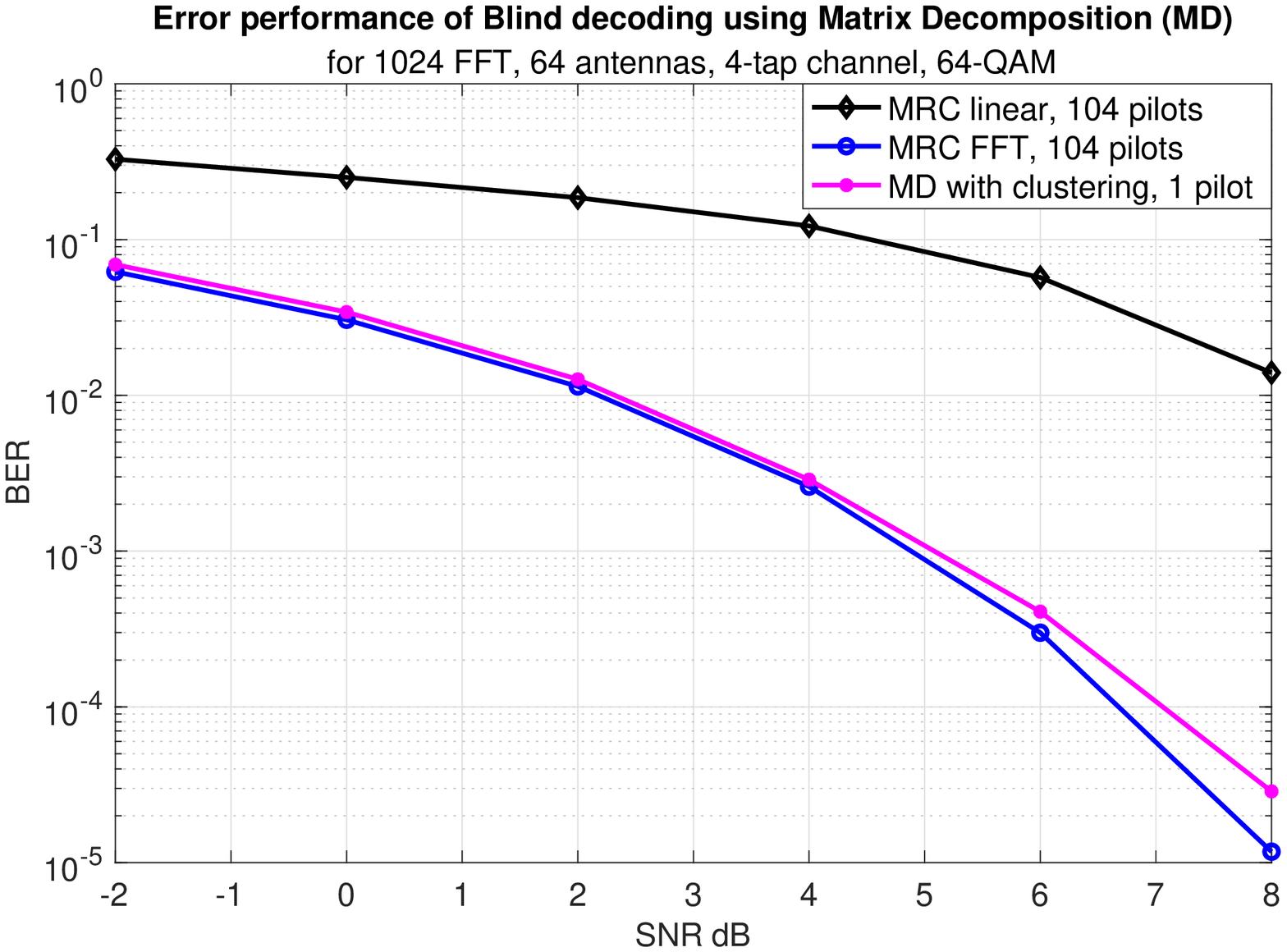}
    \caption{Comparison of uncoded BER of MRC against 20 iterations of alternating minimization for $\eta=1$ using the clustering method in Algorithm \ref{ClusterAlgo}.}
    \label{BER_cluster}
\end{figure}

To improve the performance of our method, instead of increasing the rotational pilots, the clustering method described in Algorithm \ref{ClusterAlgo} with $\eta=1$ rotational pilot is used to estimate the residual scaling parameter. From Fig. \ref{BER_cluster} we observe that the performance of the proposed MD method with just 1 pilot now matches that of MRC with 104 pilots. 

\begin{figure}
    \centering
    \includegraphics[width=0.9\columnwidth]{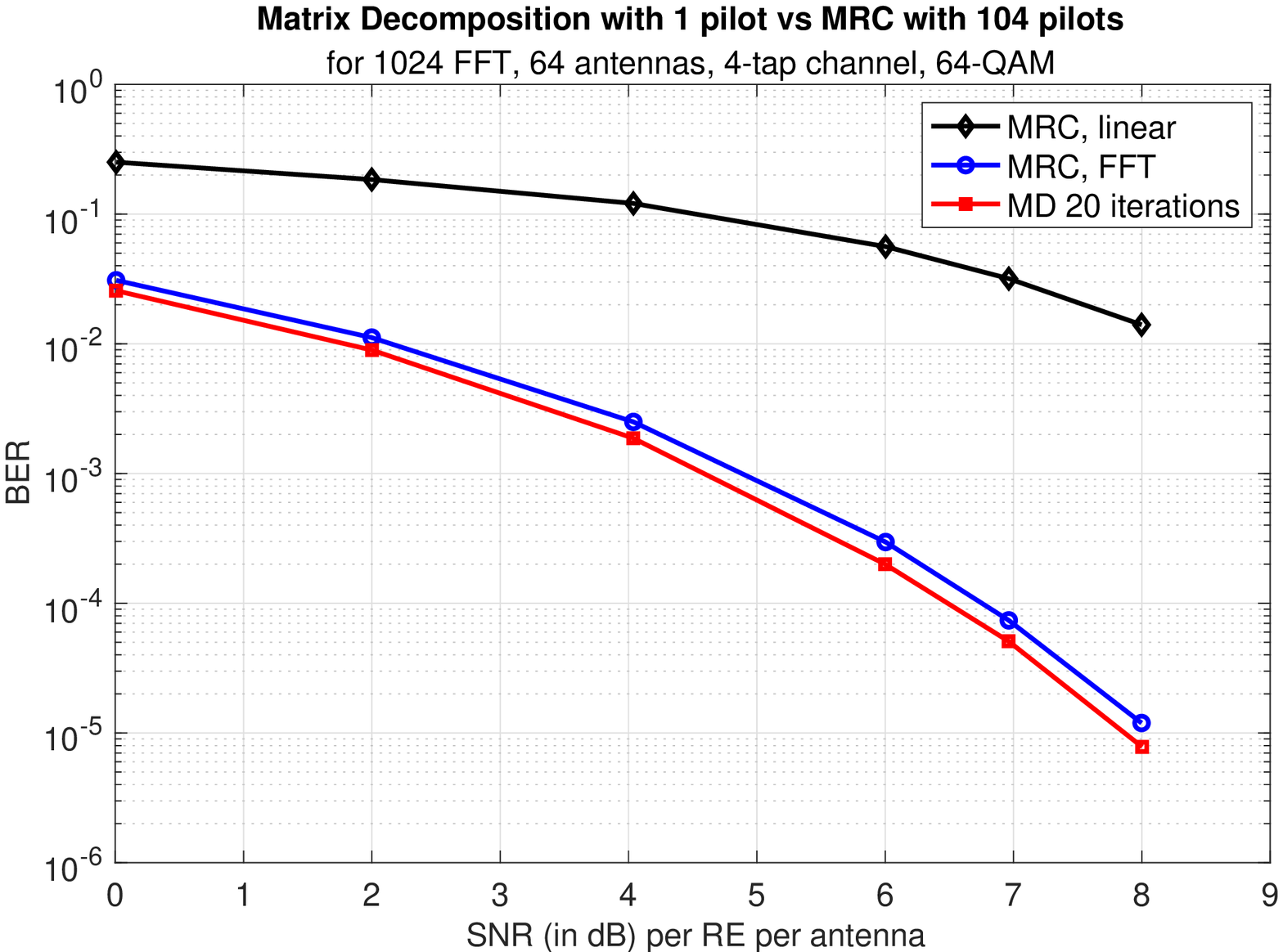}
    \caption{BER after 10 iterations of the modified alternating minimization method with 1 pilot compared against MRC with 104 pilots.}
    \label{BERmodifiedAM}
\end{figure}

We now examine the performance of the modified alternating minimization method with the built-in rotational correction described in Algorithm \ref{BlindalgoSU}. Fig. \ref{BERmodifiedAM} shows the BER for 10 iterations of the modified algorithm that uses 1 rotational pilot against MRC with 104 (10\%) pilots. We observe that the new algorithm performs slightly better than MRC across all SNRs.

\begin{figure}
    \centering
    \includegraphics[width=0.9\columnwidth]{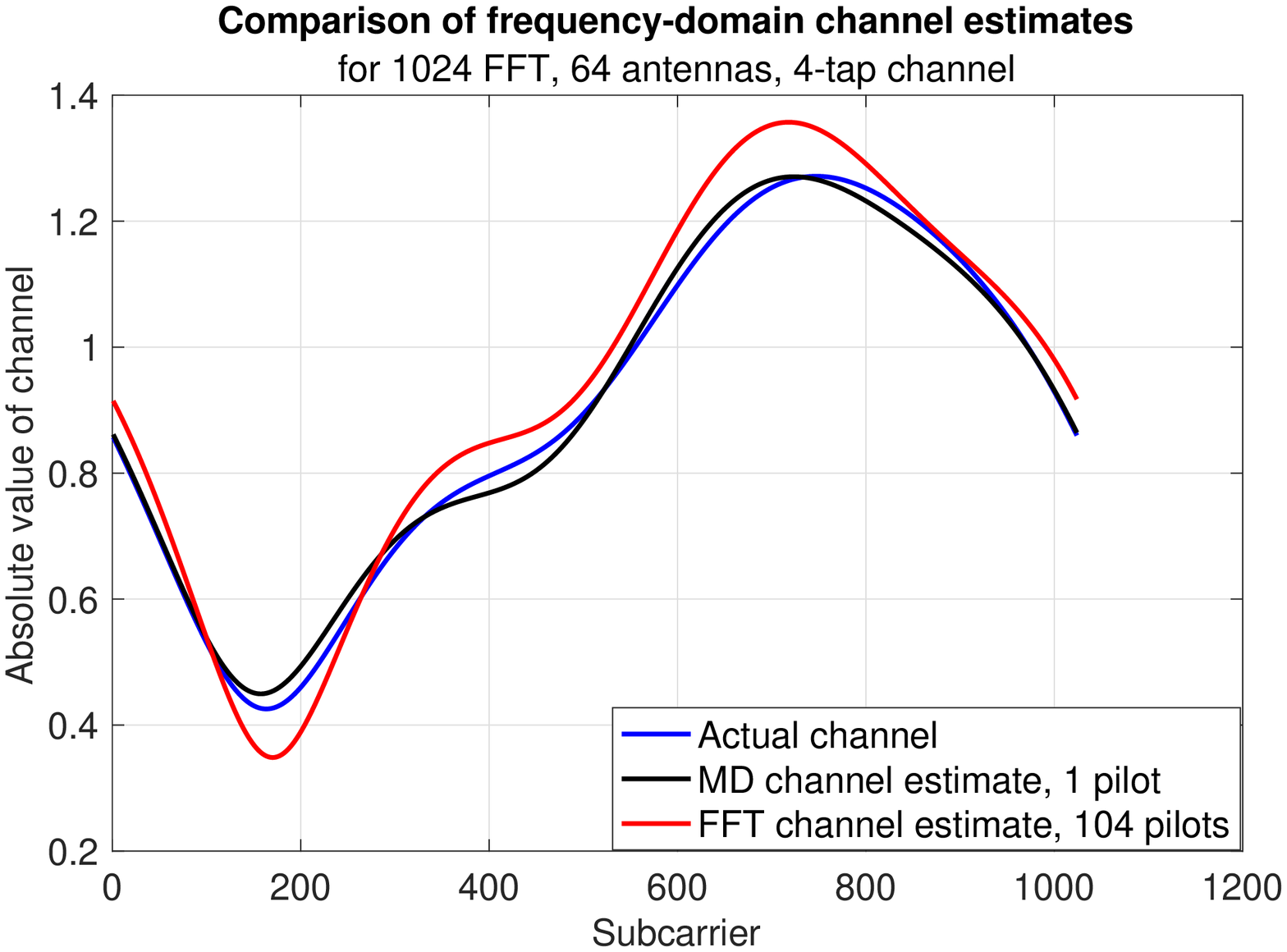}
    \caption{Comparison of the frequency-domain channel magnitudes estimated using the proposed Algorithm \ref{BlindalgoSU} from 1 pilot against the FFT interpolation method from 104 pilots. Actual channel shown in blue for reference.}
    \label{Hest}
\end{figure}

A comparison of the frequency-domain channels estimated using both the proposed Algorithm \ref{BlindalgoSU} and the conventional pilot-based method with FFT interpolation from 104 pilots is given in Fig. \ref{Hest}. We observe that the channel estimated using the proposed method with just 1 pilot follows the variations in the actual channel (in blue) more closely than the channel estimated from 104 pilots using the FFT interpolation method. This explains why the BER of our method is lower than that of MRC using the FFT-interpolated channel in Fig. \ref{BERmodifiedAM}.

\begin{figure}
    \centering
    \includegraphics[width=0.9\columnwidth]{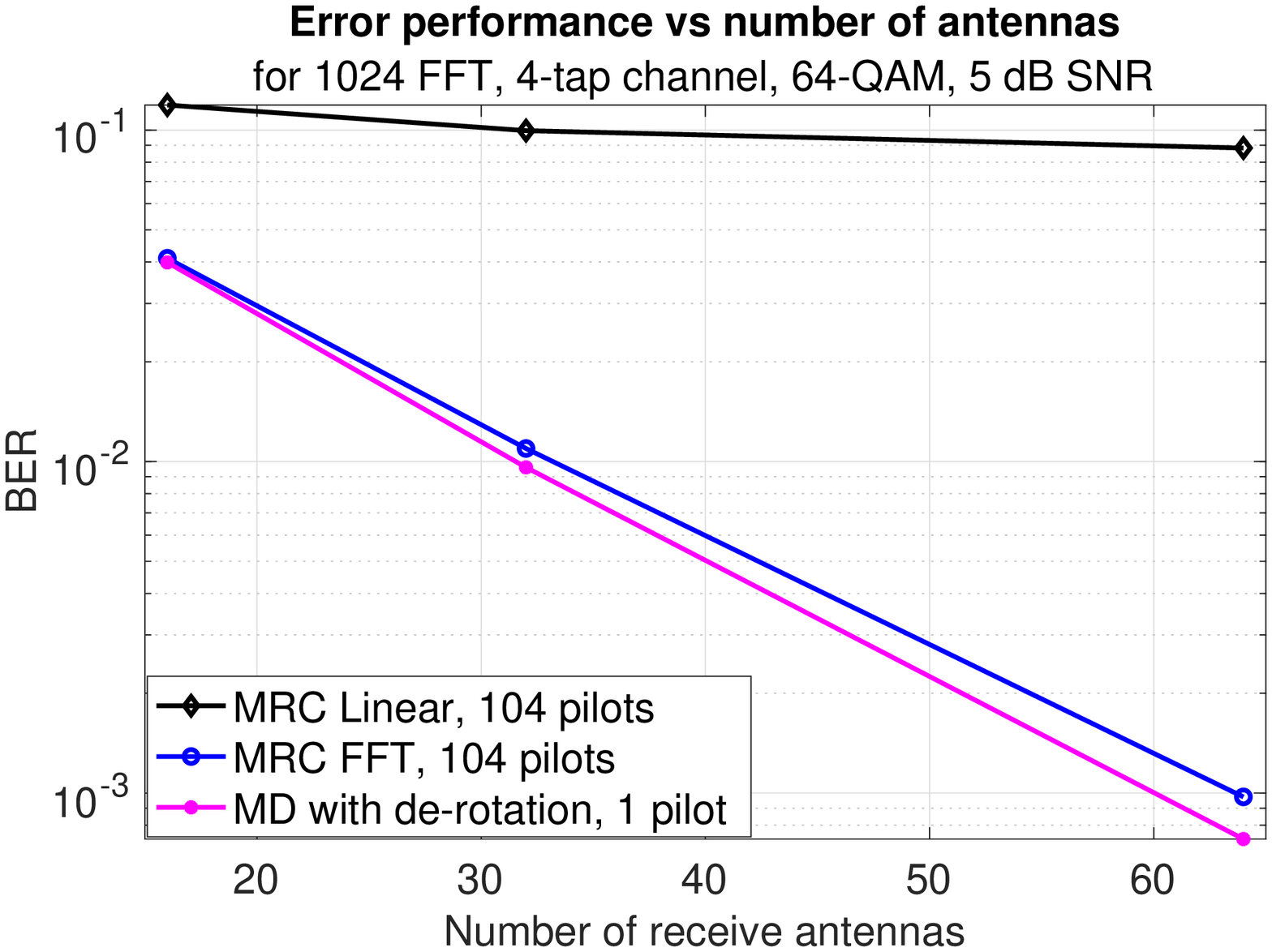}
    \caption{BER of matrix decomposition and MRC at 10 dB SNR for 16, 32, and 64 receive antennas. Algorithm \ref{BlindalgoSU} with 10 iterations and 1 pilot compared against MRC with 104 pilots.}
    \label{BERvsAnt}
\end{figure}

We next examine the effect of the number of receive antennas on the blind decoding. Fig. \ref{BERvsAnt} shows the BER plotted as a function of the number of receive antennas $N_r$. As expected, the BER reduces with increasing $N_r$ due to the increase in array gain. 

\begin{figure*}[!t]
\centering
\subfloat[]{\includegraphics[width=0.9\columnwidth]{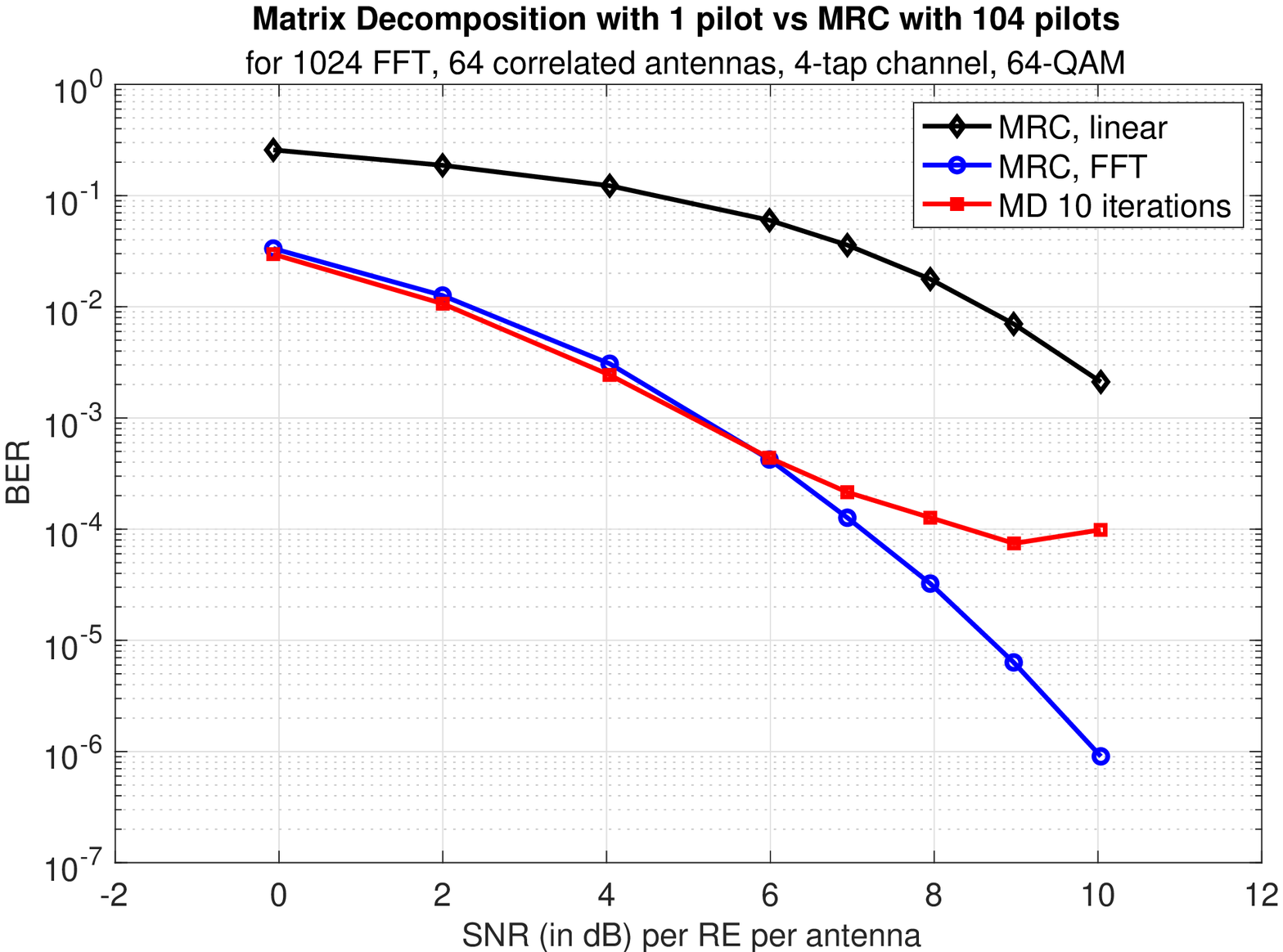}%
\label{Blind10corrSU}}
\hfil
\subfloat[]{\includegraphics[width=0.9\columnwidth]{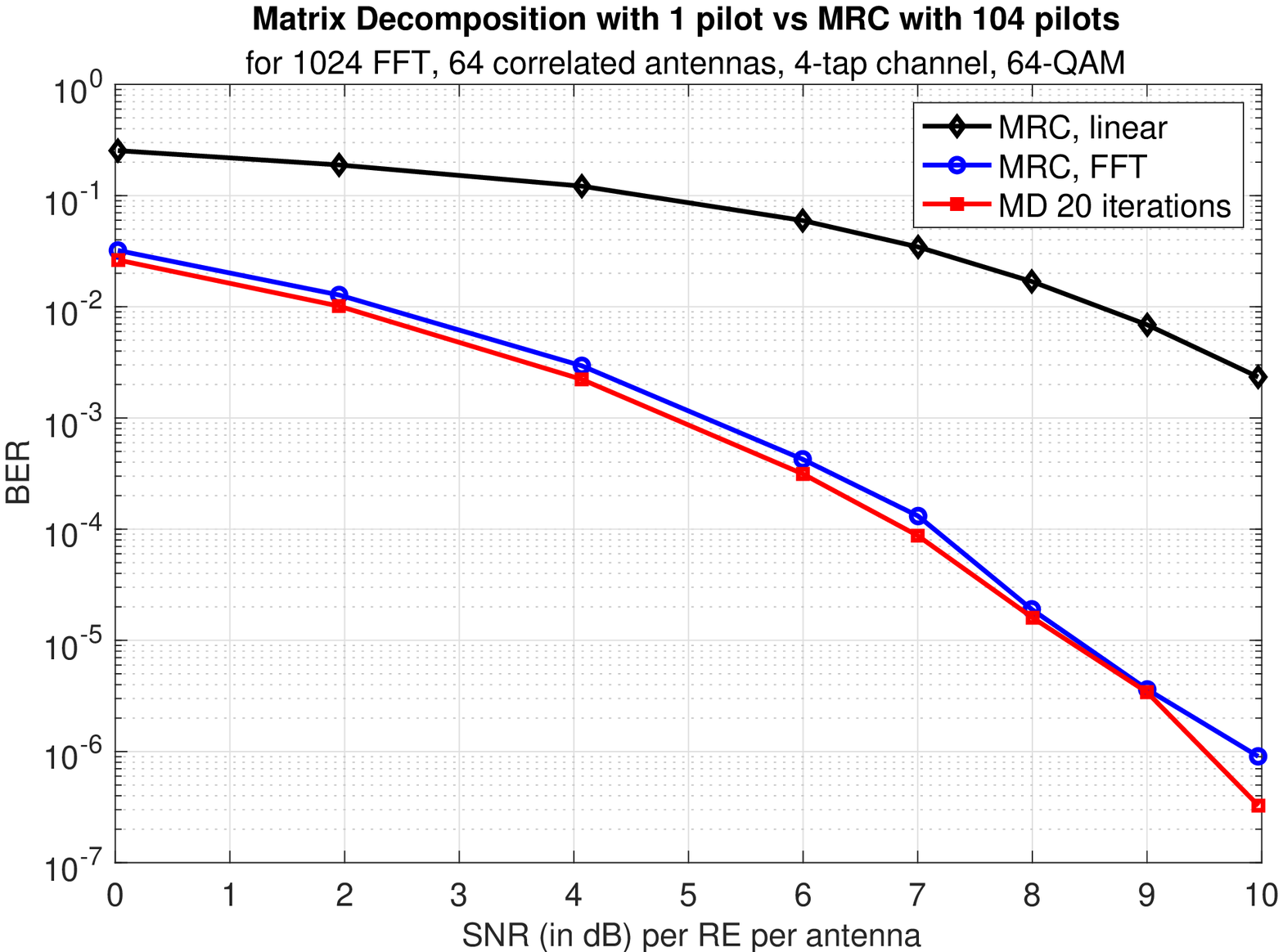}%
\label{Blind20corrSU}}
\caption{Uncoded BER for receive antenna correlation coefficient of 0.7 with (a) 10 iterations, and (b) 20 iterations of alternating minimization in Algorithm \ref{BlindalgoSU}. More iterations are needed to match the MRC performance in the presence of antenna correlation.}
\label{BlindcorrSU}
\end{figure*}

We also evaluate the effect of receive antenna correlation on our method. For this, we use the exponential correlation model for uniform linear array given in \cite{loyka2001}. While the results discussed so far assume zero antenna correlation, Fig. \ref{Blind10corrSU} shows the BER for the proposed method for 10 iterations of Algorithm \ref{BlindalgoSU} with receive antenna correlation coefficient of 0.7. We observe that the method is unable to match the MRC performance at high SNR unlike the earlier zero correlation case. However, when the number of iterations is increased to 20 in Fig. \ref{Blind20corrSU}, we observe that the proposed method matches the MRC performance across all SNRs even in the presence of correlation. This is similar to what is observed in the case of fronthaul compression using this decomposition in the presence of antenna correlation \cite{aswathy2023ojcoms}.

\begin{figure*}[!t]
\centering
\subfloat[]{\includegraphics[width=0.9\columnwidth]{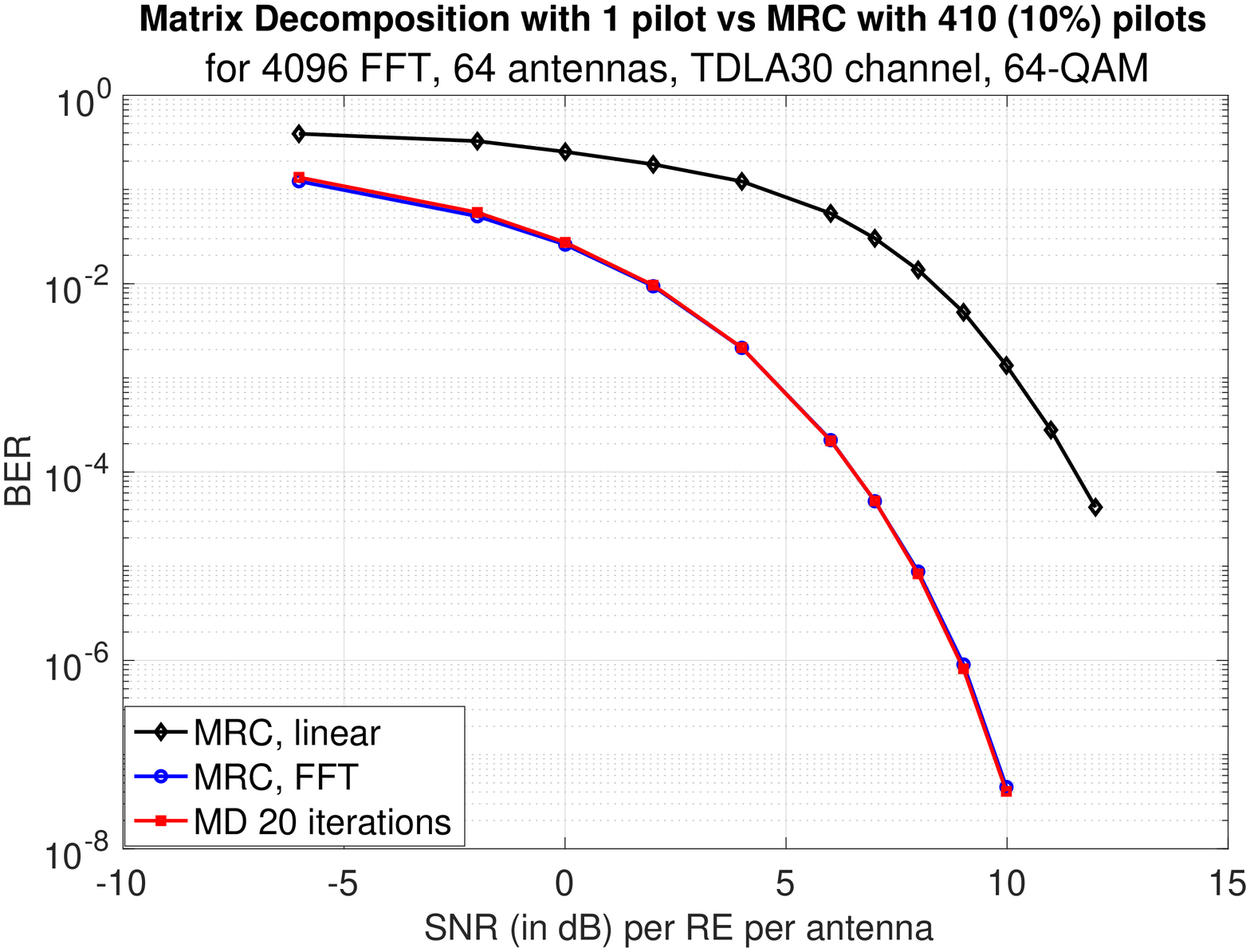}%
\label{BlindTDLA4096nocorr}}
\hfil
\subfloat[]{\includegraphics[width=0.9\columnwidth]{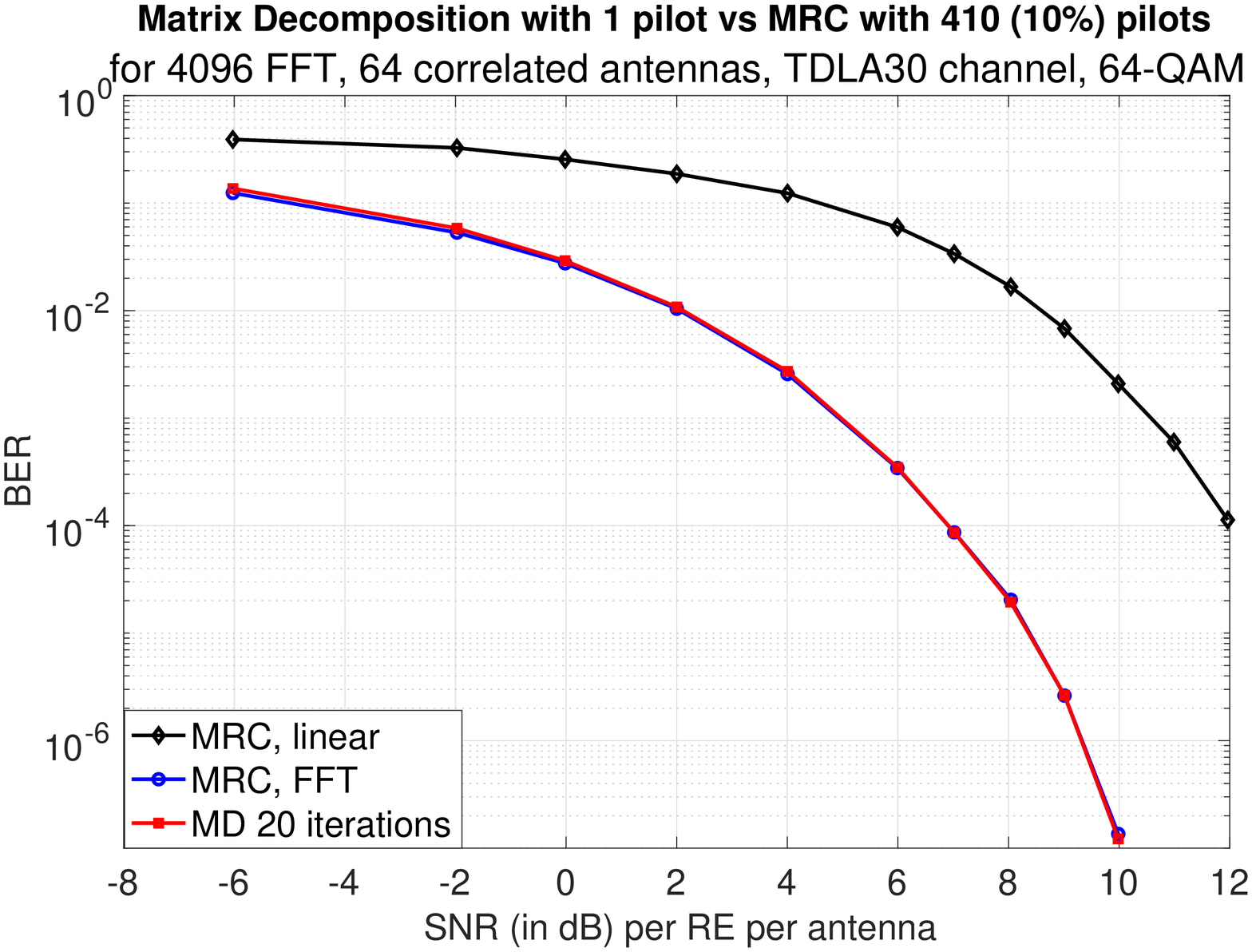}%
\label{BlindTDLA4096corr}}
\caption{BER of 20 iterations of the proposed blind matrix decomposition method (Algorithm \ref{BlindalgoSU}) for 4096-FFT in TDLA30 channel with 12 taps with (a) no receive antenna correlation, and (b) receive antenna correlation of 0.7.}
\label{BlindTDLA4096}
\end{figure*}

We now apply the algorithm to a realistic multi-path fading environment provided in the 3GPP technical specifications for 5G NR \cite{38104}. We examine the algorithm performance for a user with a standard 100 MHz bandwidth, which requires an FFT size of 4096, transmitting to the base station through a TDLA30 channel, which has a delay spread of 30 ns (maximum Doppler frequency of 5 Hz for a subcarrier spacing of 30 kHz). This channel model has 12 taps for 4096-FFT and 30 kHz subcarrier spacing. Fig. \ref{BlindTDLA4096nocorr} shows the BER for 20 iterations of the algorithm in this case when the receive antenna correlation is zero. We can observe that the proposed method with just 1 pilot again matches the performance of the conventional method with 400$\times$ (10\% of the OFDM symbol) more pilots. Fig. \ref{BlindTDLA4096corr} proves that the method can still match the performance of the conventional method even in the presence of receive antenna correlation.

\section{Blind Estimation for Multi-User MIMO}\label{MU-MIMO}

We now extend this method for use in the multi-user MIMO scenario, where more than one user is assigned the same time-frequency resources by the base station. The $N$ subcarriers of the OFDM symbol contain data from multiple users in an overlapping manner. Suppose there are $N_u$ single-antenna users scheduled simultaneously in one OFDM symbol, then the received signal matrix in the frequency domain becomes
\begin{equation}
    \mathbf{Y_f} = \Sigma_{u=1}^{N_u} \mathbf{X_{f}}(u)\mathbf{F_{L}H_{t}}(u) + \mathbf{W_f},
    \label{eq.18}
\end{equation}
where $\mathbf{X_{f}}(u)$ denotes the $N\times N$ diagonal matrix with the frequency domain ($M$-QAM) symbols of user $u$ as its diagonal, and $\mathbf{H_{t}}(u)$ is the $L\times N_r$ time-domain channel matrix of user $u$. We have assumed a maximum of $L$ multi-paths for each user. 

\par{Conventionally, orthogonal pilot sequences would be assigned to these $N_u$ users for estimation of their respective channels and consequent decoding of the data. Moreover, the subcarriers assigned for the pilots of any user cannot be assigned for data of any other user to avoid pilot-data interference, resulting in wastage of spectrum. Our objective, again, is to blindly estimate $\mathbf{X_{f}}(u)$ and $\mathbf{H_{t}}(u)$ for $u = 1, 2, ..., N_u$ without such pilot sequences. We introduce the following notation to express this objective function in a form similar to $\eqref{eq.3}$ used for the single-user case. We rewrite \eqref{eq.18} as
\begin{equation}
    \mathbf{Y_f}= [\mathbf{X_{f}}(1) \hspace{1mm}\mathbf{X_{f}}(2) \hspace{1mm}... \hspace{1mm}\mathbf{X_{f}}(N_u)] \diag(\mathbf{F_L})
    \begin{bmatrix}
        \mathbf{H_{t}}(1) \\
        \mathbf{H_{t}}(2) \\
        \vdots \\
        \mathbf{H_{t}}(N_u) \\
    \end{bmatrix} + \mathbf{W_f},
    \label{eq.19}
\end{equation}
where $\diag(\mathbf{F_L})$ is the block matrix 
\begin{equation*}
\diag(\mathbf{F_L})=
  \begin{bmatrix}
   \mathbf{F_{L}} & 0 & . & . & . & 0 \\
    0 & \mathbf{F_{L}} & . & . & . & 0\\
    . & . & .  &   &   &. \\
    . & . &   &  . &   &. \\
    . & . &   &   & .  &. \\
    0 & 0 & . & . & . & \mathbf{F_{L}}\\
  \end{bmatrix}_{N N_{u}\times L N_{u}} .
\end{equation*}
We denote the data estimate of user $u$ by the $N\times N$ diagonal matrix $\mathbf{\hat{X}}(u)$, the time-domain channel estimate of user $u$ by the $L\times N_r$ matrix $\mathbf{\hat{H}}(u)$, and let $\mathbf{\hat{X}} = [\mathbf{\hat{X}}(1) \hspace{1mm}\mathbf{\hat{X}}(2) \hdots \mathbf{\hat{X}}(N_u)]$ and $\mathbf{\hat{H}}= [\mathbf{\hat{H}^{T}}(1)\hspace{1mm}\mathbf{\hat{H}^{T}}(2) \hdots \mathbf{\hat{H}^{T}}(N_u)]^T$. Now the objective becomes to estimate $\mathbf{\hat{X}}$ and $\mathbf{\hat{H}}$ such that $\Vert \mathbf{Y_{f} - \hat{X}}\diag(\mathbf{F_{L})\hat{H}} \Vert_{F}^{2}$ is minimized, which is similar to \eqref{eq.3}.}

\par{We again use the alternating minimization approach to solve this problem by assuming an initial value for $\mathbf{\hat{X}}$ and finding the $\mathbf{\hat{H}}$ that satisfies
\begin{equation}
    \mathbf{\hat{H}} = \argmin_{\mathbf{H}} \Vert \mathbf{Y_{f} - \hat{X}}\diag(\mathbf{F_{L})H} \Vert_{F}^{2}.
    \label{eq.20}
\end{equation}
The regularized linear least squares solution for \eqref{eq.20} is
\begin{equation}
    \mathbf{\hat{H}} = (\mathbf{\hat{X}}\diag(\mathbf{F_{L}))^{\dagger}Y_{f}},
    \label{eq.21}
\end{equation}
with the regularization term $\mu \mathbf{I_{LN_u}}$ included in the pseudo-inverse. We apply the $\mathbf{\hat{H}}$ found in \eqref{eq.21} to find the $\mathbf{\hat{X}}$ that minimizes
\begin{equation}
    \mathbf{\hat{X}} = \argmin_{\mathbf{X}} \Vert \mathbf{Y_{f} - X}\diag(\mathbf{F_{L})\hat{H}} \Vert_{F}^{2}.
    \label{eq.22}
\end{equation}
We denote the frequency-domain channel estimate of each user by $\mathbf{B}(u) = \mathbf{F_{L}\hat{H}}(u)$ so that 
\begin{equation*}
    \mathbf{B} =  \diag(\mathbf{F_L})\mathbf{\hat{H}} = [\mathbf{B^{T}}(1) \hspace{1mm} \mathbf{B^{T}}(2) \hspace{1mm} \hdots \hspace{1mm} \mathbf{B^{T}}(N_u)]^{T}.
\end{equation*}
The solution to \eqref{eq.22}, $\mathbf{\hat{X}}$ has to be in the form of $N_u$ diagonal $N\times N$ matrices stacked horizontally. Substituting the above notations in \eqref{eq.22}, we have
\begin{equation}
\begin{aligned}
    ||\mathbf{Y_f} - \mathbf{X}\diag(\mathbf{F_L})\mathbf{\hat{H}}||_{F}^2 &= ||\mathbf{Y_f} - \mathbf{X B}||_{F}^2, \\
    &= \sum_{n=1}^{N} || \mathbf{y_{n}^{T}} - \mathbf{\hat{x}^{T}(n)B_n} ||_{2}^{2},
    \label{eq.23}
\end{aligned}    
\end{equation}
where $\mathbf{y_{n}^{T}}$ is the $n$-th row of $\mathbf{Y_f}$, $\mathbf{\hat{x}^{T}(n)}$ is a row vector made up of the $n-$th diagonal element of each user data estimate $\mathbf{\hat{X}(u)}$, i.e., $\mathbf{\hat{x}^{T}(n)} = [ \hat{x}_{n}(1) \hspace{1mm} \hat{x}_{n}(2) \hspace{1mm} ... \hspace{1mm} \hat{x}_{n}(N_u)]$, and $\mathbf{B_n}$ is the $N_{u} \times N_{r}$ matrix obtained by stacking the $n$-th row of $\mathbf{B}(u)$, for $u = 1,2,...,N_u$, vertically. Here, $\mathbf{y_{n}^{T}}$ is the received signal at the $N_r$ antennas for the $n$-th subcarrier, $\mathbf{\hat{x}^{T}(n)}$ represents the estimated transmitted signal on the $n$-th subcarrier for all the users, and $\mathbf{B_n}$ is the estimated frequency-domain channel matrix for the $n$-th subcarrier for all the users. Thus, for each subcarrier $n \in 1,2,...,N$, this leads to the linear least squares solution
\begin{equation}
    \mathbf{\hat{x}^{T}(n)} = \mathbf{y_{n}^{T} (B_n)^{\dagger}}.
    \label{eq.24}
\end{equation} 
The solutions given in \eqref{eq.21} and \eqref{eq.24} are calculated alternately and iterated until the solutions converge or for a fixed number of iterations.}

\par{Although the above algorithm is similar to the multi-user fronthaul compression algorithm described in \cite{aswathy2023ojcoms}, the objective of the compression algorithm was the convergence  
$\mathbf{\hat{Y}_{f} = \hat{X}}\diag(\mathbf{F_{L})\hat{H}} \rightarrow \mathbf{Y_f}$, whereas here, we need the convergence $\mathbf{\hat{X}}(u) \rightarrow \mathbf{X_{f}}(u)$, for $u = 1, 2, ..., N_u$. Similar to the single-user case discussed in sections \ref{blind} and \ref{initpoint}, this convergence again hinges on the initial point used for $\mathbf{\hat{X}}$, the estimation of which is described below.}

\subsection{Initial Point for Multi-User MIMO}

We once again turn to the SVD of $\mathbf{Y_f}$, specifically the left singular vectors that span the column space of $\mathbf{Y_f}$ which contains the user symbols $\mathbf{X_{f}}(1), \mathbf{X_{f}}(2), \hdots, \mathbf{X_{f}}(N_u)$. Following the procedure we used to analyze the single user case, we break down the channel of each user to gain insights about the left singular vectors in the multi-user scenario.

Let the average tap powers of user $u$'s time-domain channel be $\rho_{1}^{u}, \rho_{2}^{u}, \hdots, \rho_{L}^{u}$. Then the time-domain channel of user $u$ is given by
\begin{equation}
    \mathbf{H_{t}}(u) = \begin{bmatrix}
        \sqrt{\rho_{1}^{u}} & & \\ & \ddots & \\ & & \sqrt{\rho_{L}^{u}}
    \end{bmatrix}\begin{bmatrix}
        \mathbf{h_{1}^{u}} \\ \mathbf{h_{2}^{u}} \\ \vdots \\ \mathbf{h_{L}^{u}}
    \end{bmatrix},
\end{equation}
where $\mathbf{h_{i}^{u}} = [h_{1,1}^{u}, h_{1,2}^{u}, \hdots, h_{1,N_r}^{u}]$ are user $u$'s channel coefficients for the $N_r$ antennas through path $i$, and the frequency-domain channel is given by
\begin{align}
    \mathbf{F_{L}H_{t}}(u) &= {\scriptstyle\begin{bmatrix}
        \mathbf{f_{1}} & \mathbf{f_{2}} & \hdots & \mathbf{f_{L}}
    \end{bmatrix}\begin{bmatrix}
        \sqrt{\rho_{1}^{u}} & & \\ & \ddots & \\ & & \sqrt{\rho_{L}^{u}}
    \end{bmatrix}\begin{bmatrix}
        \mathbf{h_{1}^{u}} \\ \mathbf{h_{2}^{u}} \\ \vdots \\ \mathbf{h_{L}^{u}}
    \end{bmatrix}}, \\
    &= {\scriptstyle\begin{bmatrix}
        \sqrt{\rho_{1}^{u}}\mathbf{f_{1}h_{1}^{u}} + \sqrt{\rho_{2}^{u}}\mathbf{f_{2}h_{2}^{u}} + \hdots + \sqrt{\rho_{L}^{u}}\mathbf{f_{L}h_{1}^{u}}
    \end{bmatrix}.}
\end{align}

Therefore, $\mathbf{Y_f}$ can be expressed as
\begin{align}
    \mathbf{Y_f} &= \Sigma_{u=1}^{N_u} \mathbf{X_{f}}(u)\mathbf{F_{L}H_{t}}(u) + \mathbf{W_{f}},\\
    &= \Sigma_{u=1}^{N_u} \mathbf{X_{f}}(u)(\Sigma_{i=1}^{L}
        \sqrt{\rho_{i}^{u}}\mathbf{f_{i}h_{i}^{u}}) + \mathbf{W_f}.
    \label{eq.YfMU}
\end{align}

We now analyze the relationship between the left singular vectors of $\mathbf{Y_f}$ and the expression in \eqref{eq.YfMU}. We begin with the simplest case where all the $N_u$ users have the same power and similar channel power delay profiles. If we assume that tap $k$ is the dominant tap for all the users, i.e.,
\begin{equation}
    \rho_{k}^{u} \gg \rho_{i}^{u}, i = 1, 2, \hdots, L, i \neq k, u = 1, 2, \hdots, N_{u},
\end{equation}
then there are $N_u$ directions which have a similar fraction of the power in $\mathbf{Y_f}$, given by $\sqrt{\rho_{k}^{u}}\mathbf{X_{f}}(u)\mathbf{f_{k}h_{k}^{u}}$, corresponding to each user's dominant tap. These directions are captured by the top $N_u$ singular vectors of $\mathbf{Y_f}$. However, the singular vectors will be a linear combination of these directions (due to the orthogonality constraints of the SVD). We can express this as a system of linear equations for the left singular vectors $\mathbf{u_{j}}, j = 1, 2, \hdots, N_u$ where
\begin{equation}
    \mathbf{u_{j}} \approx \alpha_{j}^{1}\mathbf{X_{f}}(1)\mathbf{f_{k}} + \alpha_{j}^{2}\mathbf{X_{f}}(2)\mathbf{f_{k}} + \hdots + \alpha_{j}^{N_u}\mathbf{X_{f}}(N_u)\mathbf{f_{k}}.
\end{equation}
We can express this compactly in the matrix notation as
\begin{equation}
    \begin{bmatrix}
        \mathbf{u_{1}^{T}} \\ \mathbf{u_{2}^{T}} \\ \vdots \\ \mathbf{u_{N_u}^{T}}
    \end{bmatrix} \approx \begin{bmatrix}
        \alpha_{1}^{1} & \alpha_{1}^{2} & \hdots & \alpha_{1}^{N_u} \\
        \alpha_{2}^{1} & \alpha_{2}^{2} & \hdots & \alpha_{2}^{N_u} \\
        \vdots & \vdots & \ddots & \vdots \\
        \alpha_{N_u}^{1} & \alpha_{N_u}^{2} & \hdots & \alpha_{N_u}^{N_u} \\
    \end{bmatrix}\begin{bmatrix}
        (\mathbf{X_{f}}(1)\mathbf{f_{k}})^{T} \\ (\mathbf{X_{f}}(2)\mathbf{f_{k}})^{T} \\ \vdots \\ (\mathbf{X_{f}}(N_u)\mathbf{f_{k}})^{T}
    \end{bmatrix}.
    \label{eq.MU_SV}
\end{equation}

\begin{figure}
    \centering
    \includegraphics[width = 0.9\columnwidth]{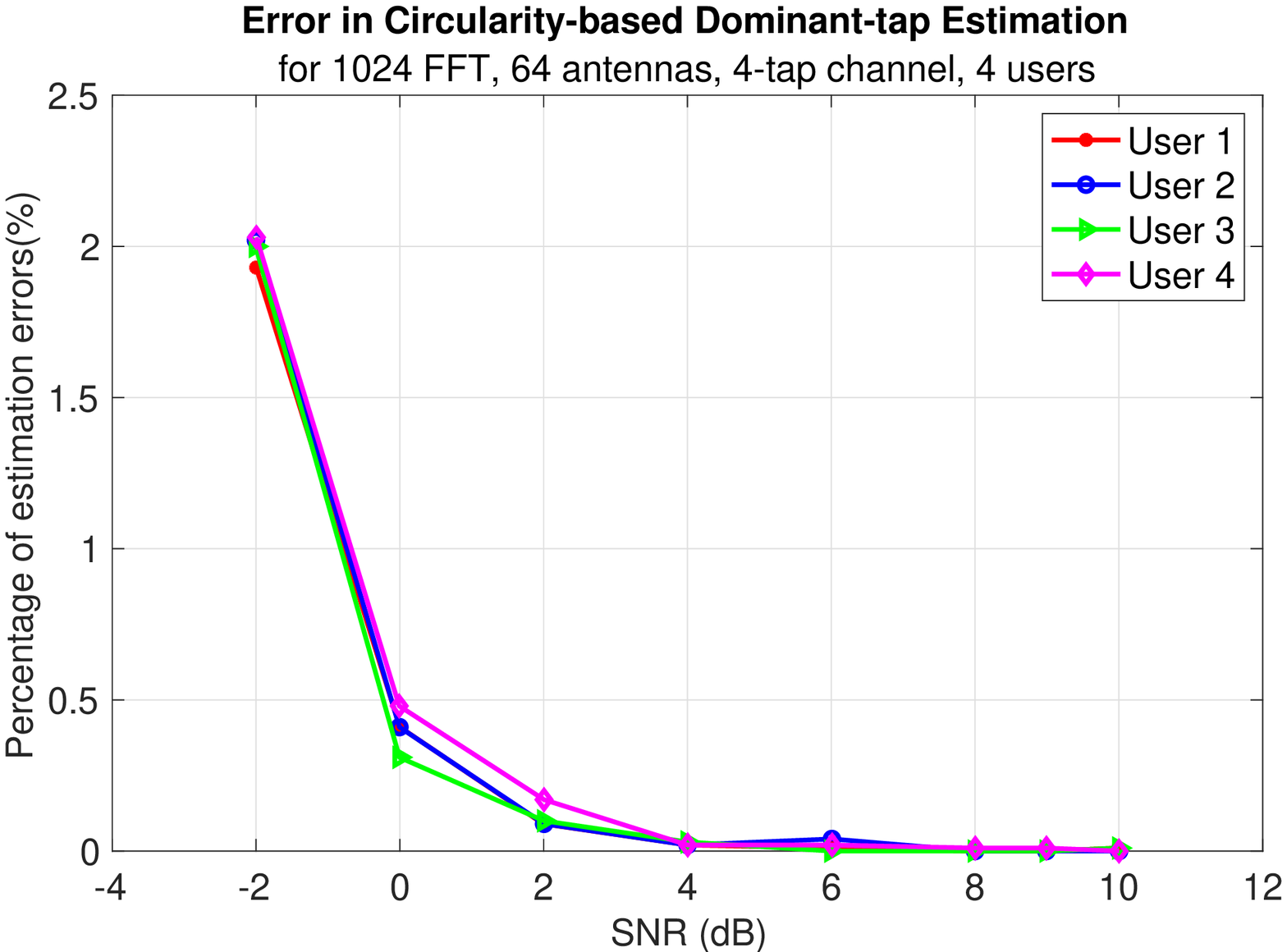}
    \caption{Percentage of errors in the estimation of the dominant tap of each user using the circularity-based method for 4 users in the system. Each user has a different channel power delay profile with 4 taps, one of which is dominant.}
    \label{Blind4udomtap}
\end{figure}

Using $\mathbf{A}$ to denote the matrix of linear coefficients (whose estimation is described in the next section), we can find 
\begin{equation}
    \begin{bmatrix}
        \mathbf{Z^{T}}(1) \\ \mathbf{Z^{T}}(2) \\ \vdots \\ \mathbf{Z^{T}}(N_u)
    \end{bmatrix} = \mathbf{A^{-1}}\begin{bmatrix}
        \mathbf{u_{1}^{T}} \\ \mathbf{u_{2}^{T}} \\ \vdots \\ \mathbf{u_{N_u}^{T}}
    \end{bmatrix},
    \label{eq.MU_IP}
\end{equation}
where $\mathbf{Z}(u) = \mathbf{\hat{X}}(u)\mathbf{f_{k}}$. From this, we can calculate the initial point for each user, $\mathbf{\hat{X}}(u)$ if the dominant tap $k$ is known. We use the circularity-based estimation method described in Algorithm \ref{BlindIPcirc} to estimate the dominant tap and the initial point for each user. The error performance of the circularity-based dominant-tap estimation method for 4 users, each with a different channel power delay profile containing 4 taps is given in Fig. \ref{Blind4udomtap}. We note that this method of initialization works even in the general case where different users have different power delay profiles and dominant taps, as long as they have similar received powers at the base station (which is generally enabled by uplink power control) and the number of scatterers/taps are low.

\subsection{Estimation of the Linear Coefficients}

We need to estimate the matrix $\mathbf{A}$ in order to find the initial points for each user from equation \eqref{eq.MU_IP}. We use a single pilot symbol per user to estimate $\mathbf{A}$. The subcarrier allocated as pilot for one user is not allocated to any other user, for data/pilot. Let $p_{u}$ be the pilot subcarrier for user $u$. Then, the pilot symbol of user $u$, $x_{p_u}(u)$ is known at the base station. We substitute this into equation \eqref{eq.MU_SV} to obtain
\begin{equation}
    {\scriptstyle\begin{bmatrix}
        u_{1}(p_1) & u_{1}(p_2) & \hdots & u_{1}(p_{N_u}) \\
        u_{2}(p_1) & u_{2}(p_2) & \hdots & u_{2}(p_{N_u}) \\
        \vdots & \vdots & \ddots & \vdots \\
        u_{N_u}(p_1) & u_{N_u}(p_2) & \hdots & u_{N_u}(p_{N_u})
    \end{bmatrix}} \approx \mathbf{A}{\scriptstyle\begin{bmatrix}
        x_{p_1}(1)\\ x_{p_2}(2) \\ \vdots \\ x_{p_{N_u}}(N_u)
    \end{bmatrix},}
\end{equation}
where $u_{j}(p_u)$ is the component $p_u$ of $j$-th left singular vector of $\mathbf{Y_f}$. Then each coefficient in $\mathbf{A}$ can be calculated as
\begin{equation}
    \alpha_{j}^{u} = u_{j}(p_u)/x_{p_u}(u), \hspace{3mm} j = 1, 2, \hdots N_{u}, \hspace{2mm} u = 1, 2, \hdots N_{u}.
    \label{eq.A}
\end{equation}

The matrix of linear coefficients, $\mathbf{A}$ calculated as above can now be substituted in equation \eqref{eq.MU_IP} to obtain the $\mathbf{Z}(u)$ corresponding to each user, $u \in 1, 2, \hdots, N_u$. The circularity-based dominant tap estimation method can be applied to each user's $\mathbf{Z}(u)$ independently, from which their initial points $\mathbf{\hat{X}}(u)$ can be readily calculated.

\subsection{De-rotation within Alternating Minimization}

In the multi-user case, the output of the alternating minimization algorithm is a set of scaled and rotated constellations, with a different scaling factor to be estimated for each user. This is again estimated using the single pilot allocated to each user. The residual rotation is corrected within the alternating minimization algorithm similar to the single user case described before (in Algorithm \ref{BlindalgoSU}). These steps are summarized in Algorithm \ref{BlindalgoMU}.

\renewcommand\footnoterule{}      
\begin{algorithm}
\caption{Blind Estimation and De-rotation for Multi- User}\label{BlindalgoMU}\footnotetext{$\mathbf{y_1}$ denotes the first column of matrix $\mathbf{Y_f}$, $\dagger$ denotes matrix pseudo-inverse, $*$ denotes complex conjugation, $a(n,r)$ denotes element at row $n$ and column $r$ of matrix $\mathbf{A}$ and $x_{k}(n)$ denotes $n-$th diagonal element of $\mathbf{\hat{X}_k}$.}
\label{alg:loop}
\begin{algorithmic}[1]
\State Input $\mathbf{Y_f}$, number of iterations $T$, initial point $\mathbf{\hat{X} = [\hat{X}}(1) \mathbf{\hat{X}}(2) \hdots \mathbf{\hat{X}}(N_u)]$, pilot subcarriers $\{p_{1}, p_{2}, \hdots, p_{N_u}\}$, pilot symbols $\{x_{p_1}(1), x_{p_2}(2), \hdots, x_{p_{N_u}}(N_u)\}$.
\State Initialize $k = 1, \mathbf{B} = [\hspace{1mm}]$
\Repeat
    \State $\mathbf{\hat{H}}\gets \mathbf{(\hat{X}\diag(\mathbf{F_L))}^{\dagger}}\mathbf{Y_f}$
    \State $\mathbf{B} \gets \diag(\mathbf{F_{L})\hat{H}}$
    \For {$n \gets 1$ to $N$}
        \State $\mathbf{\hat{x}^{T}(n)} = \mathbf{y_{n}^{T} (B_n)^{\dagger}}$
    \EndFor
    \If{$k > 3$}
        \If{$k = 4$}
            \For {$u \gets 1$ to $N_u$}
                \State $\lambda_{u} \gets \hat{x}_{p_u}(u)/x_{p_u}(u)$
                \State $\mathbf{\hat{X}}(u) = \mathbf{\hat{X}}(u)/\lambda_{u}$
            \EndFor
        \EndIf
        \For {$u \gets 1$ to $N_u$}
            \State $\mathbf{\hat{X}_q}(u)\leftarrow$ QAM constellation closest to $\mathbf{\hat{X}}(u)$
            \State $\mathbf{\hat{X}}(u) \leftarrow \mathbf{\hat{X}_q}(u)$
        \EndFor
    \EndIf
    \State $k \gets k+1$
\Until $k = T$\\
\Return $\mathbf{\hat{X}}, \mathbf{\hat{H}}$
\end{algorithmic}
\end{algorithm}

\begin{figure}
    \subfloat[$\mathbf{u_1}$]{
        \includegraphics[width=0.47\columnwidth]{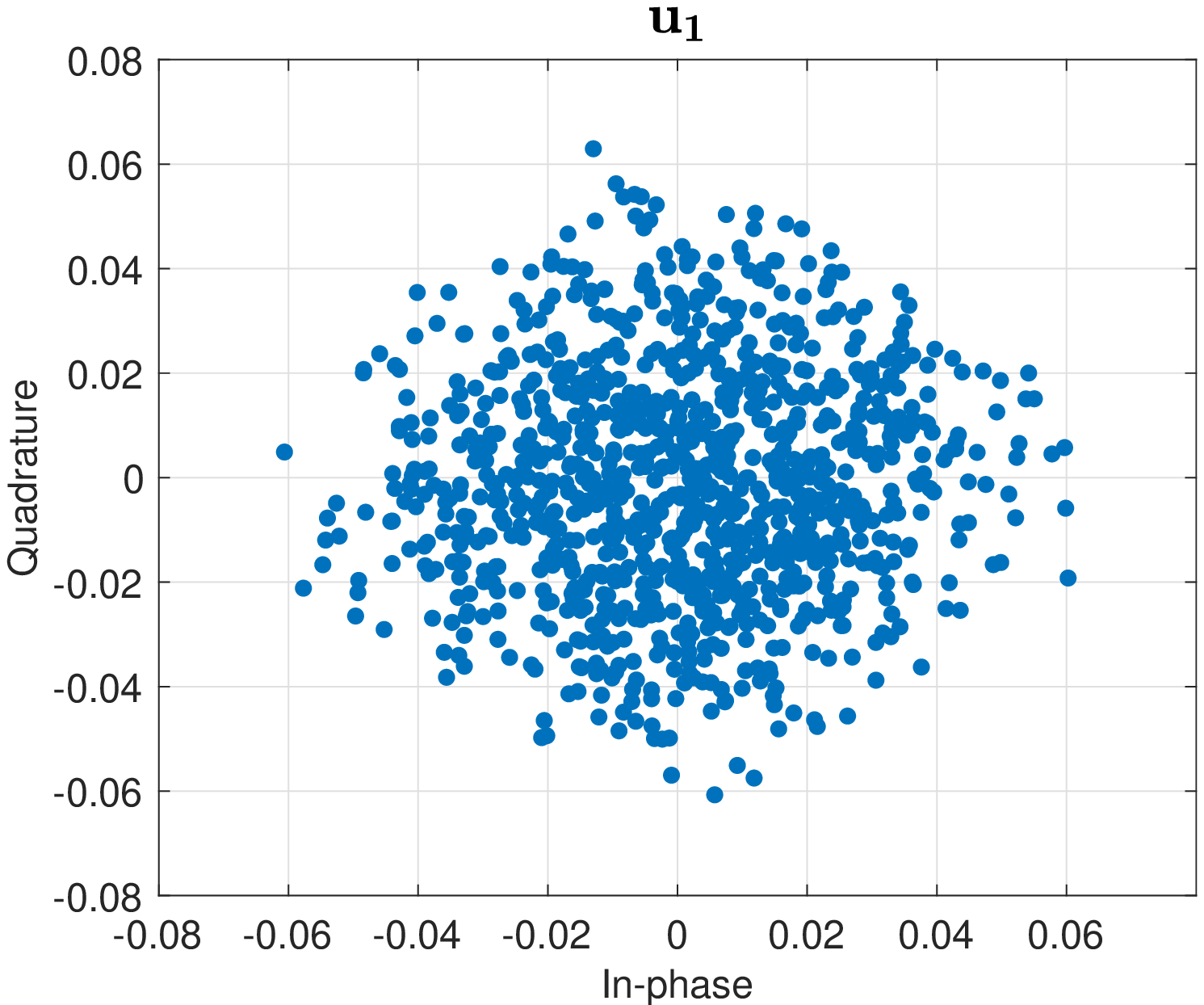}
        \label{U1}
    }
    \hfill
    \subfloat[$\mathbf{u_2}$]{
        \includegraphics[width=0.47\columnwidth]{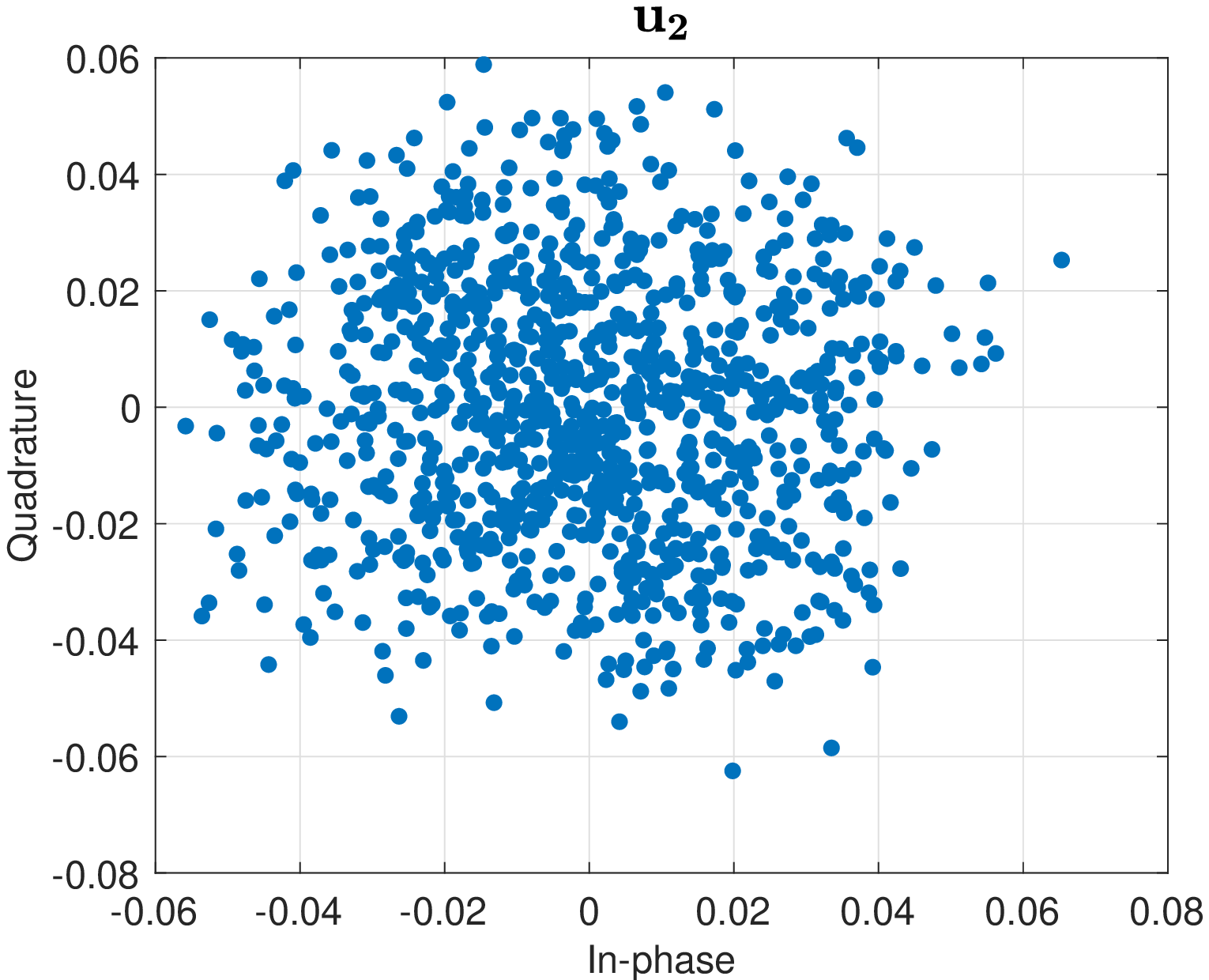}
        \label{U2}
    }
    \hfill
    \subfloat[$\mathbf{u_3}$]{
        \includegraphics[width=0.47\columnwidth]{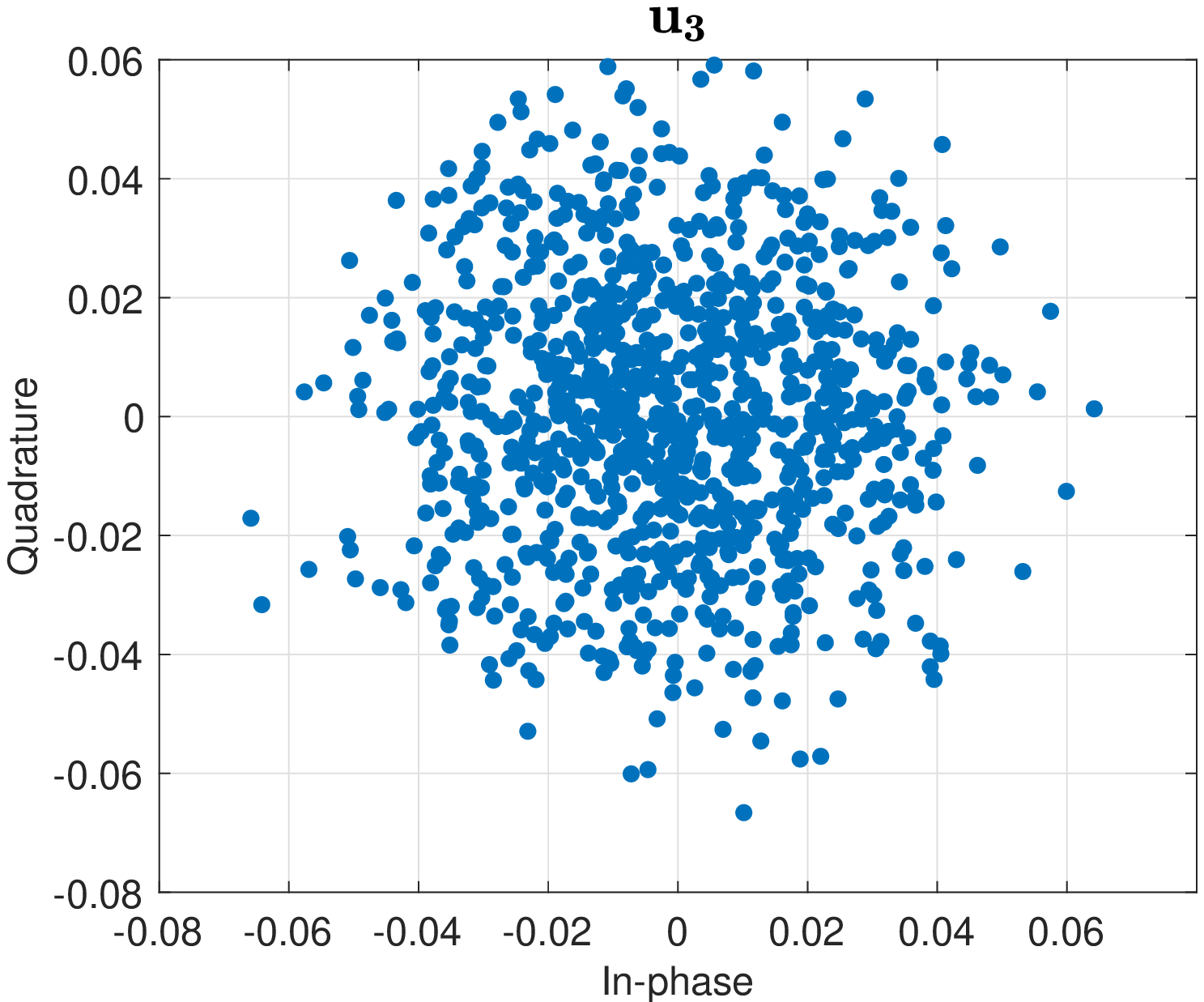}
        \label{U3}
    }
    \hfill
    \subfloat[$\mathbf{u_4}$]{
        \includegraphics[width=0.47\columnwidth]{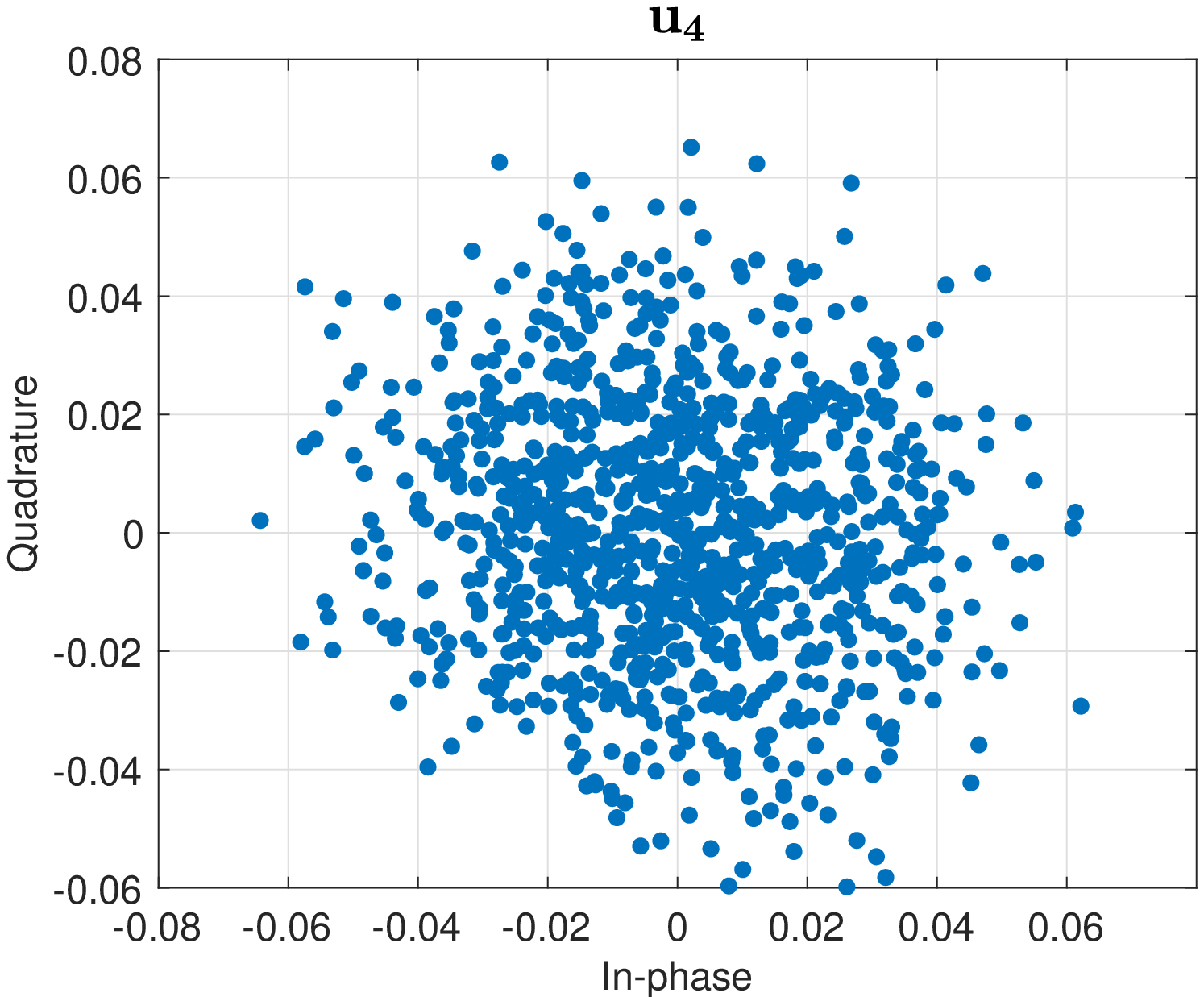}
        \label{U4}
    }
\caption{Scatter-plots of the top four left singular vectors for a 4-user system at 5dB SNR.}
\label{BlindU}
\end{figure}

\begin{figure}
    \centering
    \subfloat[Initial $\mathbf{\hat{X}}(1)$]{
        \includegraphics[width=0.65\columnwidth]{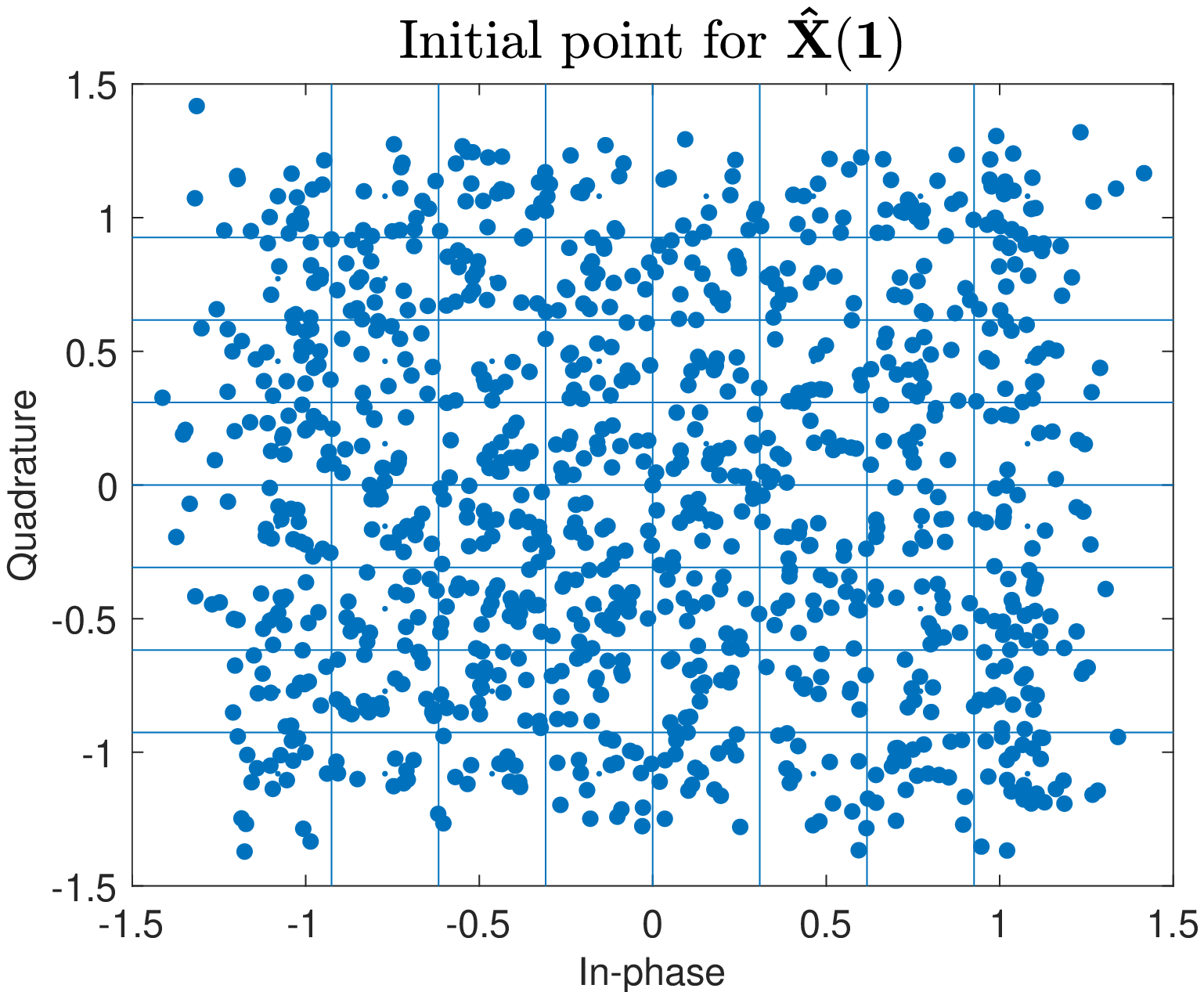}
        \label{X1_1}
    }
   \par
    \centering
    \subfloat[$\mathbf{\hat{X}}(1)$ after 4 iterations]{
        \includegraphics[width=0.65\columnwidth]{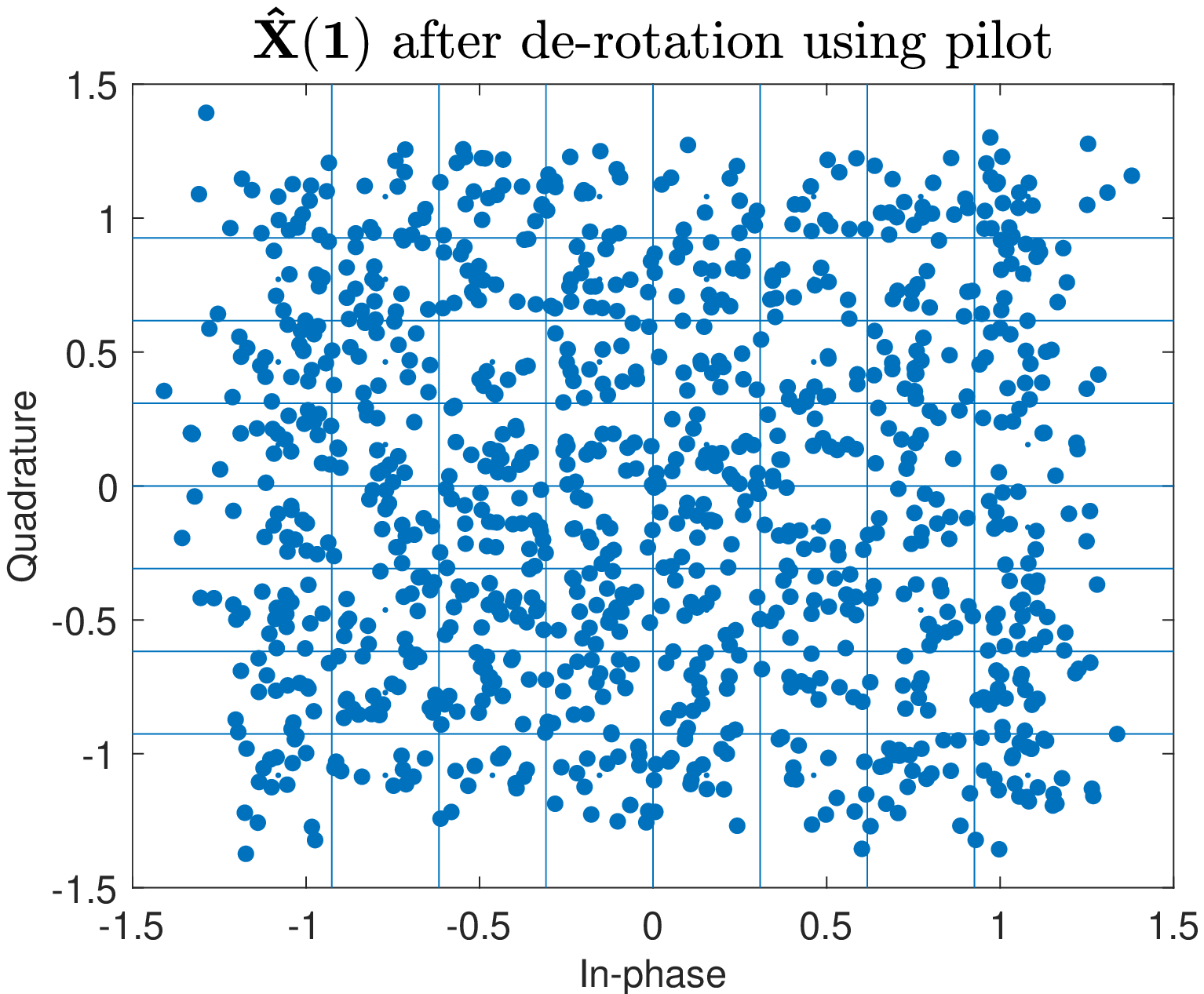}
        \label{X1_lambda}
    }
   \par
    \centering
    \subfloat[Final $\mathbf{\hat{X}}(1)$, after 10 iterations]{
        \includegraphics[width=0.65\columnwidth]{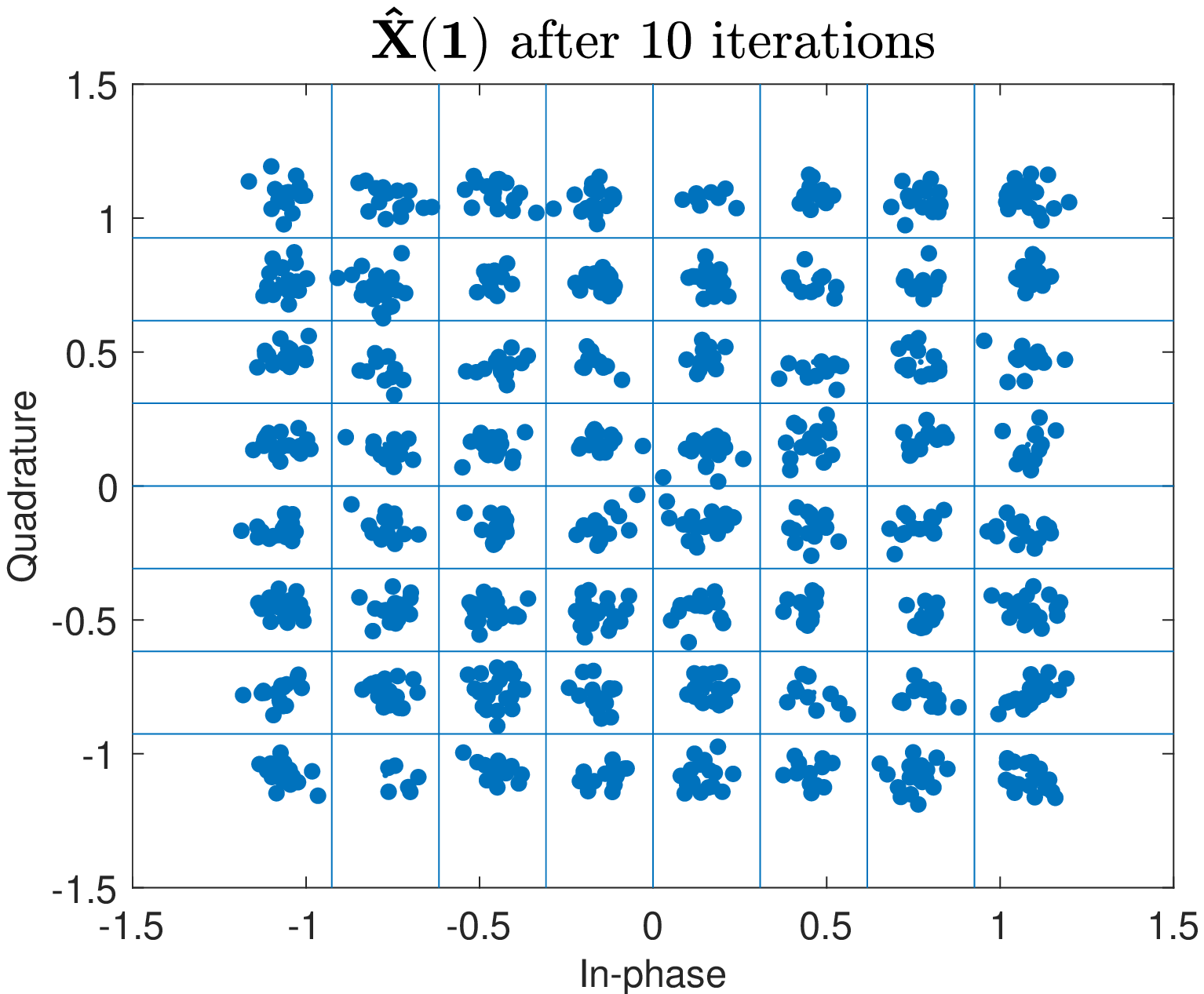}
        \label{X1_10}
    }
\caption{Scatter-plots of $\mathbf{\hat{X}}(1)$ obtained at different stages of Algorithm \ref{BlindalgoMU} at 5dB SNR for a 4-user MIMO system with 1024 FFT and 64 receive antennas.}
\label{BlindMUX1}
\end{figure}

We now take a look at the scatter-plots obtained at various steps in the above algorithm to see how it progressively arrives at the correct constellation for each user. Fig. \ref{BlindU} shows the scatter-plots of the top 4 left singular vectors, $\mathbf{u_{1}, u_{2}, u_{3},}$ and $\mathbf{u_{4}}$ of the received signal matrix, $\mathbf{Y_f}$ at an SNR of 5dB for each user in a 4-user MIMO system. From these, we estimate the dominant channel tap of each user using the circularity-based Algorithm \ref{BlindIPcirc} and calculate their initial points for $\mathbf{\hat{X}}(u)$ using equations \eqref{eq.MU_IP}-\eqref{eq.A}. Fig. \ref{X1_1} shows the initial point thus obtained for User 1. We begin alternating minimization and after 4 iterations calculate the scaling/rotation factor for each user, $\lambda_{u}$ using their pilot symbol. The de-rotated constellation of $\mathbf{\hat{X}}(1)$ (for User 1) obtained after this step is shown in Fig. \ref{X1_lambda}. We observe that there is still a slight residual rotation at this stage. We then perform a series of mapping of the $\mathbf{\hat{X}}(u)$ to the closest $M$-QAM constellation along with the alternating minimization for the rest of the iterations. The final $\mathbf{\hat{X}}(u)$ obtained after 10 such iterations for User 1 is shown in Fig. \ref{X1_10}, which is ``cleaner" and now completely de-rotated compared to the previous stage. The scatter-plots of the constellations at the corresponding stages for the rest of the users also show a similar trend.

\section{Error Performance for the Multi-User Case}\label{BER_MU}

We evaluate the performance of Algorithm \ref{BlindalgoMU} with four users in the system against that of conventional pilot-based channel estimation and demodulation methods. We use 64-QAM modulation, 1024-point FFT, 64 receive antennas at the base station, and 4-tap pedestrian channel for each user. Our method uses four pilots, i.e., one pilot per user, to resolve the scaling and rotation. 

We compare the proposed method against the conventional method which uses 10\% of the OFDM symbol for pilots, i.e, a total of 104 pilot subcarriers with 26 pilots allocated for each user in a non-overlapping manner. An FFT-based interpolation is used to estimate the frequency-domain channel of each user from these pilots. These channel estimates are then used to apply the minimum mean squared error (MMSE) equalization on the received signal matrix, $\mathbf{Y_f}$ to recover the user data.

\begin{figure}
    \centering
    \includegraphics[width = 0.9\columnwidth]{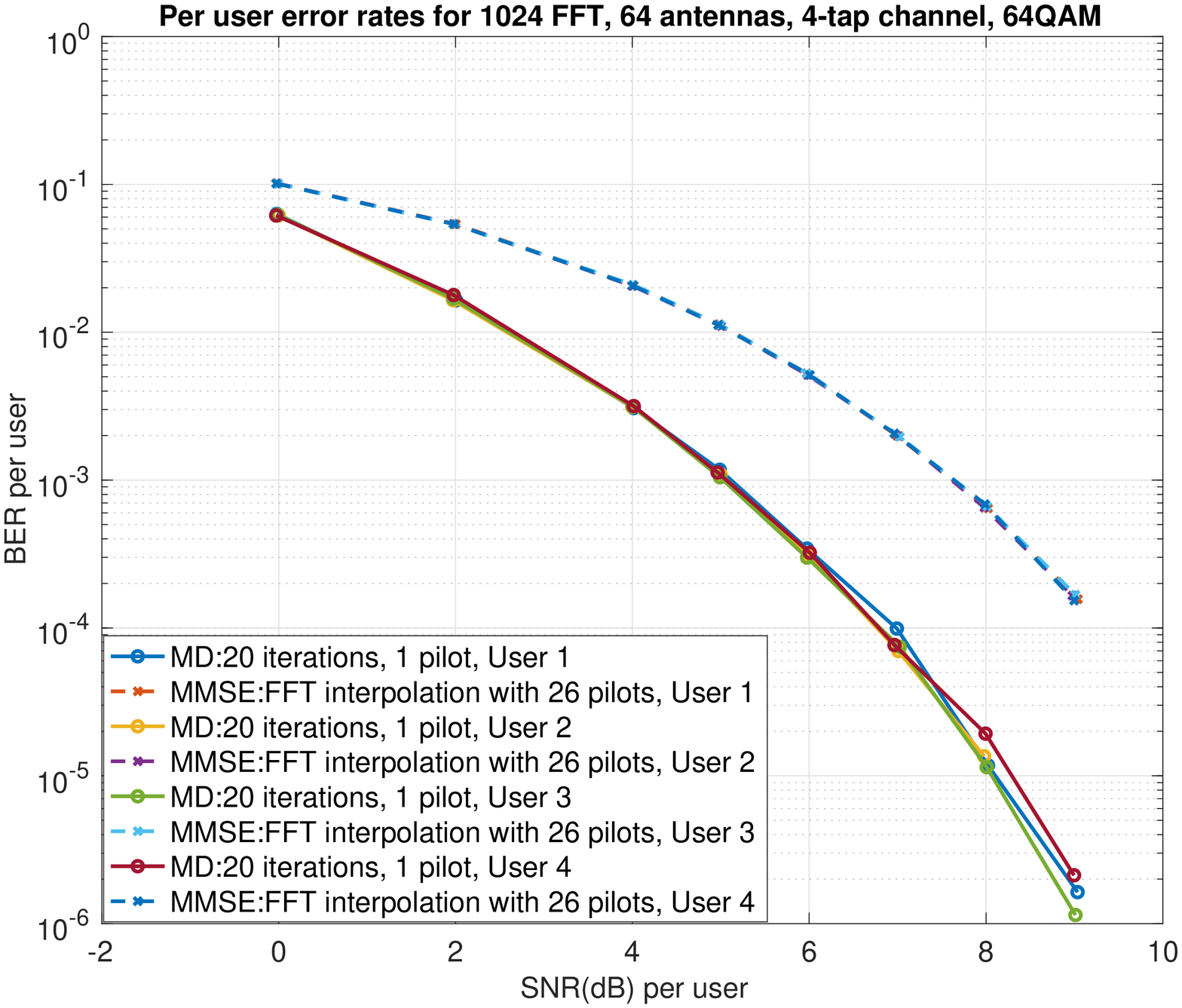}
    \caption{BER for 4 users with equal SNR and same channel power delay profile}
    \label{BER4u_samePDP}
\end{figure}

We first evaluate the case where all the four users have the same channel power delay profiles and equal SNR at the receiver. Fig. \ref{BER4u_samePDP} shows the error rates for this case where we assume that the dominant channel tap is known at the base station. The performance of our method at 20 iterations beats that of the conventional method while using only a single pilot per user by about 3 dB.

\begin{figure}
    \centering
    \includegraphics[width = 0.9\columnwidth]{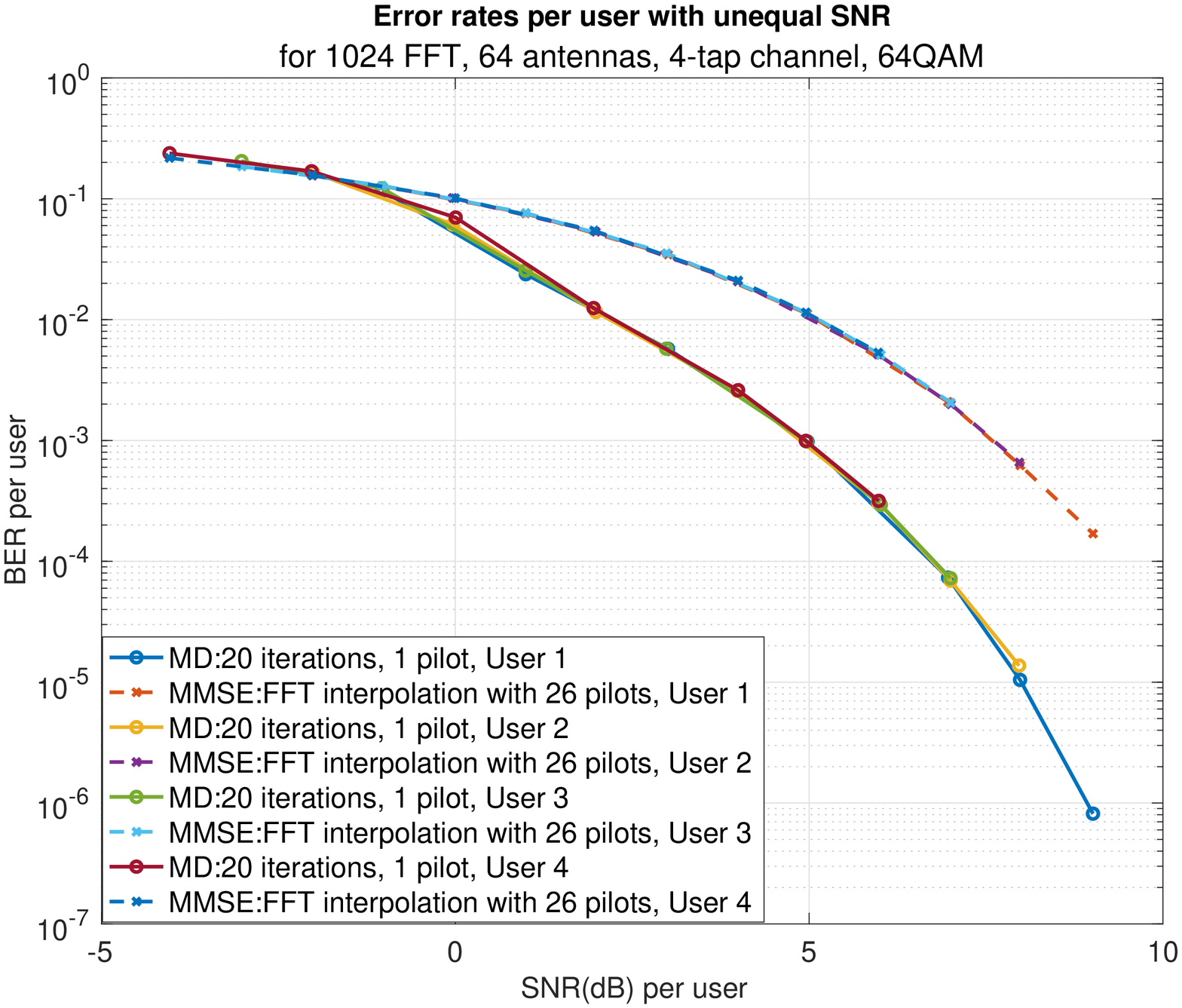}
    \caption{BER for 4 users with different SNRs and same channel profile. The maximum SNR difference between any 2 users is 3 dB.}
    \label{BER4u_uneqSNR}
\end{figure}

We next look at the scenario where four users with different SNRs are multiplexed in the same time-frequency resources and have the same power delay profiles. Fig. \ref{BER4u_uneqSNR} shows the error rates for this case, where there is a minimum of 1 dB and a maximum of 3 dB difference in SNRs between the users multiplexed together. We observe that the SNR difference between the users does not affect the per user error rates of the proposed method, which still outperforms that of the conventional method.

\begin{figure}
    \centering
    \includegraphics[width = 0.9\columnwidth]{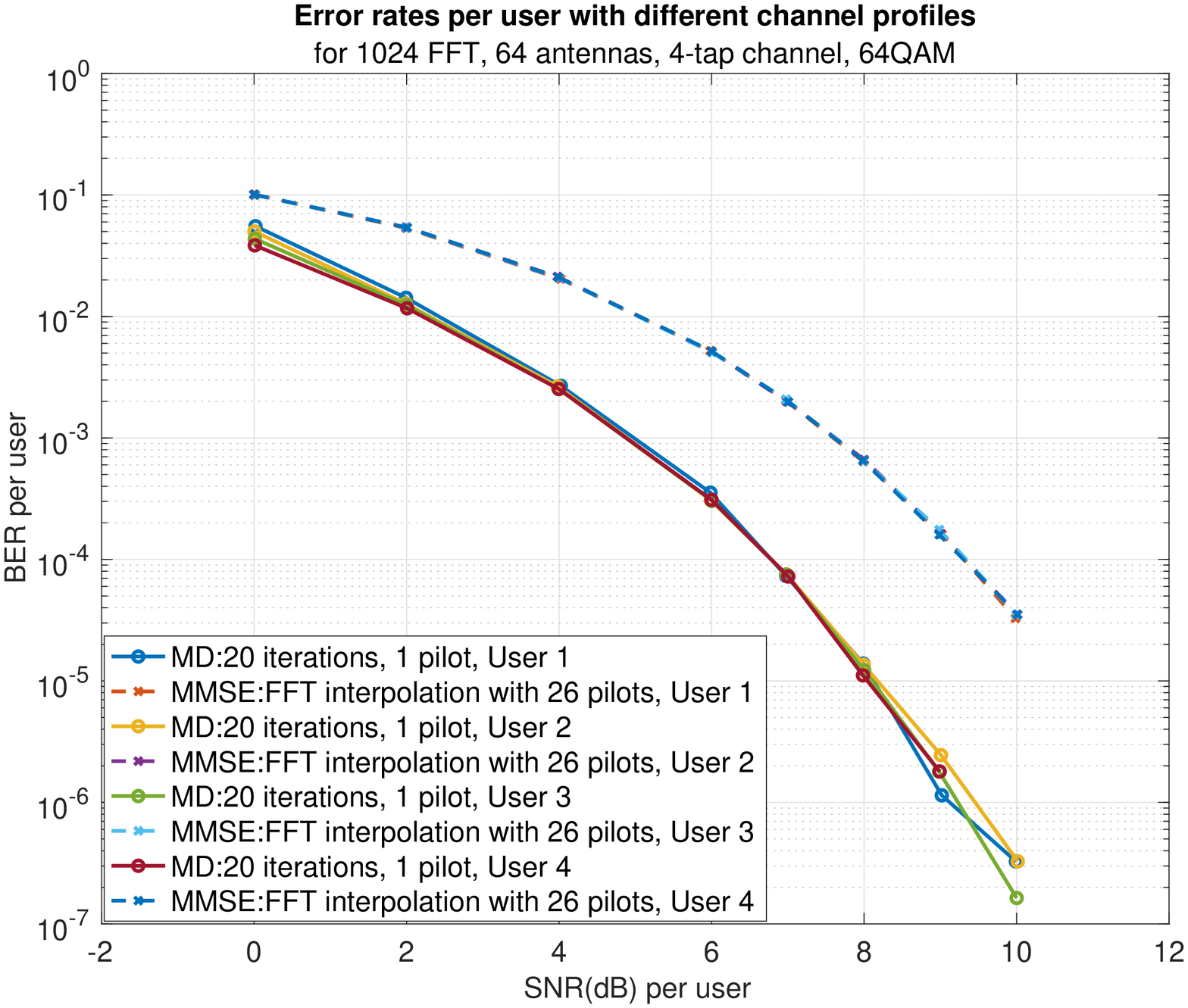}
    \caption{BER for 4 users with equal SNR but different channel power delay profiles}
    \label{BER4u_diffPDP}
\end{figure}

We next evaluate the scenario where the four users have different channel power delay profiles and the dominant tap information of the users is not available at the receiver. Fig. \ref{BER4u_diffPDP} shows the error performance of our method using the circularity-based initial point estimation for the dominant taps. We observe that there is no drop in performance of our method (20 iterations) even with the higher number of unknowns at the receiver.

\begin{figure*}[!t]
\centering
\subfloat[]{\includegraphics[width=0.9\columnwidth]{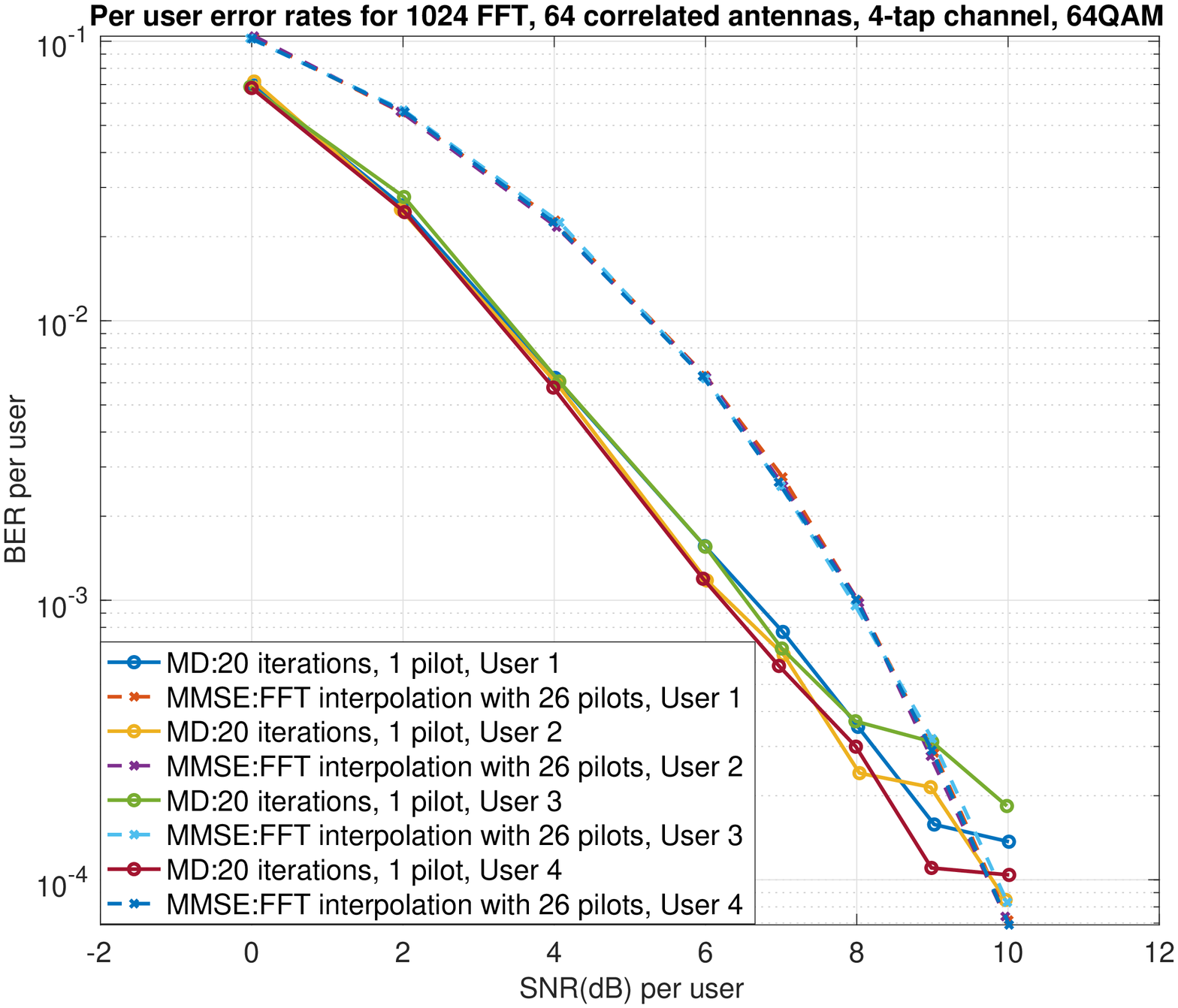}%
\label{BER4u_corr20dem}}
\hfil
\subfloat[]{\includegraphics[width=0.9\columnwidth]{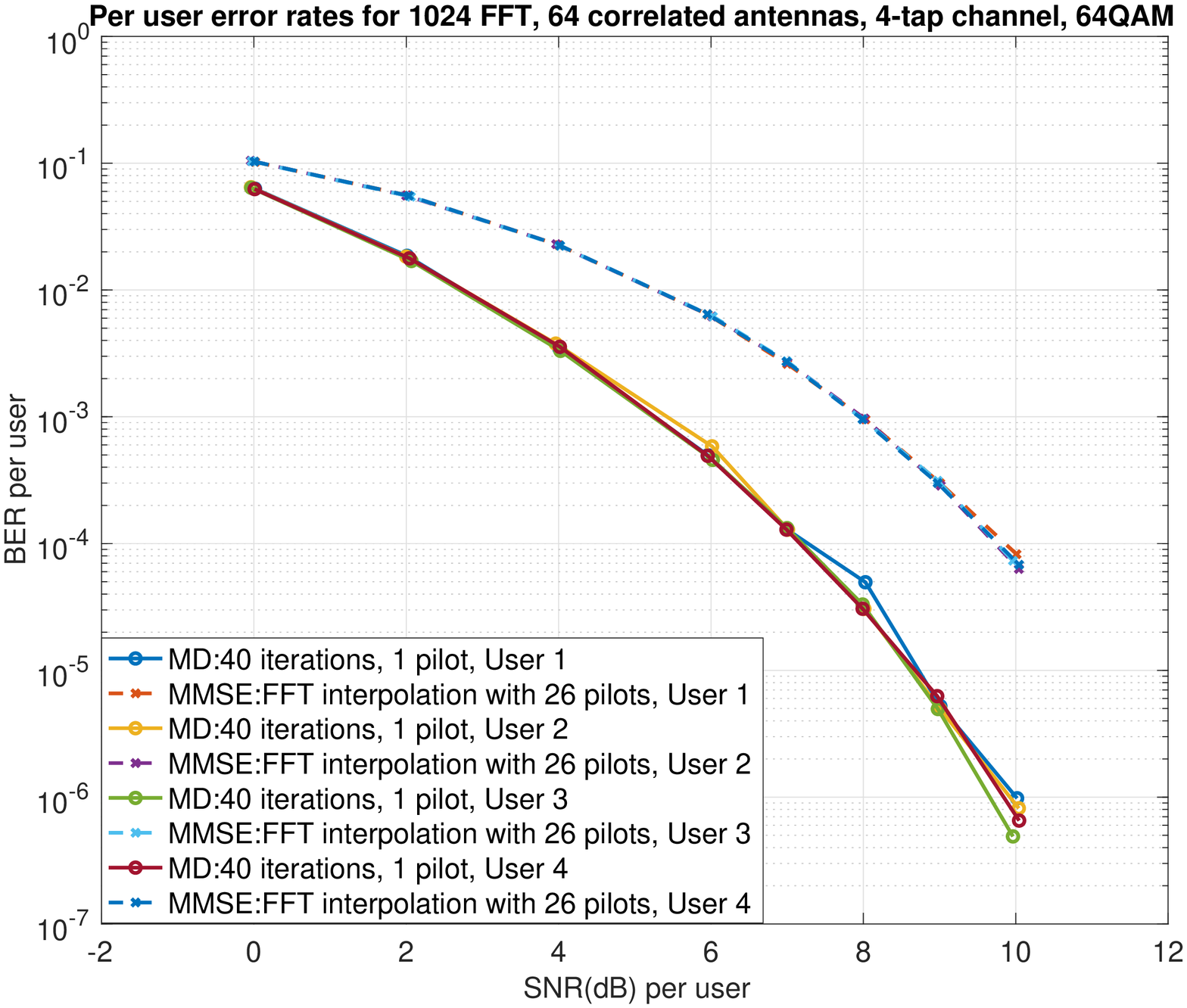}%
\label{BER4u_corr40dem}}
\caption{BER for 4 users in the system and correlated receive antennas (correlation coefficient of 0.7) with (a) 20 iterations, and (b) 40 iterations of alternating minimization in Algorithm \ref{BlindalgoMU}. The users have equal SNR. More iterations are needed to match the MRC performance in the presence of antenna correlation at higher SNRs.}
\label{BER4u_corr}
\end{figure*}

The results above are for zero correlation among the receive antennas. We next evaluate our method's performance in the presence of receive antenna correlation. We assume a uniform linear array with exponential correlation and correlation coefficient of 0.7 \cite{loyka2001}. Fig. \ref{BER4u_corr20dem} shows the BER for 20 iterations of Algorithm \ref{BlindalgoMU}, assuming the users have equal SNR at the receiver. We observe that the proposed method performs better than the conventional method at low SNRs. However, at high SNRs, the errors from residual rotation degrades its performance and its BER becomes closer to that of the conventional method. Increasing the number of iterations to 40 allows our method to improve the correction of the residual rotation, and our method performs better than the conventional method across all SNRs as demonstrated in Fig. \ref{BER4u_corr40dem}. 

\section{Spectral Usage Gains}\label{Gain}

The proposed method uses only a single pilot per user to estimate the user signal and channel, resulting in an almost 100\% utilization of the OFDM symbol/frequency spectrum for data and throughput. In the previous section, we showed that it outperforms the error rates of the conventional pilot-based channel estimation and demodulation method. In this section, we quantify the spectral usage gains/improvement in throughput obtained using the proposed algorithm when the conventional method tries to match the error performance of our method. 

We perform Monte Carlo simulations of an OFDM system with 1024-FFT, 64 antennas, 4-tap channel using both the proposed method and the conventional one to demodulate the user data. With a single-user in the system, the conventional pilot-based method with FFT interpolation for channel estimation needs at least 10.2\% of the OFDM symbol to be occupied by pilots to match the BER of the proposed method given in Algorithm \ref{BlindalgoSU}. Therefore, its spectral utilization (for data/throughput) is only 89.8\%, compared to 99.9\% for the proposed method. Thus, our method offers a 10\% gain in throughput. In a multi-user system with four users, the conventional method requires orthogonal/non-overlapping pilots for each user. To match the BER of our method which uses only four pilots (99.6\% spectral utilization), the conventional method requires at least 25\% of the OFDM symbol to be filled with only pilots to estimate the channel accurately enough using FFT-interpolation. Therefore, our method offers a 24.6\% gain in throughput. These results are summarized in Table \ref{Table_Spectrum}. 

\begin{table}
\caption{Comparison of Spectral Utilization of Matrix Decomposition (MD)-based blind estimation and Pilot-based conventional estimation in Single-User and Multi-User Systems for the same Bit Error Rate.}
\centering
\begin{tabular}{ |c||c|c|  }\hline
$N_u$ & MD-based blind decoding & Pilot-based conventional decoding\\\hline\hline
1 & 99.9\% & 89.8\% \\\hline
4 & 99.6\% & 75\% \\\hline
\end{tabular}
\label{Table_Spectrum}
\end{table}

\section{Alternating Minimization under Temporally Correlated Channels}\label{Tempcorr}

The proposed matrix decomposition technique using alternating minimization  blindly estimates the user signal in each OFDM symbol received at a massive MIMO base station individually. Since the complexity of the algorithm will scale with the dimensions of the received signal, applying many iterations of it to each OFDM symbol independently could increase the complexity of the receiver. In this section, we discuss a way to reduce the complexity of applying the algorithm on the received symbols by leveraging the underlying temporal correlations in the system. For this, we look at the case when the base station receives data spanning multiple OFDM symbols from the same user. We examine how correlated the channel is across these symbols and discuss ways to reduce the number of iterations of the algorithm required for successive symbols from the same user for different user mobility conditions.

The coherence time of the channel determines the frequency of channel estimation required. If the channel coherence time spans multiple OFDM symbols, the channel estimated from the previous symbols is used for equalization/MIMO combining. Once the channel is estimated, conventional equalization/MIMO combining techniques like zero forcing or maximal ratio combining have very low complexity compared to blind demodulation algorithms \cite{daumont2010analytical}, \cite{zhang2017blind}. Typically, blind demodulation algorithms have to be executed for every symbol since they estimate only the user symbol, which keeps changing, unlike the channel. Thus, what they save on pilot overhead is offset by the increased complexity of executing the algorithms. However, the blind estimation method using matrix decomposition proposed in this work estimates both the user symbols and the channel simultaneously. Therefore, the channel estimated from one OFDM symbol using the matrix decomposition algorithm can be re-used for the equalization of the subsequent OFDM symbols, depending on the channel coherence time. This avoids the need for executing the matrix decomposition algorithm for every OFDM symbol, whose complexity is high.

For rapidly varying channels, the pilot overhead becomes very high in conventional decoding. However, if the channel realizations, although varying, are correlated in time, our method provides a way to leverage this correlation to reduce the complexity of its execution for successive OFDM symbols in the following way. For the first OFDM symbol, the proposed blind estimation in Algorithm \ref{BlindalgoSU} with $\mathbf{\hat{X}}$ initialized using the top left singular vector of $\mathbf{Y_f}$ is used to obtain the estimates of both the user symbols and the channel. We recall that the alternating minimization algorithm can begin with an initial guess for either $\mathbf{\hat{X}}$ or $\mathbf{\hat{H}}$. Therefore, for the subsequent OFDM symbols, the channel estimated from the previous symbol can be used as initial value of $\mathbf{\hat{H}}$ for the alternating minimization. Since the channel realizations for successive OFDM symbols are correlated, this ensures that the initial points are not orthogonal to the true channel. The correlation also reduces the number of iterations of the algorithm required to reach convergence. Thus, the overall complexity of the algorithm for subsequent symbols is reduced by avoiding the need to calculate the top singular vector of $\mathbf{Y_f}$ as well as by decreasing the number of iterations of alternating minimization. 

In the following sections, we analyse the reduction in number of iterations of our algorithm required to blindly estimate successive OFDM symbols as a function of the temporal correlation of the channel, which in turn depends on the user mobility conditions.

\subsection{Temporally Correlated Channel Model}

In order to characterize the savings in algorithm complexity as a function of time, we first need to model the variation of the channel of a mobile user over time. There are three aspects that need to be factored into the model: the frequency selectivity of the channel due to its multi-path nature, the spatial correlation between the receive antennas, and the temporal correlation between the channels at different sampling instances at the receiver. 

Suppose there are $L$ independent paths in the user's channel with a power delay profile $P(\tau)$, where the fading coefficients for antenna $r$ at time $t$ from all the multi-paths is given by 
\begin{equation}
    \mathbf{g_{r}(t)} = \begin{bmatrix}
        \sqrt{\rho(\tau_1)} & & \\ & \ddots & \\ & & \sqrt{\rho(\tau_L)}
        \end{bmatrix}\begin{bmatrix} q_1 \\ \vdots \\ q_L\end{bmatrix}
        = \begin{bmatrix} g_{r}(t,\tau_1) \\ \vdots \\ g_{r}(t,\tau_L)\end{bmatrix},
\end{equation}
where $\rho(\tau_i)$ is the average power of path $i$ with delay $\tau_i$, and  $q \sim \mathcal{CN}(0,1)$ represents Rayleigh fading for each path. 

A discrete-time model for wide-sense stationary Rayleigh fading MIMO channel in the baseband is given in \cite{xiao2003discrete}. From this model, we obtain the autocorrelation of the channel coefficients at receive antenna $r$ as
\begin{equation}
    \mathbb{E}\{g_{r}(t,\tau)g_{r}^{*}(t-kT,\tau')\} = J_{0}(2\pi f_{d}kT)P(\tau)\delta(\tau-\tau'),
\end{equation}
where $T$ represents sampling duration/OFDM symbol time, $J_{0}$ is the zeroth-order Bessel function of the first kind, and $f_d$ is the Doppler shift given by
\begin{equation}
    f_d = \Big(\frac{v}{c}\Big)f_0,
\end{equation}
for carrier frequency $f_0$, speed of light $c$, and for a user moving at speed $v$. Let 
\begin{equation}
    \eta_k = J_{0}(2\pi f_dkT),
\end{equation}
represent the temporal correlation coefficient between the channel at time $t$ and the channel at time $t+kT$, i.e., correlation between channels which are $k$ OFDM symbol duration apart. Thus, the temporal correlation between the channels depends on the time gap between the symbols as well as the speed of the mobile user.

We also consider the spatial correlation between the antennas for a multi-antenna base station. We assume a single antenna user transmitting to a base station with $N_r$ receive antennas. Therefore, we need to factor in only the receive antenna correlation. We use the exponential correlation model for uniform linear array from \cite{loyka2001}, used in the previous sections. We denote this $N_r\times N_r$ spatial correlation matrix at the receiver by $\mathbf{R}$.

The Kalman filter-based channel generator given in \cite{kim2019channel} takes into account this spatial and temporal correlation to generate the channel coefficients at different channel instantiation intervals. It models the channel at different sampling instants as a first order Gauss-Markov process. The initial point is given by
\begin{equation}
    \mathbf{H_{0} = G_{0}R}^{\frac{1}{2}},
\end{equation}
where the $L\times N_r$ matrix $\mathbf{G_{0}}$ is formed as $\mathbf{G_{0} = [g_{1}(0) \hspace{1mm} g_{2}(0) \hspace{1mm} \hdots \hspace{1mm} g_{N_r}(0)]}$. The channel after $k$ OFDM symbol durations, i.e., after time $kT$ is then formulated as the Kalman filter output
\begin{equation}
    \mathbf{H_{k}} = \eta_{k}\mathbf{H_{0}} + \sqrt{1-\eta_k^{2}}\mathbf{G_{k}R}^{\frac{1}{2}},
\end{equation}\label{eq:TempcorrH}where $\eta_k$ is the temporal correlation and $\mathbf{R}$ is the spatial correlation, and $\mathbf{G_k}$ is formed similar to $\mathbf{G_0}$.

Thus, the $L\times N_r$ matrix $\mathbf{H_k}$ is the spatially and temporally correlated channel realization of the user, $k$ symbols after the first OFDM symbol of the user. The following sections will use the channel realizations generated in the above manner to characterize the reduction in complexity of the matrix decomposition algorithm over time as a function of different user speeds.

\subsection{Blind Estimation Performance over time}

The alternating minimization procedure given in Algorithm \ref{BlindalgoSU} decomposes each OFDM symbol received at the $N_r$ antennas, in the matrix $\mathbf{Y_f}$ into the user QAM symbols $\mathbf{\hat{X}}$ and the time-domain channel matrix $\mathbf{\hat{H}}$. For the first OFDM symbol from a user (designated as Symbol 0, received at time $t=0$), the alternating minimization in Algorithm \ref{BlindalgoSU} is initialized with $\mathbf{\hat{X}}$, calculated using the procedure given in Algorithm \ref{BlindIPvar}. For the subsequent OFDM symbols from the same user, the alternating minimization begins with the $\mathbf{\hat{H}}$ estimate from the previous symbol. 

We want to analyze the reduction in the number of iterations of alternating minimization required for blind estimation in the subsequent OFDM symbols initialized in this manner as a function of the time $t$ elapsed between the symbols. This depends on the temporal correlation between the channels of these symbols, modeled in the previous section, and is determined by the speed of the user. We also want to determine the span of time in which the advantage (if any) from the temporal correlations can be leveraged by the algorithm. We simulate a system with 64 exponentially correlated (with a correlation coefficient of 0.7) antennas at the base station, OFDM symbol size of 1024, multi-path channel with four taps, and 64-QAM user data. We consider a span of 10ms, constituting one frame in 5G NR, and focus on three equally spaced OFDM symbols from the same user in this frame, the first received at time $t=0$, the second at time $t = 5$ms and the third at $t = 10$ms. We benchmark our algorithm against conventional method that uses 104 pilots (10\%) for channel estimation and equalization. We run as many iterations of our algorithm as is required to match the BER obtained for the conventional method for each symbol.

\begin{figure}
    \centering
    \includegraphics[width = 0.8\columnwidth]{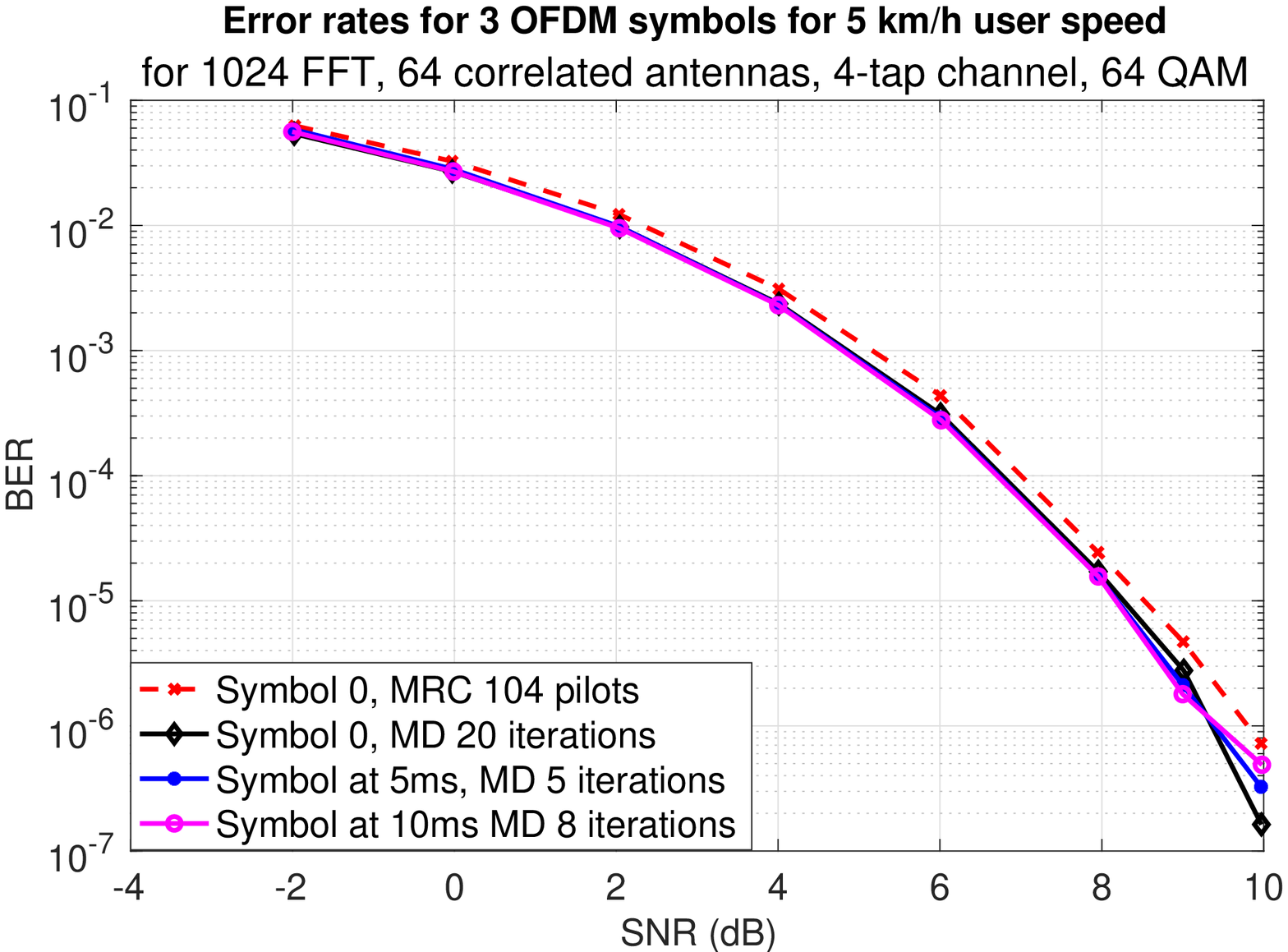}
    \caption{Uncoded BER for conventional decoding and 3 OFDM symbols (5ms apart) from a user moving at 5km/h speed, decoded using matrix decomposition method. $\mathbf{\hat{H}}$ estimate of first symbol used as initialization for the subsequent symbols. Values of the temporal correlation coefficient, $\eta_k$ are 0.96 and 0.87 for the symbols at 5ms and 10ms, respectively.}
    \label{Tempcorr_esti5kmph}
\end{figure}

Fig. \ref{Tempcorr_esti5kmph} shows the BER curves for the conventional method and the three OFDM symbols demodulated using the proposed algorithm, for a user moving at speed 5km/h. For the symbols received at 5ms and 10ms, the alternating minimization is initialized using the $\mathbf{\hat{H}}$ obtained from the first symbol. While the first symbol requires 20 iterations of alternating minimization, we note that the symbol at 5ms needs only 5 iterations to match the performance of the conventional method. However,the symbol at 10ms requires 8 iterations due to the fact that as the time elapsed between the symbols increases, the channel correlation between them decreases.

\begin{figure}
    \centering
    \includegraphics[width = 0.8\columnwidth]{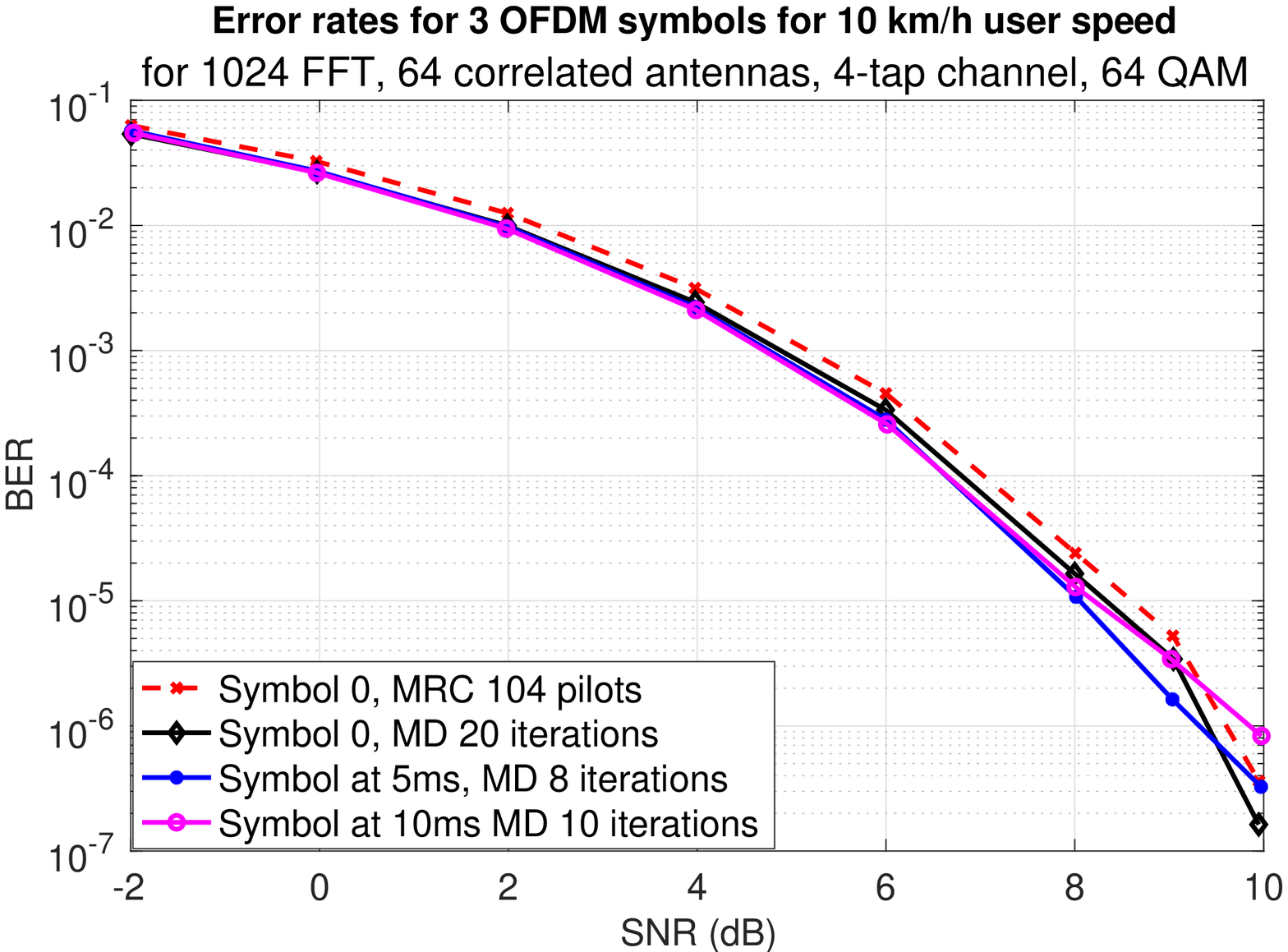}
    \caption{Uncoded BER for conventional decoding and 3 OFDM symbols (5ms apart) from a user moving at 10km/h speed, decoded using matrix decomposition method. $\mathbf{\hat{H}}$ estimate of first symbol used as initialization for the subsequent symbols. Values of the temporal correlation coefficient, $\eta_k$ are 0.87 and 0.53 for the symbols at 5ms and 10ms, respectively.}
    \label{Tempcorr_esti10kmph}
\end{figure}

Fig. \ref{Tempcorr_esti10kmph} shows the BER curves for a user speed of 10km/h. We observe that the number of iterations required for the second and third symbols has increased to 8 and 10 iterations, respectively, because increased user speed also decreases the channel correlation. These results are summarized in Table \ref{Tbl_empcorr_esti}.

\begin{table}
\centering
\caption{Number of iterations of alternating minimization for blind estimation of successive OFDM symbols as a function of user speed and time elapsed between symbols.}
\begin{tabular}{ |c|c|c| }
 \hline
 User speed & Received symbol & Number of\\
 $v$ & time $t$ & iterations\\
 \hline \hline
 \multirow{3}{*}{5 km/h} & 0 & 20\\\cline{2-3}
  & 5 ms & 5 \\\cline{2-3}
  & 10 ms & 8 \\\hline
 \multirow{3}{*}{10 km/h} & 0 & 20\\\cline{2-3}
  & 5 ms & 8 \\\cline{2-3}
  & 10 ms & 10 \\\hline
\end{tabular}
\label{Tbl_empcorr_esti}
\end{table}

The above results demonstrate that we can obtain up to a 50\%-75\% reduction in complexity for the matrix decomposition method of blindly estimating OFDM symbols when the receiver stores the channel estimate obtained for the first received symbol. Thus, the proposed method removes the need for pilot signals in multiple OFDM symbols from a user for an entire frame (spanning 10ms) in a 5G MIMO system. 

\section{Computational Complexity}\label{complexity}

We analyze the computational complexity of the proposed method, specifically Algorithm \ref{BlindalgoSU} for the single user case, and Algorithm \ref{BlindalgoMU} for the multi-user case, in terms of the parameters of the system, i.e, the OFDM symbol size $N$, the number of receive antennas $N_r$, the number of channel multi-paths $L$, as well as the number of multiplexed users $N_u$. The following analysis is for the regime $N \gg N_{r} \gg L > N_{u}$, since $N$ is in the thousands in 5G NR, $N_r$ is in the hundreds in massive MIMO, $L$ is in the order of tens of taps and determines the channel rank \cite{aswathy2019qr}, and $N_u$ needs to be less than the channel rank for multiplexing users in the same time-frequency resources.

\subsection{Single User Case}

We first derive the complexity of the alternating minimization framework (steps 4 to 7 in Algorithm \ref{BlindalgoSU}) when $N_u = 1$. The complexity matrix pseudo-inverse in step 4 is $\mathcal{O}(LNN_r)$ complex computations, as derived in \cite{aswathy2023ojcoms}. This is the single user channel estimation step, therefore, we denote its complexity as
\begin{equation}
    C_{\mathbf{\hat{H}}}^{SU} \approx \mathcal{O}(LNN_r)  \hspace{3mm} \text{operations.}
\end{equation}
The conversion from the $L\times N_r$ time-domain channel to frequency-domain channel as $\mathbf{B = F_{L}\hat{H}}$ in step 5 involves $LNN_r$ complex operations. The MRC equation in step 7 involves $N_r$ complex operations in the numerator as well as the denominator, which is repeated for the $N$ subcarriers, resulting in a total of $2NN_r$ operations. Therefore, the combination of steps 5-7, which estimates $\mathbf{\hat{X}}$ has a complexity of
\begin{align}
     C_{\mathbf{\hat{X}}}^{SU} &= LNN_r + 2NN_r, \\
     &\approx \mathcal{O}(LNN_r) \hspace{3mm} \text{operations.}
\end{align}
   
The de-rotation component of the algorithm consists of steps 11, 12 and 14. Computation of the scaling factor from one pilot in step 11 is a single complex division operation. Applying the scaling factor to the constellation $\mathbf{\hat{X}}$, the $N$ complex divisions in step 12 can be parallelized as the scaling of each constellation point of $\mathbf{\hat{X}}$ can be performed independently of the others, again adding only one complex operation to the algorithm in terms of computation time. Similarly, the mapping of each point in $\mathbf{\hat{X}}$ to the nearest point in the $M$-QAM constellation in step 14 can also be carried out independently for each point and therefore parallelized. Since the Voronoi diagram for a QAM constellation partitions the complex plane into regular rectangular regions, each mapping in step 14 consists of two complex comparisons with the cell boundaries of the partition. Therefore, we observe that the complexity of the de-rotation component of the algorithm is trivial compared to that of the alternating minimization component. Thus, the complexity of each iteration in Algorithm \ref{BlindalgoSU} is dominated by and can be approximated to $\mathcal{O}(LNN_r)$.

\subsection{Multi-User Case}

We now derive the computational complexity of the proposed method for the multi-user case, given in Algorithm \ref{BlindalgoMU}. The algorithm begins with the channel estimation in step 4 which can be split in a manner similar to the one used in \cite{aswathy2023ojcoms}. This allows us to make use of the diagonal nature of $\mathbf{\hat{X}}$ and $\diag(\mathbf{F_L})$ to cut down on the number of computations involved. We use the notation $\mathbf{\hat{X}}\diag(\mathbf{F_L}) = \mathbf{A}$ to simplify the expression in step 4 as follows:
\begin{equation}
    \mathbf{\hat{H}} = \mathbf{A}^{\dagger}\mathbf{Y_f} 
    = (\mathbf{A}^{H}\mathbf{A} + \mu \mathbf{I_{N_{u}L}})^{-1}(\mathbf{A^{H}Y_f}).
    \label{eq.step4MU}
\end{equation}
The computation in \eqref{eq.step4MU} can be carried out in the following three stages:

\subsubsection{Compute $\mathbf{A}$}

Compute and store the $N\times N_{u}L$ matrix,
\begin{align}
    \mathbf{A} &=\mathbf{\hat{X}}\diag(\mathbf{F_L}), \\ &= [\mathbf{\hat{X}(1)F_L} \hspace{2mm} \mathbf{\hat{X}(2)F_L} \hspace{2mm} \hdots \hspace{2mm} \mathbf{\hat{X}(N_u)F_L}]. 
\end{align} Each $\mathbf{\hat{X}(u)F_L}$ for $u = 1, 2, \hdots, N_u$, involves $NL$ complex operations. Therefore, the total complexity of step 4 in stage 1 for $N_u$ users is
\begin{equation}
    C_{4,1} = N_{u}NL \hspace{3mm} \text{operations.}
\end{equation} 

\subsubsection{Compute the two terms in equation \eqref{eq.step4MU}}

Compute $(\mathbf{A}^{H}\mathbf{A} + \mu \mathbf{I_{N_{u}L}})^{-1}$ and $\mathbf{A^{H}Y_f}$ in parallel to cut down on computation time. 

In the first term, $\mathbf{A^{H}A}$ requires approximately $N_{u}^{2}L^{2}N$ complex operations, adding the regularization term to it involves $N_{u}^{2}L^{2}$ complex additions, and the inverse of the resultant $N_{u}L\times N_{u}L$ matrix involves $\mathcal{O}(N_{u}^{3}L^{3})$ complex operations. Therefore, the complexity of step 4 in stage 2 for computing the first term $(\mathbf{A}^{H}\mathbf{A} + \mu \mathbf{I_{N_{u}L}})^{-1}$ in \eqref{eq.step4MU} is
\begin{equation}
    C_{4,2}(1) = N_{u}^{2}L^{2}N + N_{u}^{2}L^{2} + N_{u}^{3}L^{3} \approx N_{u}^{2}L^{2}N
\end{equation} operations, by approximating to the dominant term, when $N \gg L > N_u$.

The second term in \eqref{eq.step4MU} involves the multiplication of the $N_{u}L\times N$ matrix, $\mathbf{A^{H}}$ and the $N\times N_r$ matrix, $\mathbf{Y_f}$. Therefore, the complexity of step 4 in stage 2 for the second term is 
\begin{equation}
    C_{4,2}(2) = N_{u}LNN_{r} \hspace{3mm} \text{operations.}
\end{equation}
Since the two terms in \eqref{eq.step4MU} are being computed in parallel, the resulting time complexity of step 4 in stage 2 will be
\begin{equation}
    C_{4,2} = \max \{C_{4,2}(1), C_{4,2}(2)\}.
\end{equation}
The quantity $N_{u}L$ can be of the same order as $N_r$ in a massive MIMO setting and we can make the approximation $C_{4,2}(1) \approx C_{4,2}(2)$; thus, both these terms take up approximately the same amount of operations and time when run in parallel. Therefore,
we can approximate
\begin{equation}
    C_{4,2} \approx N_{u}LNN_{r} \hspace{3mm} \text{operations.}
\end{equation}

\subsubsection{Compute $\mathbf{\hat{H}}$}

In the third stage of step 4, multiplication of the first term (an $N_{u}L\times N_{u}L$ matrix) and the second term (an $N_{u}L\times N_{r}$ matrix) in equation \eqref{eq.step4MU} results in a complexity of 
\begin{equation}
    C_{4,3} = N_{u}^{2}L^{2}N_{r} \hspace{3mm} \text{operations.}
\end{equation}

Summing up the number of operations in the above three stages, the total computational complexity of the channel estimation in step 4 is
\begin{align}
    C_{4} &= C_{4,1}+C_{4,2}+C_{4,3}, \\
    &\approx N_{u}NL + N_{u}LNN_{r} + N_{u}^{2}L^{2}N_{r}.
\end{align}
We observe that the dominant term here arises from the second stage. Therefore, we approximate the complexity of the multi-user channel estimation in step 4 to
\begin{equation}
    C_{\mathbf{\hat{H}}}^{MU} \approx \mathcal{O}(N_{u}LNN_{r}).
    \label{eq.C_H}
\end{equation}

Once the time-domain channel is estimated, the frequency-domain channel is calculated in step 5, as $\mathbf{B} = \diag(\mathbf{F_{L})\hat{H}}$, in Algorithm \ref{BlindalgoMU}. For each user, $u \in 1, 2, \hdots, N_{u}$, computing the frequency-domain user channel in $\mathbf{B}$, $\mathbf{F_{L}\hat{H}(u)}$ involves approximately $NLN_r$ complex operations. Thus, the total complexity of step 5 for all the $N_u$ users is
\begin{equation}
    C_{5} = N_{u}NLN_{r} \hspace{3mm} \text{operations.}
\end{equation}

The zero-forcing or MMSE channel equalization in step 7, $\mathbf{x^{T}(n) = y^{T}(n)(B_{n})^{\dagger}}$ for each subcarrier, $n \in 1, 2, \hdots, N$ requires $N_{u}^{2}N_{r}$ complex operations for the $N_{r}\times N_{u}$ matrix pseudo-inverse, $\mathbf{(B_{n})^{\dagger}}$, and $N_{r}N_{u}$ complex operations for the product $\mathbf{y^{T}(n)(B_{n})^{\dagger}}$. Thus, the complexity of step 7 is
\begin{equation}
    C_{7} = N(N_{r}N_{u} + N_{r}N_{u}^{2}) \hspace{3mm} \text{operations.}
\end{equation}

Combining steps 5 and 7, the complexity of the estimation of $\mathbf{\hat{X}}$ for multi-user is
\begin{equation}
    C_{\mathbf{\hat{X}}}^{MU} = NN_{r}(N_{u}L + N_{u} + N_{u}^{2}) \hspace{3mm} \text{operations.}
\end{equation}
Both $L$ and $N_{u}$ can be of the same order (since $L$ determines channel rank and the rank determines the number of multiplexed users), therefore, we can make the following approximation:
\begin{equation}
    C_{\mathbf{\hat{X}}}^{MU} \approx \mathcal{O}(N_{u}LNN_{r}).
    \label{eq.C_X}
\end{equation}

The de-rotation component of the algorithm, in steps 9 to 20 is applied to each user independently. Therefore, similar to the single user case, the additional complexity from the de-rotation part is trivial compared to $C_\mathbf{\hat{H}}^{MU}$ or $C_\mathbf{\hat{X}}^{MU}$. We also observe from \eqref{eq.C_H} and \eqref{eq.C_X} that $C_\mathbf{\hat{H}}^{MU}$ and $C_\mathbf{\hat{X}}^{MU}$ are of the same order, hence the complexity of Algorithm \ref{BlindalgoMU} can be approximated to $\mathcal{O}(N_{u}LNN_{r})$, which is $N_u$ times the approximate complexity of Algorithm \ref{BlindalgoSU} for the single user.

\section{Conclusion}\label{concl}

In this paper, we proposed an almost pilot-free demodulation method for a massive MIMO receiver. This technique blindly decomposes the received OFDM signal into a user signal component and a channel component using an alternating minimization algorithm. The output of the algorithm, which are scaled/rotated versions of the transmitted signal and channel, can be de-rotated using a single pilot. We discussed the appropriate initial point and methods for de-rotating the final outputs. We showed that the proposed method can be adapted to both single-user and multi-user cases and analyzed its performance for different practical scenarios. Link level simulations prove that the method with just 1 pilot per user can match the performance of the conventional method with 25-100$\times$ more pilots for channel estimation, thus reducing the pilot overhead drastically. We also demonstrated how the algorithm can be applied to multiple OFDM symbols received from a mobile user and use the inherent temporal correlations in the channel to greatly reduce its complexity by cutting down on the initialization time as well as decreasing the number of iterations required. Therefore, the proposed method can greatly improve the throughput of massive MIMO communication systems and save precious spectral resources.

\section{Acknowledgements}
We thank Prof. Jeffrey Andrews (University of Texas at Austin) for valuable input to the discussion on temporally correlated MIMO channel models. This work was funded in part by the Ministry of Electronics and Information Technology (MeiTY), Government of India, through the project, “Next Generation Wireless Research and Standardization on 5G and Beyond”, and in part by the Department of Telecommunications (DoT), Government of India, through the 5G testbed project.

\bibliographystyle{IEEEtran}

\bibliography{ref}

\newpage

\end{document}